%% file: main-text.tex
\documentclass[a4paper,11pt]{report}

\usepackage{graphicx,amssymb,amstext,amsmath}
\usepackage{subfigure}
\usepackage[retainorgcmds]{IEEEtrantools}
\usepackage[section] {placeins}
\usepackage{ctable}
\usepackage[cm]{fullpage}
\usepackage{booktabs}
\usepackage{multirow}
\usepackage{tabularx}
\usepackage{hyperref}  % hyperref
\hypersetup{bookmarks=true, pdfauthor={Pei Zhang},colorlinks=true,linkcolor=black,citecolor=blue,breaklinks=true}

\begin{document}

\renewcommand{\thesection}{\Roman{section}}
\renewcommand{\thesubsection}{\thesection.\arabic{subsection}}
\renewcommand{\thesubsubsection}{\thesubsection.\arabic{subsubsection}}
\newcommand{\nocontentsline}[3]{}
\newcommand{\tocless}[2]{\bgroup\let\addcontentsline=\nocontentsline#1{#2}\egroup}

\title{\textbf{Eigenmode Simulations \\of Third Harmonic Superconducting Accelerating Cavities \\for FLASH and the European XFEL}}
%\thanks{Work supported partly by European Commission under FP7 Research Infrastructures grant agreement No.227579}%

\author{Pei~Zhang$^{\dagger \ddagger \ast}$, Nicoleta~Baboi$^\ddagger$, Roger~M.~Jones$^{\dagger \ast}$\\
%\mbox{$^\dagger$The University of Manchester, Manchester, U.K.}\\
\mbox{$^\dagger$School of Physics and Astronomy, The University of Manchester, Manchester, U.K.}\\
\mbox{$^\ddagger$Deutsches Elektronen-Synchrotron (DESY), Hamburg, Germany}\\
\mbox{$^\ast$The Cockcroft Institute, Daresbury, Warrington, U.K.}}

\date{\today}

\maketitle

\begin{abstract}
The third harmonic nine-cell cavity (3.9~GHz) for FLASH and the European XFEL has been investigated using simulations performed with the computer code CST Microwave Studio\textregistered. The band structure of monopole, dipole, quadrupole and sextupole modes for an ideal cavity has been studied. The higher order modes for the nine-cell structure are compared with that of the cavity mid-cell. The $R/Q$ of these eigenmodes are calculated.
\end{abstract}

\renewcommand{\abstractname}{Acknowledgements}

%-----------------------------------------------------------
\tableofcontents
%-----------------------------------------------------------
\include{chap-main}
%-----------------------------------------------------------
%%%%%%%%%%%%%%%%  Section  %%%%%%%%%%%%%%%%%%%%%%%%%
%                                                                                                                                          %
%                                                Summary                                                                         %
%                                                                                                                                          %
%%%%%%%%%%%%%%%%%%%%%%%%%%%%%%%%%%%%%%%%%%%%%
\chapter{Summary}
The passband structure of the third harmonic superconducting cavity has been studied. The monopole, dipole, quadrupole and sextupole modes have been simulated with the CST Microwave Studio\textregistered. The results of the cavity mid-cell have been related to the nine-cell results and the consistency has been observed. The intention of this report is to provide a guide for the electromagnetic mode distributions of the third harmonic cavity at FLASH and for the European XFEL.
%-----------------------------------------------------------
%%%%%%%%%%%%%%%% %%%%%%%%%%%%%%%%%%%%%%%%%%%%%
%                                                                                                                                          %
%                                                Acknowledgments                                                                 %
%                                                                                                                                          %
%%%%%%%%%%%%%%%%%%%%%%%%%%%%%%%%%%%%%%%%%%%%%
%-----------------------------------------------------------
\begin{abstract}
We would like to thank Dr.~Ian~Shinton for the useful discussions and Dr.~Martin~Dohlus for carefully reading this manuscript. This work received support from the European Commission under the FP7 Research Infrastructures grant agreement No.227579.
\end{abstract}
%-----------------------------------------------------------
\addcontentsline{toc}{chapter}{\numberline{}Bibliography}
\bibliography{refs}             % this causes the references to be listed
\bibliographystyle{ieeetr}

%%%%%%%%%%%%%%%%  Section  %%%%%%%%%%%%%%%%%%%%%%%%%
%                                                                                                                                          %
%                                                Appendices                                                                         %
%                                                                                                                                          %
%%%%%%%%%%%%%%%%%%%%%%%%%%%%%%%%%%%%%%%%%%%%%
\appendix
\include{app-list}

\end{document}

%% file: chap-main.tex
%%%%%%%%%%%%%%%  Section 0  %%%%%%%%%%%%%%%
%                                                                                                           %
%                                            Introduction                                              %
%                                                                                                            %
%%%%%%%%%%%%%%%%%%%%%%%%%%%%%%%%%%%%
\chapter{Introduction}\label{sec:intro}
FLASH \cite{flash-1,flash-2} is a free-electron laser facility at DESY. It uses ultra-short electron bunches with high peak current to generate high brilliance coherent light pulses. FLASH is a user facility for photon science and a test facility for various accelerator studies. The beam is accelerated by superconducting TESLA 1.3~GHz cavities \cite{tesla-1,tesla-2}. Third harmonic 3.9~GHz cavities \cite{acc39-2} are used to linearize the curvature of bunch's energy spread caused by the sinusoidal 1.3~GHz RF field \cite{acc39-0}.
% ACC39 
\begin{figure}[h]\center
\includegraphics[width=0.45\textwidth]{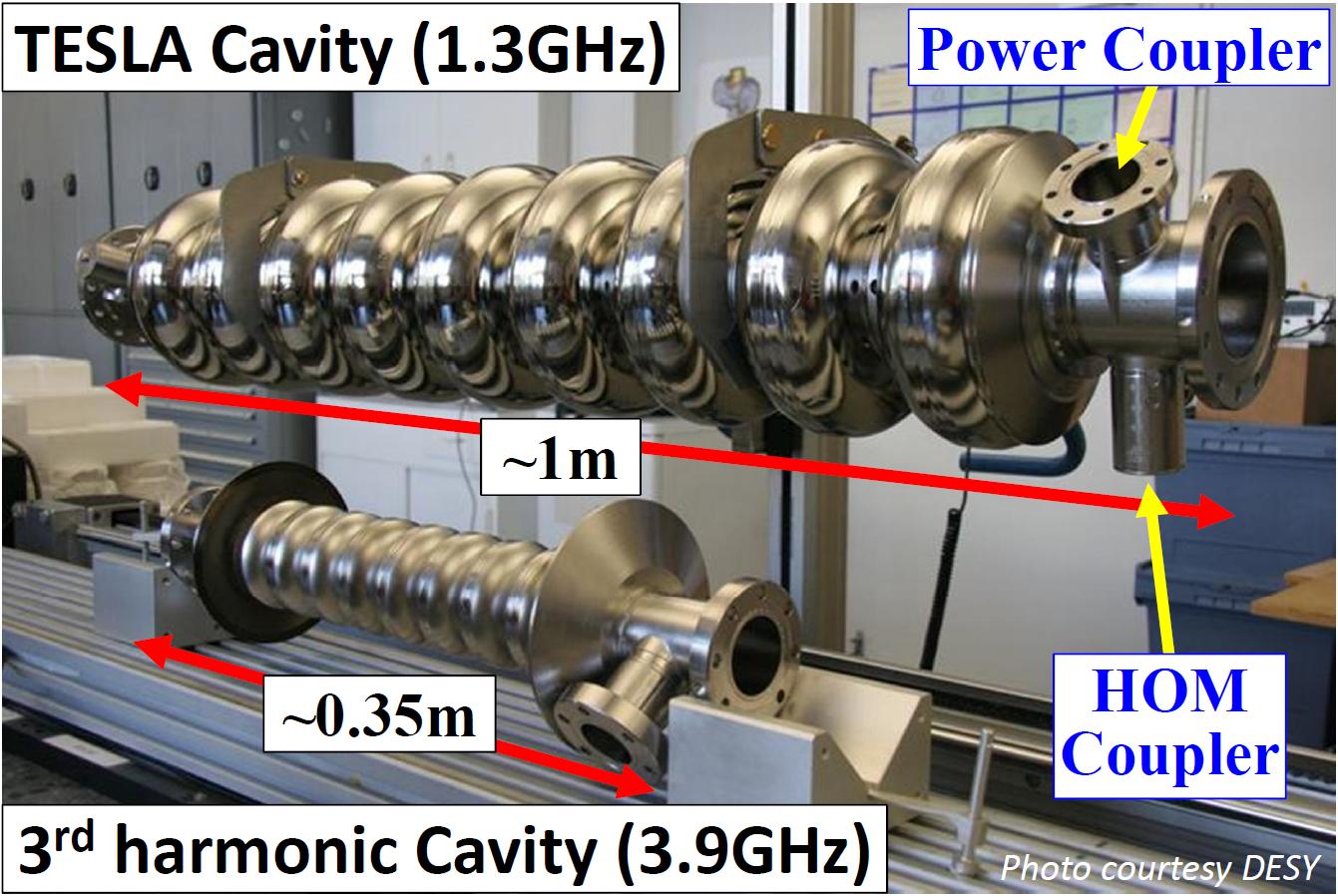}
\caption{A TESLA style cavity operating at 1.3~GHz (top) and the corresponding third harmonic cavity (bottom).}
\label{acc39-acc1}
\end{figure}

\begin{figure}[h]\centering
\includegraphics[width=0.7\textwidth]{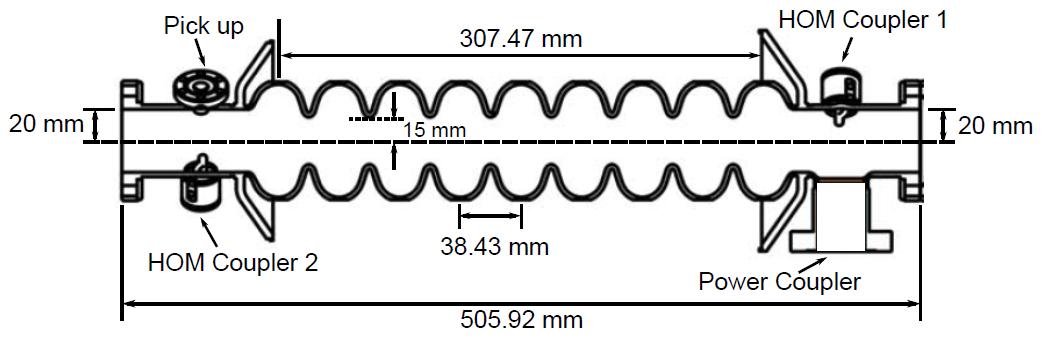}
\caption{Schematic of a third harmonic cavity with one power coupler, one pick up probe and two HOM couplers.}
\label{acc39-cavity}
\end{figure}

\begin{figure}[h]\center
\includegraphics[width=0.9\textwidth]{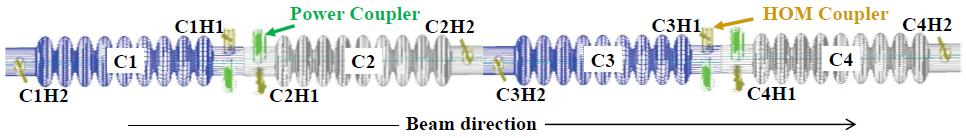}
\caption{Schematic of the four cavities within ACC39 module.}
\label{cavity-cartoon}
\end{figure}

The third harmonic cavity inherits a similar design of to the 1.3~GHz TESLA cavity (with some modif{}ications) as shown in Fig.~\ref{acc39-acc1}. A schematic of the third harmonic cavity is illustrated in Fig.~\ref{acc39-cavity} along with the main dimensions. It has one power coupler and one pick-up probe installed on the beam pipe connecting end-cells. It is also equipped with two higher order mode (HOM) couplers installed on each side of the connecting beam pipes with dif{}ferent rotations and dif{}ferent designs \cite{acc39-3}. In FLASH there are four 3.9~GHz cavities as shown in Fig.~\ref{cavity-cartoon}.
% acc39 cavity drawing

% 4 cavity string cartoon
The wakef{}ields in the third harmonic cavity are signif{}icantly stronger than those in the 1.3~GHz TESLA cavity due to a much smaller iris radius: 15~mm compared with 35~mm \cite{tesla-2}. Unlike the 1.3~GHz TESLA cavity, most HOMs in the third harmonic cavity are above the cutof{}f frequencies of the connecting beam pipes in order to achieve a better damping of the HOMs \cite{acc39-4}. However, this allows HOMs to propagate amongst cavities in the module, and thus gives rise to a dense coupled modal spectrum in the third harmonic cavity.

In this report, the third harmonic cavity is firstly treated as a periodic structure with an infinite number of repetitions of the mid-cell. The dispersion curves of monopole, dipole, quadrupole and sextupole passbands are described in Chapter~\ref{sec:mid-cell}. The beam pipes connecting the cavities are modeled as circular waveguides and described in Chapter~\ref{sec:bp-guide}. The eigenmodes obtained for an ideal third harmonic cavity without couplers are presented in Chapter~\ref{sec:eigen-39cav}. A list of modes simulated up to 11~GHz for an ideal third harmonic cavity is shown in Appendix~\ref{sec:mode-list}. The parameters used in the CST Microwave Studio{\textregistered} for these eigenmode simulations are listed in Appendix~\ref{sec:para-list}. Extensive electric field distributions for monopole, dipole, quadrupole and sextupole modes with both electric (EE) and magnetic (MM) boundary conditions are presented in Appendix~\ref{sec:dist}.   

%%%%%%%%%%%%%%%  Section 1  %%%%%%%%%%%%%%%
%                                                                                                           %
%                                 The passband of the mid-cell                                 %
%                                                                                                            %
%%%%%%%%%%%%%%%%%%%%%%%%%%%%%%%%%%%%
\chapter{The Third Harmonic Cavity as a Periodic Structure}\label{sec:mid-cell}
\begin{figure}[h]
\centering
\includegraphics[width=0.6\textwidth]{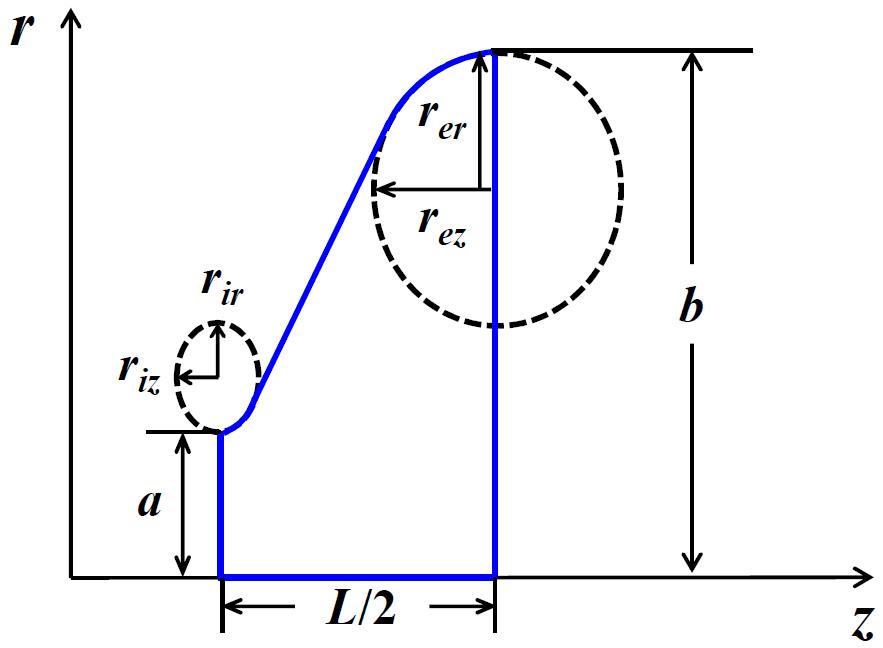}
\caption{Parameterization of cell geometry. The blue curve represents the cell wall.}
\label{simu-cell-geo}
\end{figure}

\begin{figure}[h]
\centering
\subfigure[Mid-cell]{
\includegraphics[width=0.2\textwidth]{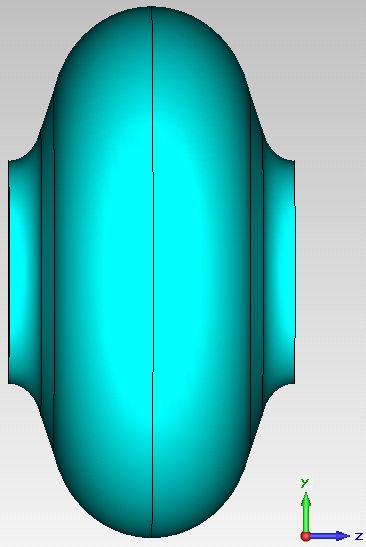}
\label{1cell-structure}
}
\quad\quad
\subfigure[Mesh view]{
\includegraphics[width=0.2\textwidth]{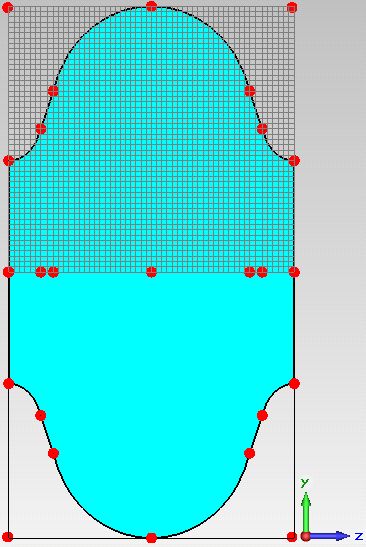}
\label{1cell-mesh}
}
\quad\quad
\subfigure[$\pi$ mode]{
\includegraphics[width=0.221\textwidth]{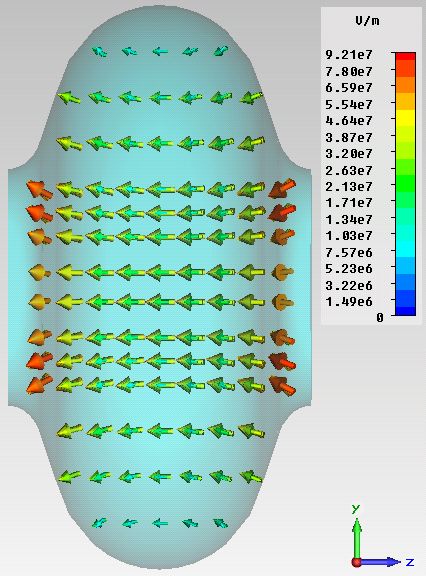}
\label{1cell-E-field}
}
\caption{The mid-cell of the third harmonic cavity as modeled in CST Microwave Studio\textregistered. (c) shows the accelerating mode with a phase advance of 180 degrees per cell ($\pi$ mode).}
\label{simu-1cell}
\end{figure}

\begin{figure}[h]
\centering
\subfigure[Monopole]{
\includegraphics[width=0.45\textwidth]{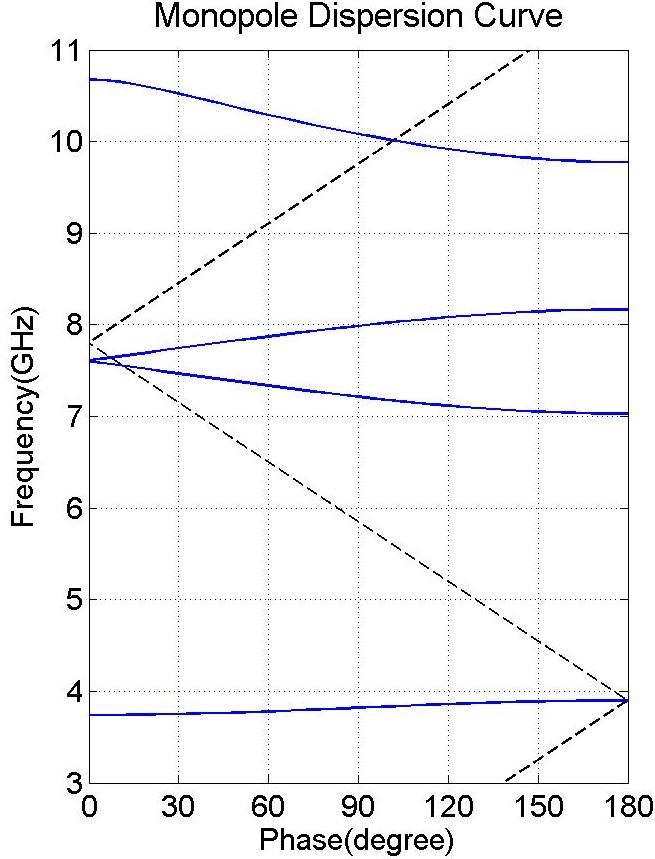}
\label{disp-1cell-mono}
}
\quad
\subfigure[Dipole]{
\includegraphics[width=0.45\textwidth]{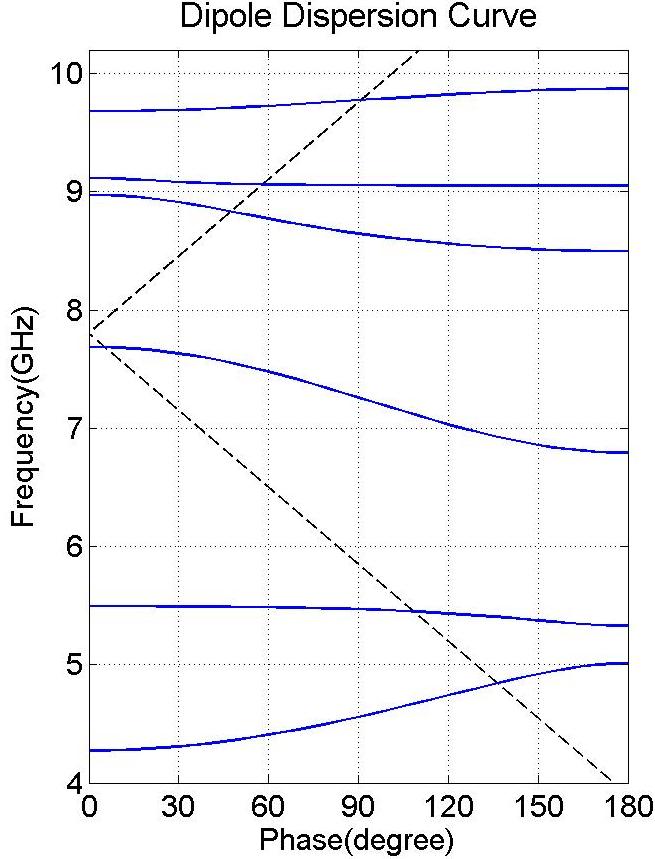}
\label{disp-1cell-dipole}
}
\subfigure[Quadrupole]{
\includegraphics[width=0.45\textwidth]{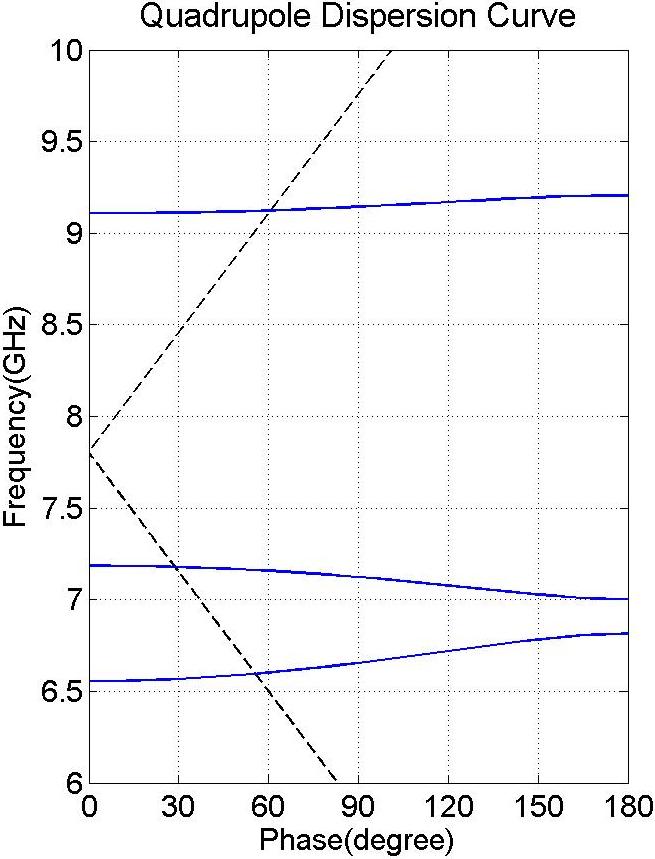}
\label{disp-1cell-quad}
}
\quad
\subfigure[Sextupole]{
\includegraphics[width=0.45\textwidth]{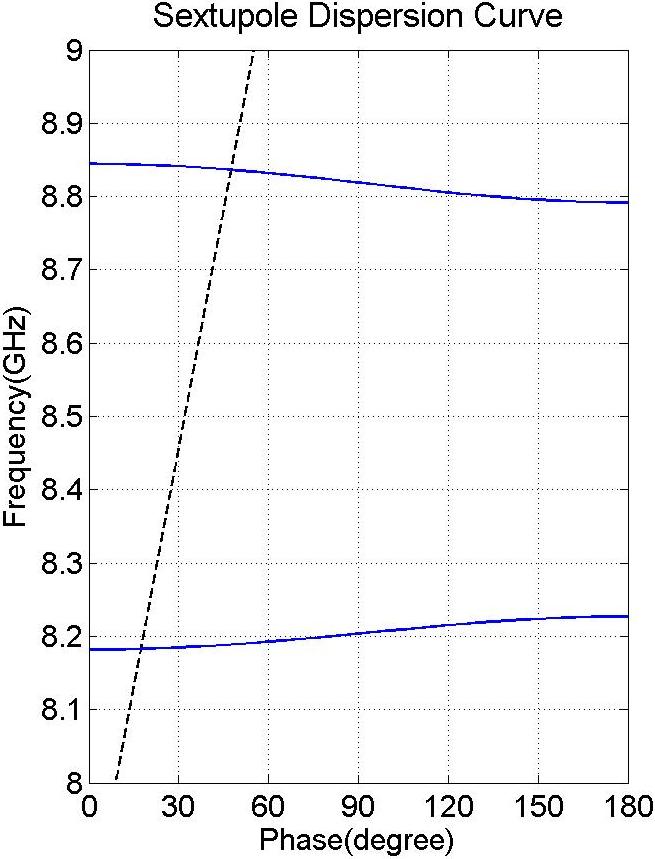}
\label{disp-1cell-sextu}
}
\caption{The band structure (blue) of a 3.9~GHz cavity mid-cell. The light line is dashed.}
\label{disp-1cell}
\end{figure}

A sketch of the cell geometry is given in Fig.~\ref{simu-cell-geo} for the third harmonic cavity. The cell is rotationally symmetric around the $z$ axis. The iris and the equator both have an elliptical shape. The mid-cells have different shape from the end-cells, and the parameters are listed in Table~\ref{table-cell-geo}. The iris of the end-cup is larger than that of the mid-cell. Fig.~\ref{1cell-structure} shows a mid-cell built in CST Microwave Studio\textregistered \cite{cst}. A hexahedral mesh was used in the calculation of the electromagnetic field as shown in Fig.~\ref{1cell-mesh}. The mesh lines were chosen such that the iris radius and the equator radius were exactly matched by mesh lines. Symmetry planes were applied on the structure to save simulation time so that only a quarter of the structure was simulated. Approximately 130,000 mesh cells for a quarter of the structure and a maximum mesh step of 0.85~mm were set. Electric (EE) boundary conditions were used on the surface of the mid-cell, while periodic boundary conditions were set on both ends of the cell. The modes of an infinite periodic chain of cavities can be obtained from single cell calculations using periodic boundary conditions:
% equation: floquet
\begin{equation}
\mathbf{E}(r,z+L)=\mathbf{E}(r,z)e^{i\phi},
\label{eq:floquet}
\end{equation}
where $\phi$ is the phase advance per cell, and $L$ is the cell length. Fig.~\ref{1cell-E-field} shows the electric field of a mode with a phase advance of 180 degrees (or $\pi$) per cell. The frequencies of several passbands are shown in the form of dispersion curves in Fig.~\ref{disp-1cell} for monopole, dipole, quadrupole and sextupole modes.
% cell geometry

% table: cavity geometry
\begin{table}[h]
\centering
\caption{Parameters of the cell geometry of the third harmonic cavity.}
\begin{tabular}{l|c|c|c}
%\toprule
\hline \hline 
& & mid-cell & end-cell \\
\hline 
Iris radius, \textbf{\textit{a}} & mm & 15.0 & 20.0 \\
\hline 
Equator radius, \textbf{\textit{b}} & mm & 35.787 & 35.787 \\
\hline 
Half cell length, \textbf{\textit{L}/2} & mm & 19.2167 & 19.2167 \\
\hline
Equator horizontal axis, \boldmath$r_{ez}$ & mm & 13.6 & 14.4 \\
\hline
Equator vertical axis, \boldmath$r_{er}$ & mm & 15.0 & 15.0 \\
\hline
Iris horizontal axis, \boldmath$r_{iz}$ & mm & 4.5 & 4.5 \\
\hline
Iris vertical axis, \boldmath$r_{ir}$ & mm & 6.0 & 6.0 \\
\hline \hline 
\end{tabular}
\label{table-cell-geo}
\end{table}

% 1-cell structure

% 1-cell dispersion (mono)(dipole)

\begin{figure}[h]
\centering
\includegraphics[width=0.9\textwidth]{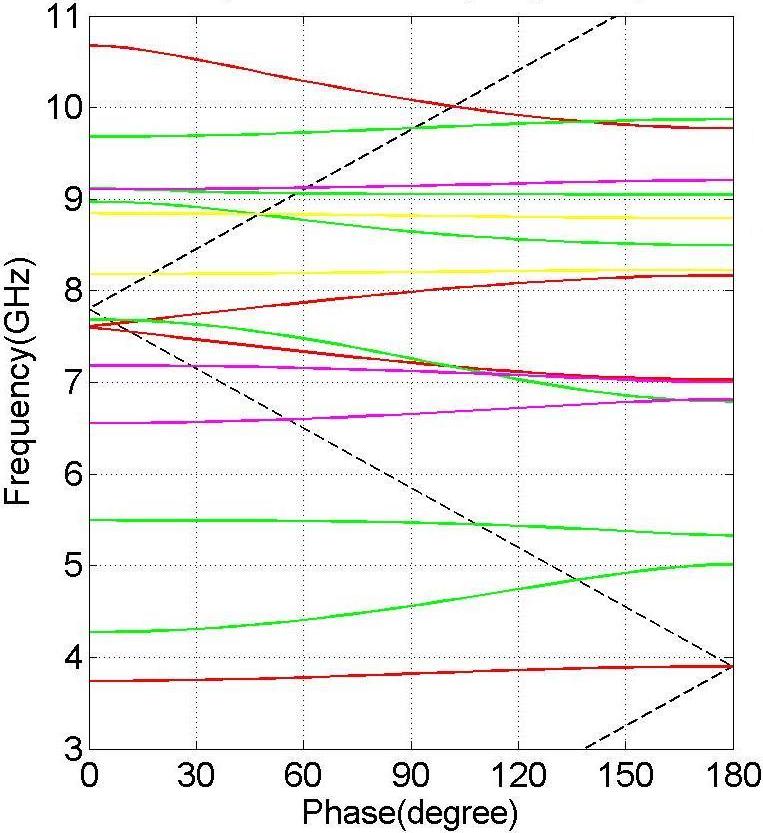}
\caption{Dispersion curve for monopole (red), dipole (green), quadrupole (magenta) and sextupole (yellow) modes. The light line is dashed.}
\label{disp-1cell-all}
\end{figure}

A beam excites strongest those modes which are synchronous to the beam, i.e. with a phase velocity equal to the speed of the accelerated particles. For FLASH, this is the speed of light:
% equation: light-line
\begin{equation}
c=v_{phase}=\frac{\omega}{k_z}=2\pi L\frac{f}{\phi},
\label{eq:light-line-1}
\end{equation}
where $k_z$ is the longitudinal wave number, $\phi=k_z L$ is the phase advance per cell, which is used as an horizontal axis in the plots of the dispersion curves. The light line is therefore the straight line:
% equation: light-line(straight line)
\begin{equation}
f(\phi)=\frac{c}{2\pi L}\phi.
\label{eq:light-line-2}
\end{equation}
It is folded into the phase range from 0 to 180 degrees in the dispersion plots due to the periodicity of the structure. By design, the light line intersects the $\pi$ mode of the first monopole passband (frequency $\approx$ 3.9~GHz), which is used for particle acceleration. Fig.~\ref{disp-1cell-all} show the dispersion curves for the monopole, dipole, quadrupole and sextupole bands up to 11~GHz.
% 1-cell dispersion (all bands)
\newpage
%%%%%%%%%%%%%%%  Section 2  %%%%%%%%%%%%%%%
%                                                                                                           %
%                            The Beam Pipe as a Circular Waveguide               		  %
%                                                                                                            %
%%%%%%%%%%%%%%%%%%%%%%%%%%%%%%%%%%%%
\chapter{The Beam Pipe as a Circular Waveguide}\label{sec:bp-guide}
The third harmonic cavities are connected with beam pipes. To study the propagation of the modes amongst cavities, the beam pipes are treated as circular waveguides. Generally, the TE and TM modes can be distinguished from the characterization of the electric and magnetic fields. The cutoff frequencies of the TE and TM modes for a circular waveguide are \cite{tesla-2}:
% equation: cutoff
\begin{subequations}
\label{eq:cutoff}
\begin{eqnarray}
f_{c_{mn}}^{TM}=c\frac{p_{mn}}{2\pi a},\label{eq:cutoff-TM}
\\
f_{c_{mn}}^{TE}=c\frac{p'_{mn}}{2\pi a},\label{eq:cutoff-TE}
\end{eqnarray}
\end{subequations}
where $m$=0, 1, 2, 3 corresponds to monopole, dipole, quadrupole and sextupole modes, $p_{mn}$ is the $n^{th}$ root of the $m^{th}$ Bessel function $J_m$, $p'_{mn}$ is the $n^{th}$ root of the derivative of the $m^{th}$ Bessel function $J'_m$, $a$ is the radius of the waveguide. The first TE mode to propagate is the mode with the smallest $p'_{mn}$, which from Table~\ref{table-bassel} is seen to be TE$_{11}$ mode. The first TM mode to propagate is then the TM$_{01}$ mode. The cutoff frequencies of TE$_{11}$ and TM$_{01}$ modes are listed in Table~\ref{table-cutoff} for a circular waveguide with a radius of 15~mm and of 20~mm. These correspond to the iris radius of a mid-cell and an end-cell of a third harmonic cavity (see Table~\ref{table-cell-geo}). By choosing a radius much larger than 1/3 of that of the 1.3~GHz cavity, the cutoff frequency is lowered so that most higher order modes propagate amongst cavities and therefore are better damped.
% table-Bessel
\begin{table}[h]
\centering
\caption{Values of $p_{mn}$ and $p'_{mn}$ \cite{tesla-2,rf-1}.}
\begin{tabular}{c|c|c|c|c||c|c|c}
\hline \hline
& & \multicolumn{3}{c||}{$p_{mn}$ (TM modes)} & \multicolumn{3}{c}{$p'_{mn}$ (TE modes)}\\ 
\hline
& m & n=1 & n=2 & n=3 & n=1 & n=2 & n=3 \\
\hline
monopole & 0 & \textbf{2.405} & 5.520 & 8.654 & 3.832 & 7.016 & 10.174\\
\hline
dipole & 1 & 3.832 & 7.016 & 10.174 & \textbf{1.841} & 5.331 & 8.536 \\
\hline
quadrupole & 2 & 5.136 & 8.417 & 11.620 & 3.054 & 6.706 & 9.970 \\
\hline
sextupole & 3 & 6.380 & 9.761 & 13.015 & 4.201 & 8.015 & 11.346 \\
\hline \hline 
\end{tabular}
\label{table-bassel}
\end{table}

% table-cutoff
\begin{table}[h]
\centering
\caption{Cutoff frequencies for the lowest order of TE and TM modes in a circular waveguide with a radius of 15~mm and 20~mm.}
\begin{tabular}{c|c|c}
\hline \hline 
& $\mathbf{a}$=15~mm & $\mathbf{a}$=20~mm \\
\hline 
$f_{c_{11}}^{TE}$ (GHz) & 5.86& 4.39\\
\hline
$f_{c_{01}}^{TM}$ (GHz) & 7.65& 5.74\\
\hline \hline 
\end{tabular}
\label{table-cutoff}
\end{table}

\newpage

%%%%%%%%%%%%%%%  Section 3  %%%%%%%%%%%%%%%
%                                                                                                           %
%                            Eigenmodes in an ideal 9-cell 3.9~GHz cavity               %
%                                                                                                            %
%%%%%%%%%%%%%%%%%%%%%%%%%%%%%%%%%%%%
\chapter{Eigenmodes in the Ideal Third Harmonic Cavities}\label{sec:eigen-39cav}
\begin{figure}[h]
\centering
\includegraphics[width=0.92\textwidth]{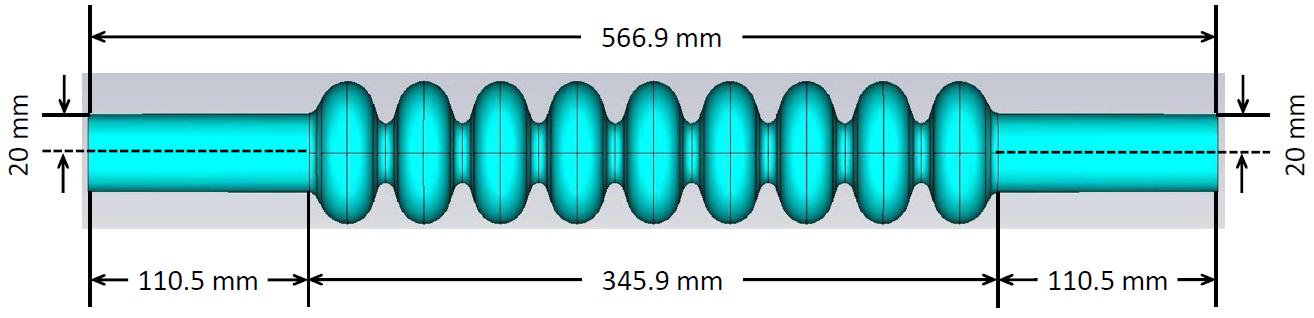}
\caption{CST Microwave Studio{\textregistered} generated geometry of the third harmonic cavity.}
\label{simu-cav-geo}
\end{figure}

\begin{figure}[h]
\centering
\includegraphics[width=0.7\textwidth]{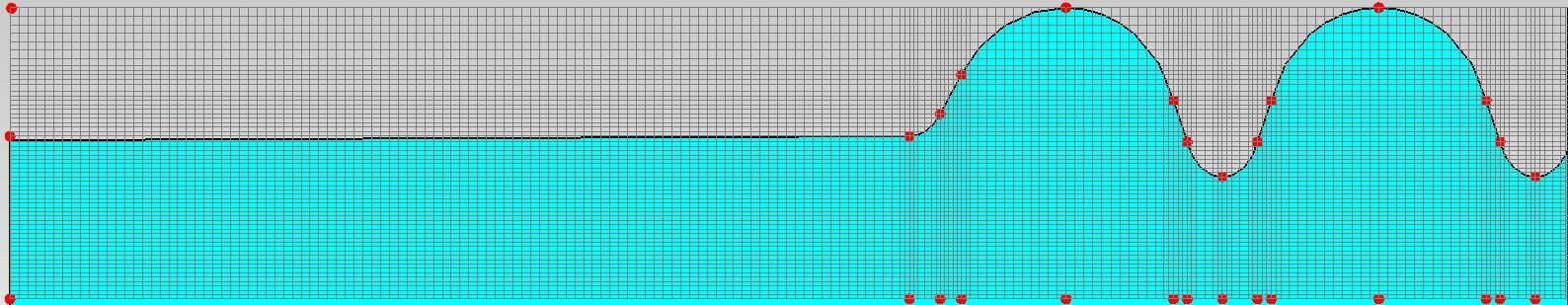}
\caption{A typical mesh used for selected cells in the third harmonic cavity.}
\label{simu-9cell-mesh}
\end{figure}

The geometry of an ideal third harmonic cavity without couplers as modeled with CST Microwave Studio\textregistered \cite{cst} is shown in Fig.~\ref{simu-cav-geo}. The shape of an individual mid-cell is shown in Fig.~\ref{1cell-structure} and the parameters are listed in Table~\ref{table-cell-geo}. The end-cups have an increased iris radius (20~mm) and are connected with beam pipes at both ends. The simulations were conducted with the Eigenmode Solver of the CST Microwave Studio\textregistered. A solver accuracy of 10$^{-6}$ in terms of the eigensystem's relative residual was used. The cavity geometry was approximated by hexahedral mesh cells. As shown in Fig.~\ref{simu-9cell-mesh}, the mesh lines were chosen such that the iris radius and the equator radius were exactly matched by mesh lines. A quarter of the structure with symmetry planes was used in order to reduce the simulation time. For the accelerating mode, a maximum mesh step of 1.1~mm, corresponding to approximately 2.1~million mesh cells for a quarter of the structure, was used. Electric (EE) boundary conditions were used. 
% cavity geometry
% mesh-9cell

\begin{figure}[h]
\centering
\includegraphics[width=0.9\textwidth]{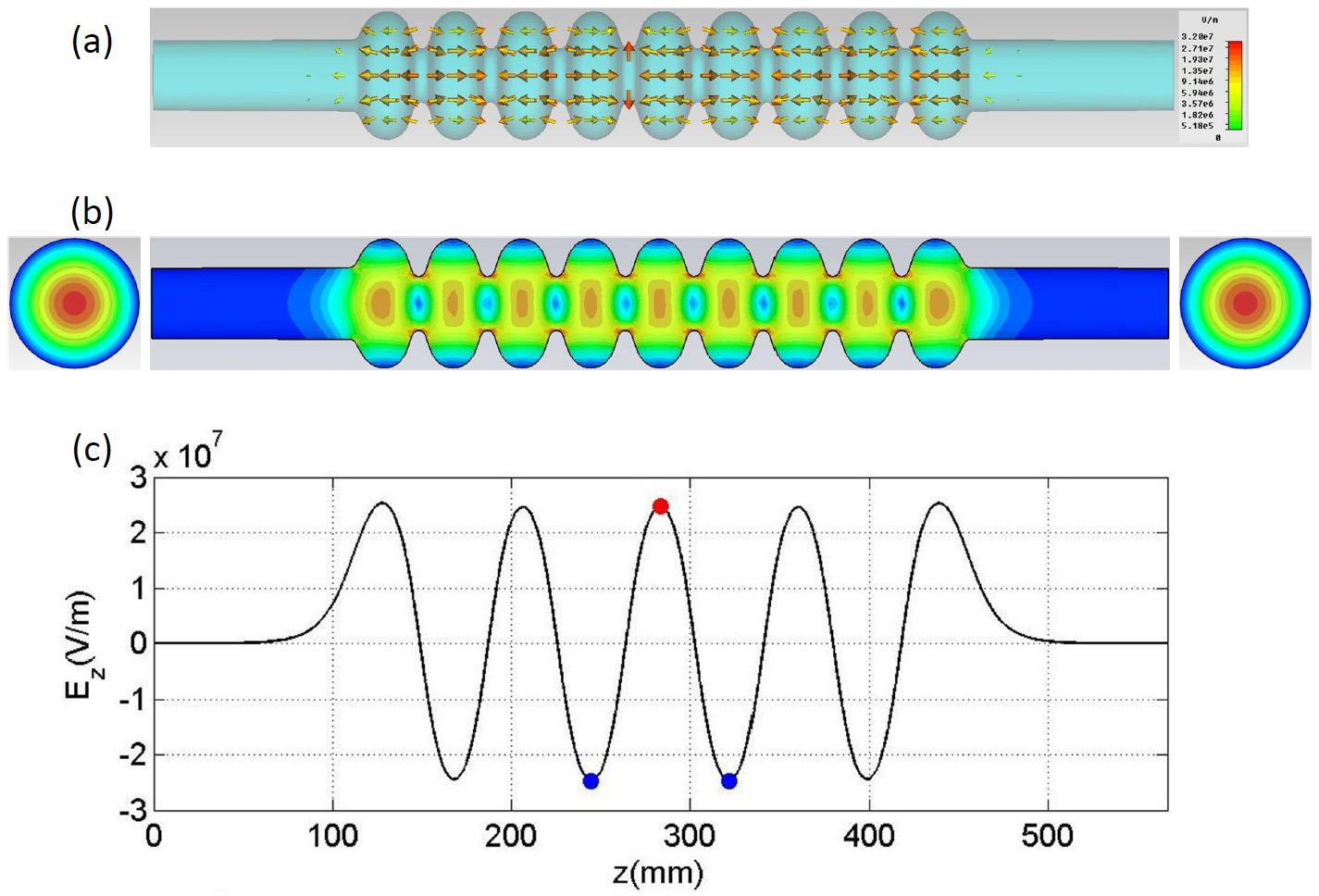}
\caption{(a) The electric field (arrows) of the accelerating mode (3.9~GHz) in the third harmonic cavity. (b) The electric field magnitude of the accelerating mode (frequency: 3.9008~GHz, $R/Q$: 373.113~$\Omega$). Electric (EE) boundary conditions were used in the simulation. (c) Longitudinal electric field $E_z$ of the accelerating mode (3.9~GHz, $\pi$ mode) on the axis of a third harmonic cavity. The red and blue dots are corresponding to the position marked in Fig.~\ref{phase-adv} for $E_L$, $E_M$ and $E_R$ respectively.}
\label{simu-EM-M1-9}
\end{figure}

The electric field for the accelerating mode (3.9~GHz) is shown in Fig.~\ref{simu-EM-M1-9}(a)(b). The $R/Q$ is defined in \cite{tesla-2} and the unit is [$\Omega/cm^2$ per cavity] throughout this report, while the ``per cavity'' is often omitted. The longitudinal component of the electric field of the accelerating mode on the axis of the third harmonic cavity is shown in Fig.~\ref{simu-EM-M1-9}(c). A good field flatness can be observed.
% E-field (M1-9-EE)

\begin{figure}[h]
\centering
\includegraphics[width=0.8\textwidth]{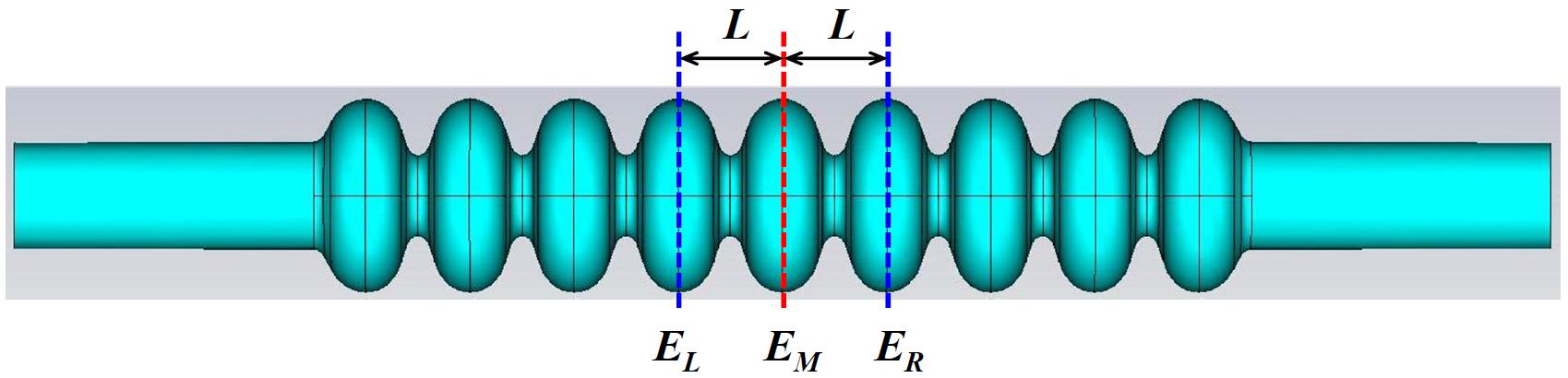}
\caption{Calculation of the phase advance per cell. $E_L$, $E_M$ and $E_R$ are longitudinal electric field at certain positions.}
\label{phase-adv}
\end{figure}

The phase advance per cell can be calculated using the electric field determined from the simulations. Based on Eq.~\ref{eq:floquet} for periodic structures, the phase advance per cell can be derived as
% equation: phase-adv (derivation)
\begin{equation}
E_M=E_z(r,z),
\label{eq:phase-adv-1}
\end{equation}
\begin{equation}
E_L=E_Me^{-i\phi}=E_z(r,z-L),
\label{eq:phase-adv-2}
\end{equation}
\begin{equation}
E_R=E_Me^{i\phi}=E_z(r,z+L),
\label{eq:phase-adv-3}
\end{equation}
\begin{equation}
E_L+E_R=2E_Mcos(\phi),
\label{eq:phase-adv-4}
\end{equation}
where $E_z(r,z)$ is the longitudinal electric field obtained from the simulations, $E_L$, $E_M$ and $E_R$ are defined as shown in Fig.~\ref{phase-adv}. Then the phase advance per cell can be calculated as
% final equation: phase advance
\begin{equation}
\phi=arccos(\frac{E_L+E_R}{2E_M}).
\label{eq:phase-adv-5}
\end{equation}
The phase advance per cell calculated for the accelerating mode is 177 degrees. This reflects the field flattness that is adjusted by the geometry of the end-cells.  %The mode frequency also deviates slightly from that of the 180 degrees obtained from the periodic solution. 
% phase-adv

\begin{figure}[h]
\centering
\subfigure[Monopole]{
\includegraphics[width=0.45\textwidth]{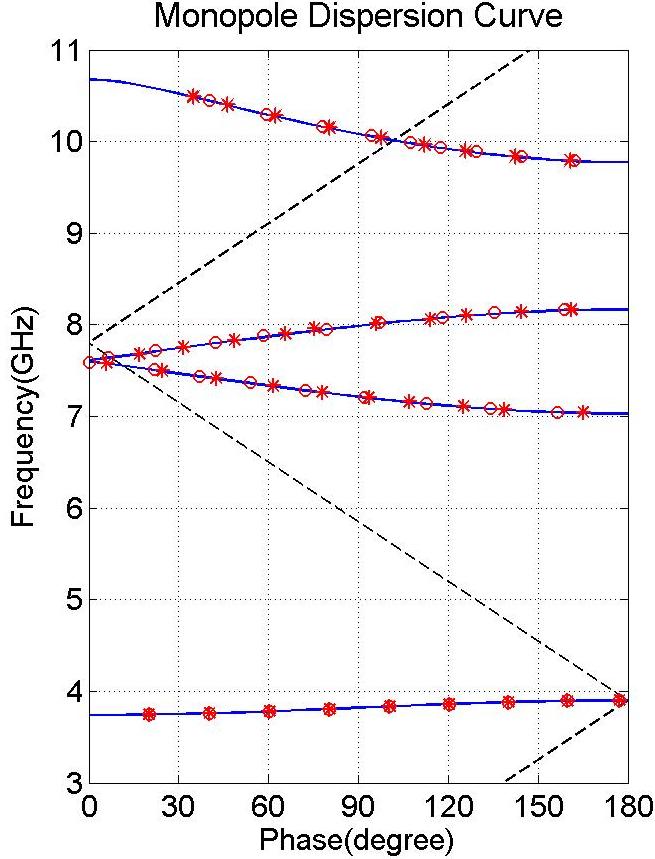}
\label{disp-1cell-mono-9cell}
}
\quad
\subfigure[Dipole]{
\includegraphics[width=0.45\textwidth]{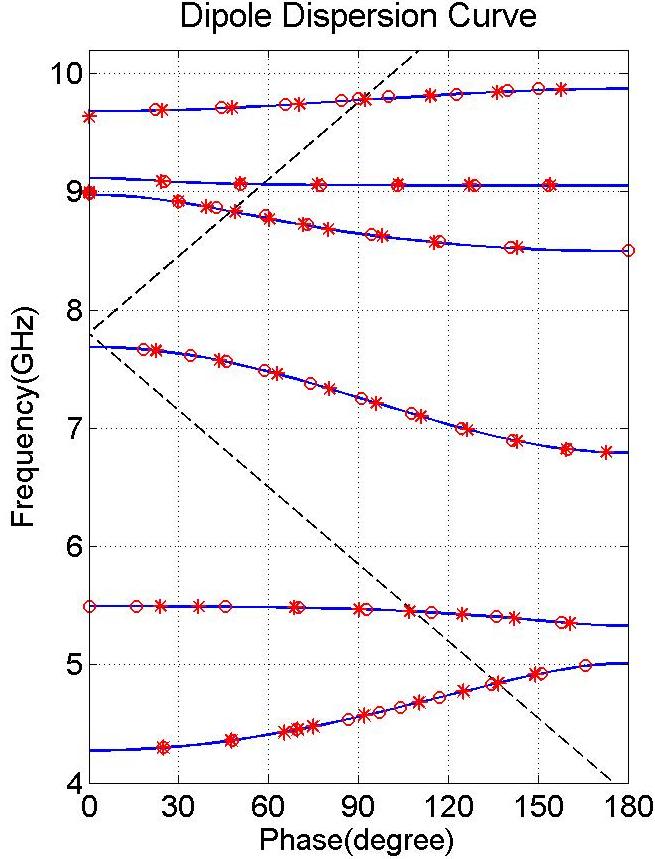}
\label{disp-1cell-dipole-9cell}
}
\subfigure[Quadrupole]{
\includegraphics[width=0.45\textwidth]{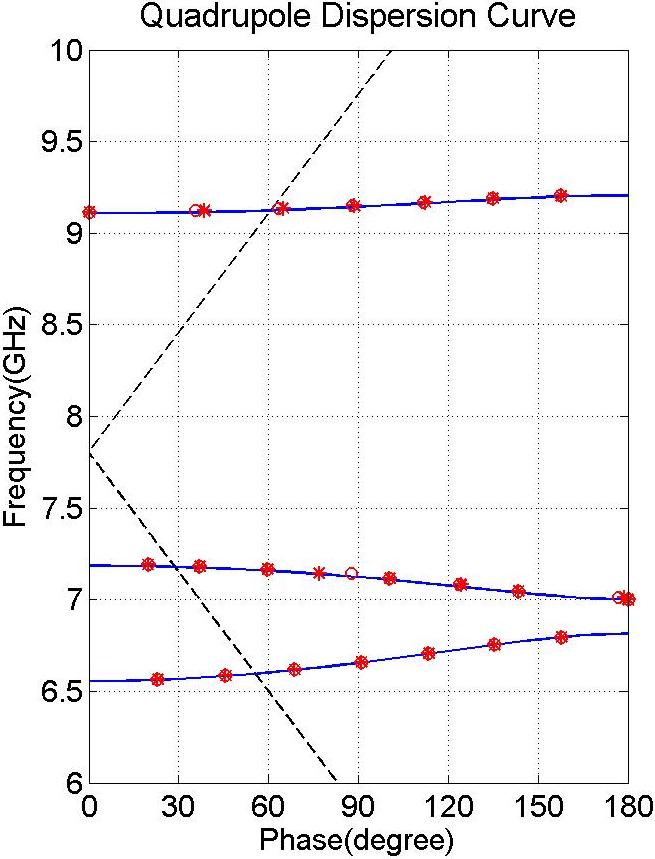}
\label{disp-1cell-quad-9cell}
}
\quad
\subfigure[Sextupole]{
\includegraphics[width=0.45\textwidth]{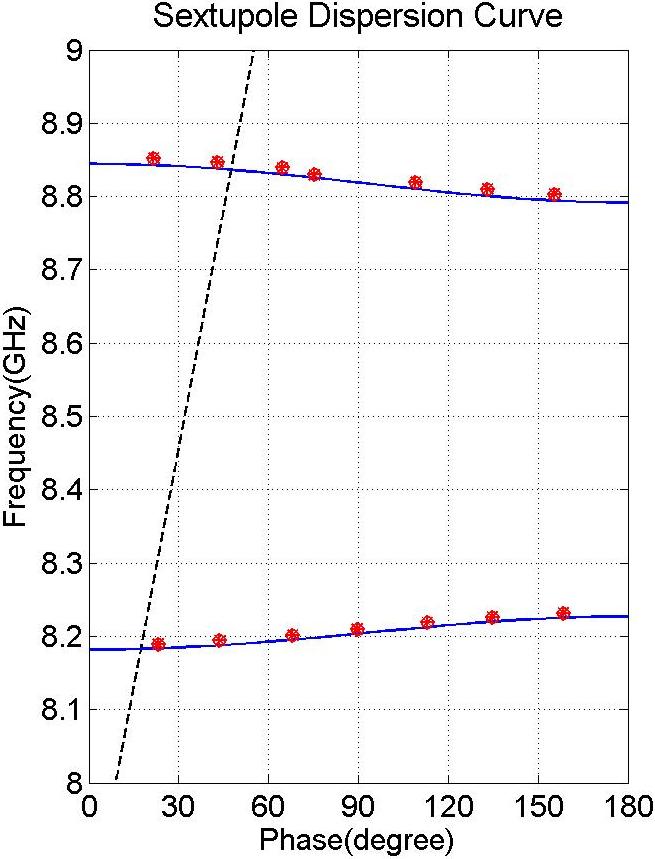}
\label{disp-1cell-sextu-9cell}
}
\caption{Monopole, dipole, quadrupole and sextupole band structure (blue) of a mid-cell and the modes in an ideal 9-cell 3.9~GHz cavity. The circles represent the modes calculated with electric (EE) boundary conditions and the asterisks represent magnetic (MM) boundary conditions. The light line is dashed.}
\label{disp-1cell-9cell}
\end{figure}

\begin{figure}[h]
\centering
\includegraphics[width=0.8\textwidth]{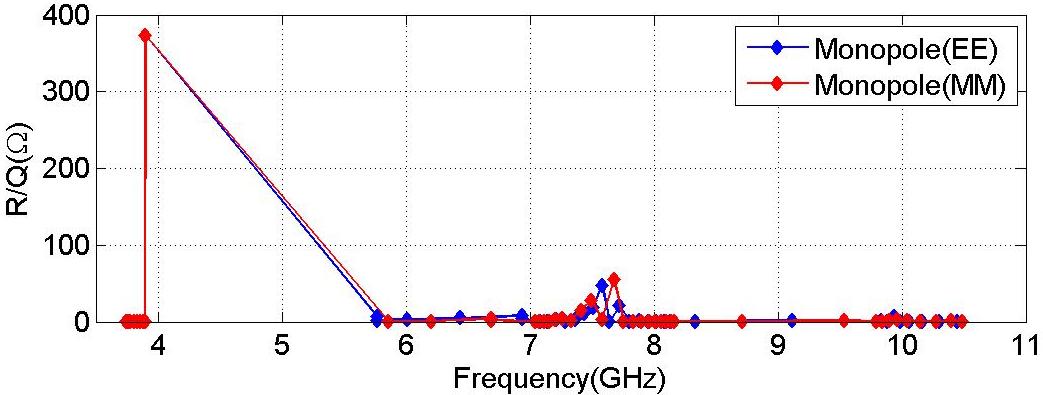}
\caption{The $R/Q$ parameter for monopole modes of a 9-cell third harmonic cavity plotted versus the frequency of the mode. The circles in blue represent the modes calculated with electric (EE) boundary conditions and the asterisks in red represent magnetic (MM) boundary conditions.}
\label{simu-mono}
\end{figure}

\begin{figure}[h]
\centering
\includegraphics[width=0.8\textwidth]{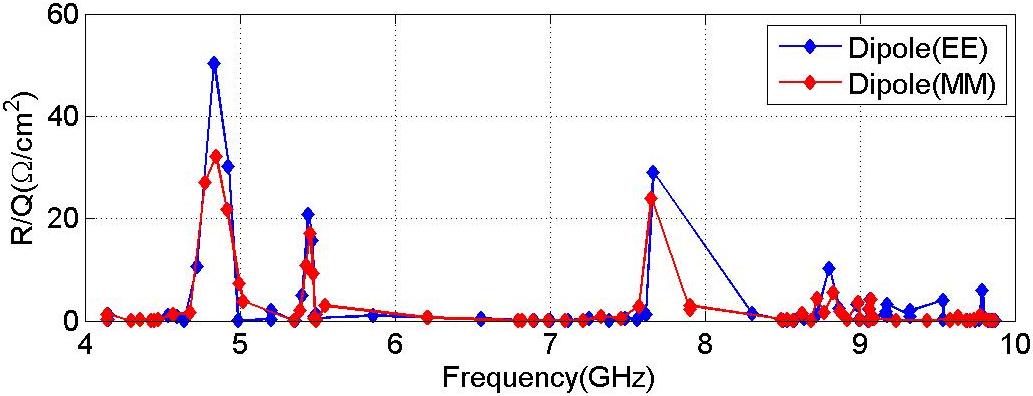}
\caption{The $R/Q$ parameter for dipole modes of a 9-cell third harmonic cavity plotted versus the frequency of the mode. The circles in blue represent the modes calculated with electric (EE) boundary conditions and the asterisks in red represent magnetic (MM) boundary conditions.}
\label{simu-dipole}
\end{figure}

\begin{figure}[h]
\centering
\includegraphics[width=0.8\textwidth]{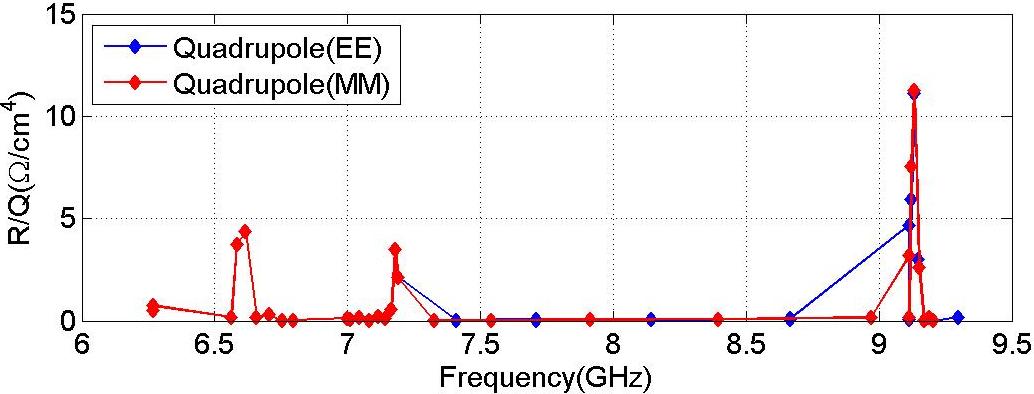}
\caption{The $R/Q$ parameter for quadrupole modes of a 9-cell third harmonic cavity plotted versus the frequency of the mode. The circles in blue represent the modes calculated with electric (EE) boundary conditions and the asterisks in red represent magnetic (MM) boundary conditions.}
\label{simu-quad}
\end{figure}

\begin{figure}[h]
\centering
\includegraphics[width=0.8\textwidth]{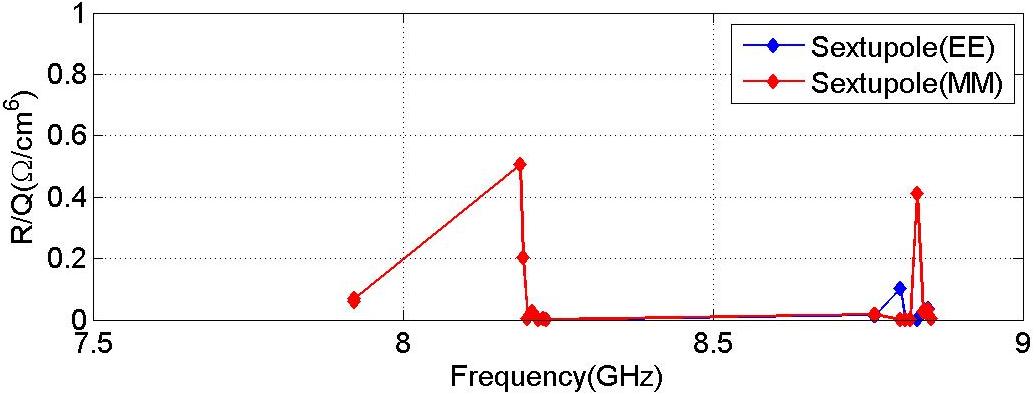}
\caption{The $R/Q$ parameter for sextupole modes of a 9-cell third harmonic cavity plotted versus the frequency of the mode. The circles in blue represent the modes calculated with electric (EE) boundary conditions and the asterisks in red represent magnetic (MM) boundary conditions.}
\label{simu-sextu}
\end{figure}

\begin{figure}[h]
\centering
\includegraphics[width=0.8\textwidth]{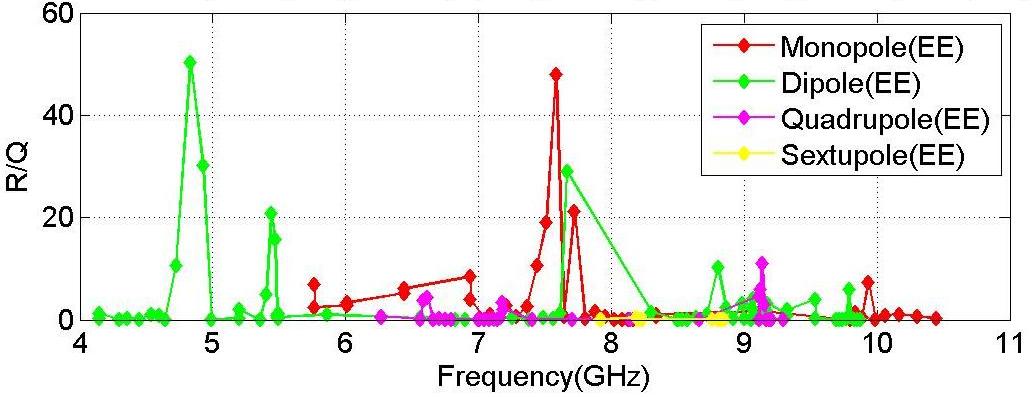}
\caption{The $R/Q$ parameter for HOMs except the fundamental modes of a 9-cell third harmonic cavity plotted versus the frequency of the mode. The modes were calculated with electric (EE) boundary conditions. The units of the $R/Q$ parameter are: $\Omega$ (monopole), $\Omega/cm^2$ (dipole), $\Omega/cm^4$ (quadrupole) and $\Omega/cm^6$ (sextupole).}
\label{simu-HOM-ee}
\end{figure}

\begin{figure}[h]
\centering
\includegraphics[width=0.8\textwidth]{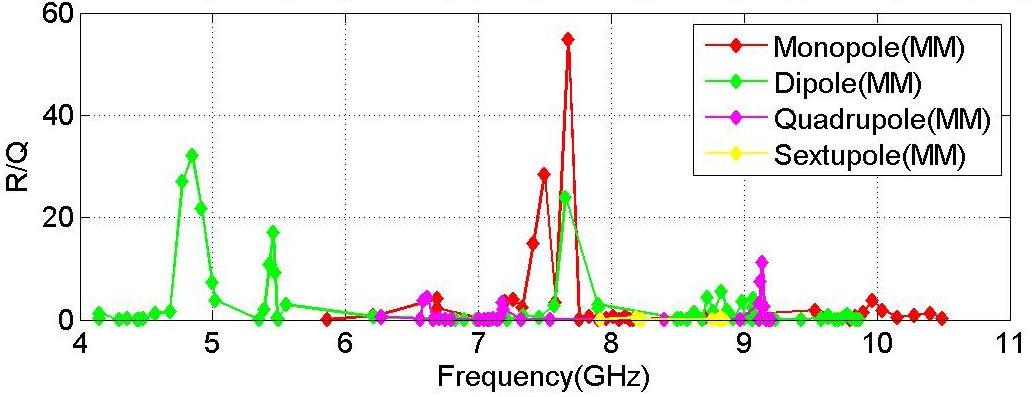}
\caption{The $R/Q$ parameter for HOMs except the fundamental modes of a 9-cell third harmonic cavity plotted versus the frequency of the mode. The modes were calculated with magnetic (MM) boundary conditions. The units of the $R/Q$ parameter are: $\Omega$ (monopole), $\Omega/cm^2$ (dipole), $\Omega/cm^4$ (quadrupole) and $\Omega/cm^6$ (sextupole).}
\label{simu-HOM-mm}
\end{figure}

Eq.~\ref{eq:phase-adv-5} has been used to calculate the phase advance per cell for all modes which have been simulated. The results for monopole, dipole, quadrupole and sextupole bands are shown in Fig.~\ref{disp-1cell-9cell} along with dispersion curves of the mid-cell presented in Section~\ref{sec:mid-cell}. Results of both electric (EE) and magnetic (MM) boundary conditions are presented. The frequencies of the first monopole passband are below the cutoff frequency of the beam pipe (see Table~\ref{table-cutoff}), therefore modes in this band do not depend on the boundary conditions. This can be seen in Fig.~\ref{disp-1cell-mono-9cell} as the overlap of asterisks and circles. Some dipole modes in the fifth dipole band and the first two modes in the first dipole band are trapped within the cavity, which explains the consistency of results from EE and MM boundary conditions in Fig.~\ref{disp-1cell-dipole-9cell}. Large deviations for different boundary conditions can be clearly seen in other dipole bands as they are propagating among cavities. Fig.~\ref{simu-mono}--\ref{simu-sextu} show the $R/Q$ value versus the frequency of each mode for both EE and MM boundary conditions. The coupling strengths for all HOMs beyond the fundamental passband are shown in Fig.~\ref{simu-HOM-ee} and Fig.~\ref{simu-HOM-mm}.
% 1-cell dispersion (with 9cell)

% f vs. R/Q (monopole)

% f vs. R/Q (dipole)

% f vs. R/Q (quadrupole)

% f vs. R/Q (sextupole)

% f vs. R/Q (HOM-EE)

% f vs. R/Q (HOM-MM)

\begin{figure}[h]
\centering
\includegraphics[width=0.9\textwidth]{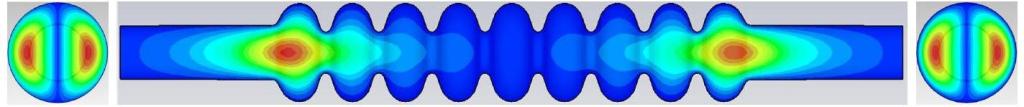}
\caption{The electric field distribution of one dipole beam-pipe mode (frequency: 4.1491~GHz, $R/Q$: 1.318~$\Omega$/cm$^2$). Electric (EE) boundary conditions were used in the simulation.}
\label{simu-Efield-DBP1-2}
\end{figure}

\begin{figure}[h]
\centering
\includegraphics[width=0.9\textwidth]{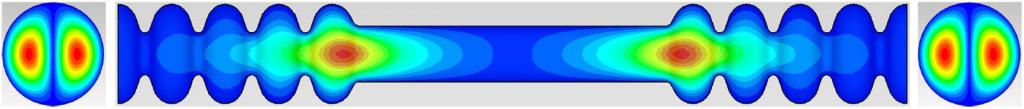}
\caption{The electric field distribution of one dipole beam-pipe mode (frequency: 4.1481~GHz, $R/Q$: 1.544~$\Omega$/cm$^2$). Electric (EE) boundary conditions were used in the simulation.}
\label{simu-Efield-DBP1-2-half}
\end{figure}

\FloatBarrier
Beside the cavity modes shown in the passbands, there are also beam-pipe modes, whose electromagnetic energy mainly deposits in the beam pipes and the end-cells of the cavity. These modes are trapped within both beam-pipe ends of the structure. One of these modes is shown in Fig.~\ref{simu-Efield-DBP1-2} and Fig.~\ref{simu-Efield-DBP1-2-half}. The dipole character of this mode can be seen clearly in the projection on the transverse plane in the middle plane of each end-cell.   
% DBP1-2(EE)
% DBP1-2(EE)(4.5cell)

\begin{figure}[h]
\centering
\includegraphics[width=0.9\textwidth]{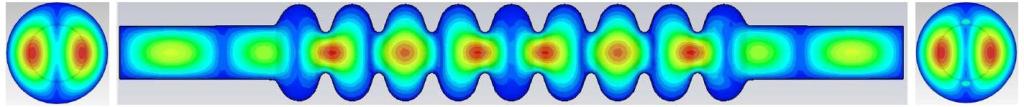}
\caption{The electric field distribution of one cavity mode from the first dipole band (frequency: 4.8327~GHz, $R/Q$: 50.307~$\Omega$/cm$^2$). Electric (EE) boundary conditions were used in the simulation.}
\label{simu-Efield-D1-8}
\end{figure}

\begin{figure}[h]
\centering
\includegraphics[width=0.99\textwidth]{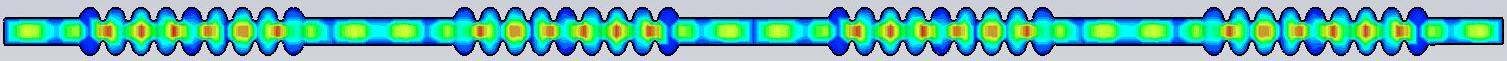}
\caption{The electric field distribution of strongest coupled cavity mode from the first dipole band (frequency: 4.8076~GHz, $R/Q$: 125.762~$\Omega$/cm$^2$ per module). Electric (EE) boundary conditions were used in the simulation.}
\label{simu-Efield-D1-8-string}
\end{figure}

\FloatBarrier
The propagating feature of one dipole cavity mode can be seen in Fig.~\ref{simu-Efield-D1-8} and Fig.~\ref{simu-Efield-D1-8-string}. The mode can couple to adjacent cavities through the attached beam pipes, and has a strong coupling to the beam represented by the large $R/Q$ value.
% D1-8(EE)
% D1-8 (string)

\begin{figure}[h]
\centering
\includegraphics[width=0.9\textwidth]{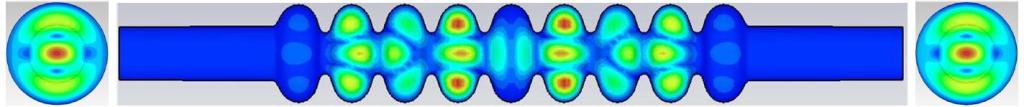}
\caption{The electric field distribution of one cavity mode from the fifth dipole band (frequency: 9.0581~GHz, $R/Q$: 2.171~$\Omega$/cm$^2$). Electric (EE) boundary conditions were used in the simulation.}
\label{simu-Efield-D5-4}
\end{figure}

\FloatBarrier
One trapped cavity mode from the fifth dipole band is shown in Fig.~\ref{simu-Efield-D5-4}. Compared with other trapped modes in this band, this mode has stronger coupling to the beam (larger $R/Q$ value).
% D5-4(EE)

\begin{figure}[h]
\centering
\includegraphics[width=0.9\textwidth]{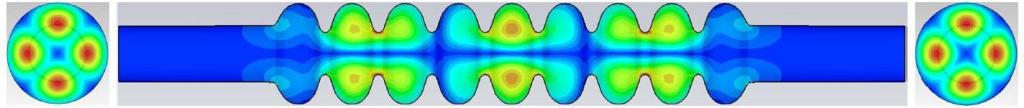}
\caption{The electric field distribution of one cavity mode from the first quadrupole band (frequency: 6.6167~GHz, $R/Q$: 4.358~$\Omega$/cm$^4$). Electric (EE) boundary conditions were used in the simulation.}
\label{simu-Efield-Q1-3}
\end{figure}

\begin{figure}[h]
\centering
\includegraphics[width=0.9\textwidth]{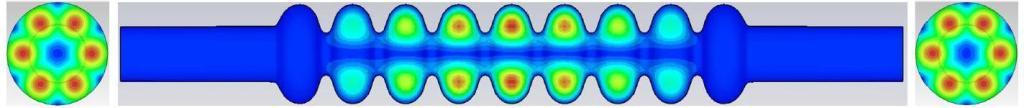}
\caption{The electric field distribution of one cavity mode from the first sextupole band (frequency: 8.1894~GHz, $R/Q$: 0.506~$\Omega$/cm$^6$). Electric (EE) boundary conditions were used in the simulation.}
\label{simu-Efield-S1-1}
\end{figure}

\FloatBarrier
One quadrupole mode and one sextupole mode are also shown in Fig.~\ref{simu-Efield-Q1-3} and Fig.~\ref{simu-Efield-S1-1}. The $R/Q$ values are in general small for these modes.
 % Q1-3(EE)
% S1-1(EE)

\begin{figure}[h]
\centering
\subfigure[Electric (EE) boudary conditions]{
\includegraphics[width=0.45\textwidth]{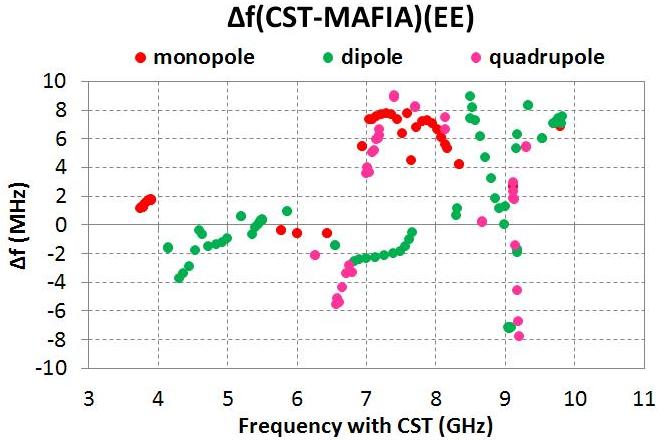}
\label{simu-9cell-cst-mafia-ee}
}
\quad
\subfigure[Magnetic (MM) boudary conditions]{
\includegraphics[width=0.45\textwidth]{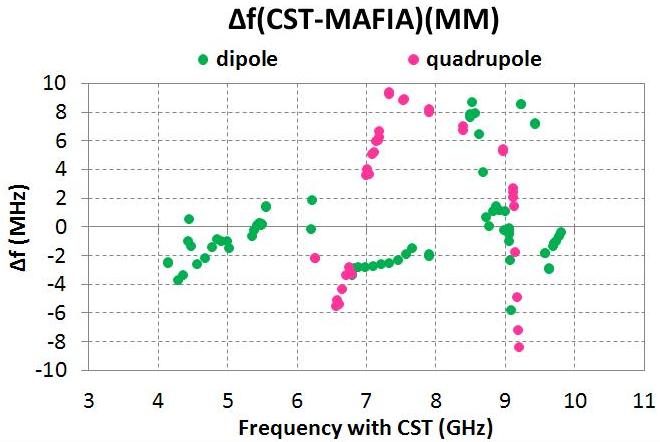}
\label{simu-9cell-cst-mafia-mm}
}
\caption{Frequency differences of modes simulated with CST{\textregistered} and MAFIA{\textregistered}. $\Delta f$ is calculated as $\Delta f=f_{CST}-f_{MAFIA}$.}
\label{simu-9cell-cst-mafia}
\end{figure}

\begin{figure}[h]
\centering
\subfigure[Electric (EE) boudary conditions]{
\includegraphics[width=0.45\textwidth]{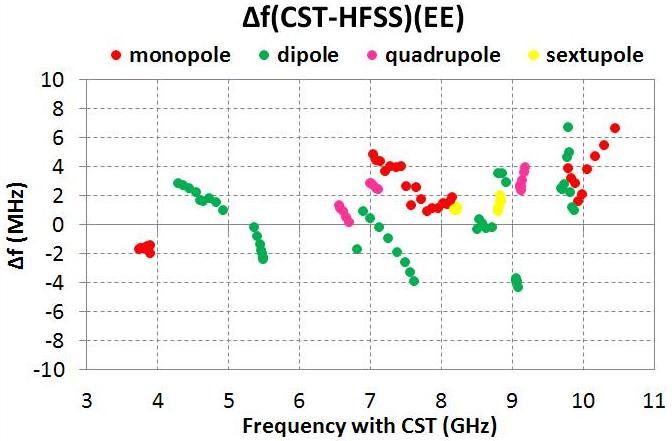}
\label{simu-9cell-cst-hfss-ee}
}
\quad
\subfigure[Magnetic (EE) boudary conditions]{
\includegraphics[width=0.45\textwidth]{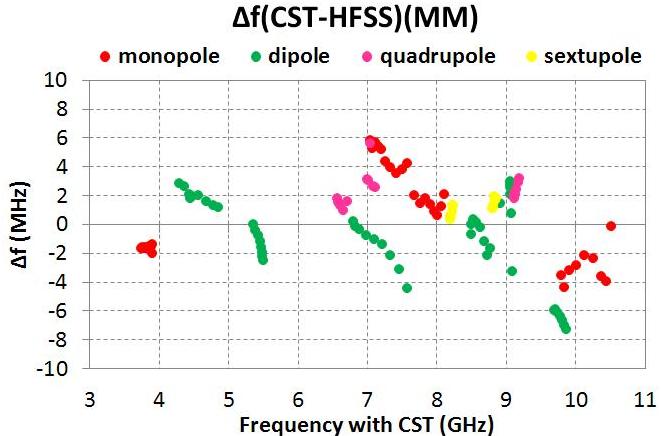}
\label{simu-9cell-cst-hfss-mm}
}
\caption{Frequency differences of modes simulated with CST{\textregistered} and HFSS{\textregistered}. $\Delta f$ is calculated as $\Delta f=f_{CST}-f_{HFSS}$.}
\label{simu-9cell-cst-hfss}
\end{figure}

\FloatBarrier
Compared to the simulations using MAFIA{\textregistered} \cite{mafia-0} with eigenvalue solver and HFSS{\textregistered} \cite{hfss} with eigenmode solver, the frequencies of the modes are shifted. A direct comparison between CST{\textregistered} and MAFIA{\textregistered} is shown in Fig.~\ref{simu-9cell-cst-mafia}, while a comparison between CST{\textregistered} and HFSS{\textregistered} is shown in Fig.~\ref{simu-9cell-cst-hfss}. The MAFIA{\textregistered} simulation results are from \cite{mafia} while the HFSS{\textregistered} simulations are from \cite{hfss-1}. The differences are within 10~MHz for both boundary conditions from both simulation codes.
% cst vs. mafia 
% cst vs. hfss 

%% file: app-list.tex
%%%%%%%%%%%%%%%%  Section 1  %%%%%%%%%%%%%%%%%%%%%%%%
%                                                                                                                                          %
%                                                  Mode List                                                                        %
%                                                                                                                                          %
%%%%%%%%%%%%%%%%%%%%%%%%%%%%%%%%%%%%%%%%%%%%%
\chapter{\label{sec:mode-list}List of Monopole, Dipole, Quadrupole and Sextupole Modes}
The frequencies and $R/Q's$ of the eigenmodes simulated on the ideal 9-cell third harmonic cavity are shown in this section. The modes are grouped in bands: ``M'' denotes monopole, ``M1'' denotes the first monopole band, ``M1-1'' denotes the first mode in M1, ``D'' denotes dipole, ``Q'' denotes quadrupole and ``S'' denotes sextupole. Beam-pipe modes in each table are denoted as ``BP''. 
%\section{\label{sec:mode}Mode List}
% mode list (Monopole) (part 1)
%\subsection{\label{sec:mode-mono}Monopole Modes}
\begin{table}[h]
\setlength\tabcolsep{8pt}
\centering
\caption{Monopole modes with electric (EE) or magnetic (MM) boundaries (part 1).}
\medskip
\begin{tabular}{c|cc|cc}
\toprule
\multicolumn{1}{c}{} & \multicolumn{2}{|c|}{EE} & \multicolumn{2}{c}{MM} \\
\midrule
\textbf{Band} & \textbf{f}(GHz) & \textbf{R/Q}($\Omega$/cm$^2$)  & \textbf{f}(GHz) & \textbf{R/Q}($\Omega$/cm$^2$) \\
\midrule
M1-1	&	3.7466	&	0.008	&	3.7466	&	0.008	\\
M1-2	&	3.7601	&	0.061	&	3.7601	&	0.061	\\
M1-3	&	3.7808	&	0.090	&	3.7808	&	0.090	\\
M1-4	&	3.8065	&	0.170	&	3.8065	&	0.170	\\
M1-5	&	3.8340	&	0.307	&	3.8341	&	0.309	\\
M1-6	&	3.8602	&	0.203	&	3.8602	&	0.204	\\
M1-7	&	3.8817	&	0.468	&	3.8817	&	0.481	\\
M1-8	&	3.8958	&	0.195	&	3.8958	&	0.197	\\
M1-9	&	3.9008	&	373.113	&	3.9008	&	373.097	\\
BP1-1	&	5.7685	&	6.967	&	5.8624	&	0.051	\\
BP1-2	&	5.7685	&	2.443	&	5.8624	&	0.051	\\
BP2-1	&	6.0123	&	2.867	&	6.2095	&	0.879	\\
BP2-2	&	6.0123	&	3.335	&	6.2095	&	0.665	\\
BP3-1	&	6.4403	&	5.212	&	6.6886	&	4.119	\\
BP3-2	&	6.4403	&	6.116	&	6.6886	&	2.302	\\
BP4-1	&	6.9393	&	8.454	&			&                 \\
BP4-2	&	6.9394	&	3.971	&			&                \\
M2-1	&	7.0483	&	0.127	&	7.0449	&	0.007	\\
M2-2	&	7.0863	&	1.012	&	7.0738	&	0.034	\\
M2-3	&	7.1424	&	0.088	&	7.1145	&	0.342	\\
M2-4	&	7.2113	&	2.914	&	7.1589	&	0.224	\\
M2-5	&	7.2877	&	0.677	&	7.2058	&	3.534	\\
M2-6	&	7.3662	&	2.727	&	7.2625	&	4.008	\\
M2-7	&	7.4418	&	10.686	&	7.3331	&	2.417	\\
M2-8	&	7.5118	&	18.963	&	7.4140	&	14.981	\\
M2-9	&	7.5843	&	47.909	&	7.4995	&	28.455	\\
M2-10    &                 &                 &	7.5810	&	3.471	\\  
\bottomrule
\end{tabular}
\label{simu-table-mono-1}
\end{table}
%\newpage

% Monopole list (part 2)
\begin{table}[h]
\setlength\tabcolsep{8pt}
\centering
\caption{Monopole modes with electric (EE) or magnetic (MM) boundaries (part 2).}
\medskip
\begin{tabular}{c|cc|cc}
\toprule
\multicolumn{1}{c}{} & \multicolumn{2}{|c|}{EE} & \multicolumn{2}{c}{MM} \\
\midrule
\textbf{Band} & \textbf{f}(GHz) & \textbf{R/Q}($\Omega$/cm$^2$)  & \textbf{f}(GHz) & \textbf{R/Q}($\Omega$/cm$^2$) \\
\midrule
M3-1	&	7.6443	&	0.593	&	7.6780	&	54.835	\\
M3-2	&	7.7248	&	21.212	&	7.7577	&	0.095	\\
M3-3	&	7.8036	&	0.245	&	7.8333	&	0.619	\\
M3-4	&	7.8809	&	1.707	&	7.9021	&	0.000	\\
M3-5	&	7.9547	&	0.640	&	7.9625	&	0.275	\\
M3-6	&	8.0229	&	0.015	&	8.0148	&	0.756	\\
M3-7	&	8.0829	&	0.113	&	8.0615	&	0.117	\\
M3-8	&	8.1311	&	0.014	&	8.1044	&	0.755	\\
M3-9	&	8.1631	&	0.001	&	8.1408	&	0.002	\\
M3-10    &    	      &                 &	8.1656	&	0.078	\\
BP5-1	&	8.3376	&	0.765	&	8.7111	&	0.521	\\
BP5-2	&	8.3376	&	1.045	&	8.7113	&	0.525	\\
BP6-1	&	9.1202	&	1.670	&	9.5377	&	1.912	\\
BP6-2	&	9.1202	&	1.755	&	9.5378	&	1.945	\\
M4-1	&	9.7966	&	0.000	&	9.7907	&	0.000	\\
M4-2	&	9.8340	&	1.511	&	9.8379	&	0.072	\\
M4-3	&	9.8868	&	0.395	&	9.9124	&	0.360	\\
M4-4	&	9.9384	&	7.388	&	10.0099	&	5.067	\\
M4-5	&	9.9886	&	0.015	&	10.1270	&	0.264	\\
M4-6	&	10.0619	&	0.921	&	10.2547	&	0.723	\\
M4-7	&	10.1692	&	1.125	&	10.3692	&	2.342	\\
M4-8	&	10.3015	&	0.778	&	10.4406	&	0.312	\\
M4-9	&	10.4485	&	0.277	&	10.5080	&	0.719	\\
\bottomrule
\end{tabular}
\label{simu-table-mono-2}
\end{table}
%\newpage
%% plot f vs. R/Q (monopole)
%\begin{figure}[h]
%\centering
%\subfigure[Monopole (electric boundary)]{
%\includegraphics[width=0.9\textwidth]{simu-mono-ee}
%\label{simu-mono-ee}
%}
%\subfigure[Monopole (magnetic boundary)]{
%\includegraphics[width=0.9\textwidth]{simu-mono-mm}
%\label{simu-mono-mm}
%}
%\subfigure[Monopole]{
%\includegraphics[width=0.9\textwidth]{simu-mono}
%\label{simu-mono}
%}
%\caption{The eigenmode frequency and $R/Q$ value of monopole modes with electric and magnetic boundaries.}
%\label{simu-mono-ee-mm}
%\end{figure}

% mode list (Dipole)(part 1)
%\subsection{\label{sec:mode-dipole-1}Dipole Modes (part 1)}
\begin{table}[h]
\setlength\tabcolsep{8pt}
\centering
\caption{Dipole modes with electric (EE) or magnetic (MM) boundaries (part 1).}
\medskip
\begin{tabular}{c|cc|cc}
\toprule
\multicolumn{1}{c}{} & \multicolumn{2}{|c|}{EE} & \multicolumn{2}{c}{MM} \\
\midrule
%\multirow{2}{*}{\textbf{Band}} & \multirow{2}{*}{\textbf{f}(GHz)} & \textbf{R/Q} & \multirow{2}{*}{\textbf{f}(GHz)} & \textbf{R/Q} \\ 
%& & ($\Omega$/cm$^2$) & & ($\Omega$/cm$^2$) \\
\textbf{Band} & \textbf{f}(GHz) & \textbf{R/Q}($\Omega$/cm$^2$)  & \textbf{f}(GHz) & \textbf{R/Q}($\Omega$/cm$^2$) \\
\midrule
BP1-1	&	4.1489	&	0.234	&	4.1474	&	0.241	\\
BP1-2	&	4.1491	&	1.318	&	4.1475	&	1.299	\\
D1-1	&	4.2982	&	0.001	&	4.2979	&	0.001	\\
D1-2	&	4.3607	&	0.292	&	4.3592	&	0.263	\\
D1-3	&	4.4485	&	0.002	&	4.4306	&	0.072	\\
D1-4	&	4.5410	&	1.076	&	4.4516	&	0.000	\\
D1-5	&	4.5989	&	0.784	&	4.4770	&	0.327	\\
D1-6	&	4.6415	&	0.165	&	4.5703	&	1.213	\\
D1-7	&	4.7245	&	10.572	&	4.6804	&	1.586	\\
D1-8	&	4.8327	&	50.307	&	4.7749	&	27.165	\\
D1-9	&	4.9270	&	30.174	&	4.8455	&	32.124	\\
D1-10	&	4.9899	&	0.000	&	4.9162	&	21.833	\\
BP2-1	&	5.2014	&	0.300	&	4.9945	&	7.376	\\
BP2-2	&	5.2040	&	2.036	&	5.0233	&	3.844	\\
D2-1	&	5.3581	&	0.041	&	5.3518	&	0.055	\\
D2-2	&	5.4050	&	5.057	&	5.3923	&	2.114	\\
D2-3	&	5.4427	&	20.877	&	5.4272	&	10.770	\\
D2-4	&	5.4678	&	15.776	&	5.4528	&	17.024	\\
D2-5	&	5.4829	&	0.895	&	5.4711	&	9.368	\\
D2-6	&	5.4911	&	1.261	&	5.4834	&	0.409	\\
D2-7	&	5.4950	&	0.307	&	5.4908	&	0.343	\\
D2-8	&	5.4958	&	0.549	&	5.4944	&	0.033	\\
BP3-1	&	5.8644	&	1.028	&	5.5532	&	2.994	\\
BP3-2	&	5.8644	&	1.026	&	5.5532	&	2.995	\\
BP4-1	&	6.5593	&	0.344	&	6.2123	&	0.595	\\
BP4-2	&	6.5594	&	0.397	&	6.2144	&	0.636	\\
D3-1	&	6.8238	&	0.011	&	6.7964	&	0.068	\\
D3-2	&	6.9003	&	0.035	&	6.8242	&	0.068	\\
D3-3	&	7.0027	&	0.058	&	6.8909	&	0.140	\\
D3-4	&	7.1225	&	0.189	&	6.9880	&	0.124	\\
D3-5	&	7.2541	&	0.549	&	7.0989	&	0.108	\\
D3-6	&	7.3833	&	0.014	&	7.2140	&	0.020	\\
D3-7	&	7.4889	&	0.455	&	7.3348	&	0.825	\\
D3-8	&	7.5621	&	0.269	&	7.4598	&	0.503	\\
D3-9	&	7.6196	&	1.354	&	7.5743	&	2.862	\\
D3-10	&	7.6680	&	28.926	&	7.6566	&	23.875	\\
BP5-1	&	8.3033	&	1.543	&	7.9033	&	2.155	\\
BP5-2	&	8.3039	&	1.537	&	7.9034	&	2.961	\\
\bottomrule
\end{tabular}
\label{simu-table-dipole-1}
\end{table}
%\newpage

% mode list (Dipole)(part 2)
%\subsection{\label{sec:mode-dipole-2}Dipole Modes (part 2)}
\begin{table}[h]
\setlength\tabcolsep{8pt}
\centering
\caption{Dipole modes with electric (EE) or magnetic (MM) boundaries (part 2).}
\medskip
\begin{tabular}{c|cc|cc}
\toprule
\multicolumn{1}{c}{} & \multicolumn{2}{|c|}{EE} & \multicolumn{2}{c}{MM} \\
\midrule
%\multirow{2}{*}{\textbf{Band}} & \multirow{2}{*}{\textbf{f}(GHz)} & \textbf{R/Q} & \multirow{2}{*}{\textbf{f}(GHz)} & \textbf{R/Q} \\ 
%& & ($\Omega$/cm$^2$) & & ($\Omega$/cm$^2$) \\
\textbf{Band} & \textbf{f}(GHz) & \textbf{R/Q}($\Omega$/cm$^2$)  & \textbf{f}(GHz) & \textbf{R/Q}($\Omega$/cm$^2$) \\
\midrule
D4-1	&	8.5002	&	0.130	&	8.5292	&	0.365	\\
D4-2	&	8.5042	&	0.096	&	8.5709	&	0.023	\\
D4-3	&	8.5322	&	0.152	&	8.6273	&	1.457	\\
D4-4	&	8.5763	&	0.115	&	8.6849	&	0.042	\\
D4-5	&	8.6397	&	0.415	&	8.7257	&	4.392	\\
D4-6	&	8.7205	&	1.038	&	8.7702	&	1.693	\\
D4-7	&	8.8033	&	10.205	&	8.8301	&	5.577	\\
D4-8	&	8.8648	&	2.470	&	8.8729	&	1.895	\\
D4-9	&	8.9196	&	0.287	&	8.9196	&	0.281	\\
D4-10	&	8.9857	&	3.258	&	8.9900	&	3.623	\\
D4-11	&	8.9980	&	0.230	&	9.0011	&	0.302	\\
D5-1	&	9.0523	&	0.002	&	9.0593	&    0.004    \\
D5-2	&	9.0530	&	0.053	&	9.0599	&    0.058    \\
D5-3	&	9.0546	&	0.058	&	9.0614	&	0.076	\\
D5-4	&	9.0581	&	2.171	&	9.0645	&	2.377	\\
D5-5	&	9.0664	&	4.116	&	9.0718	&	4.158	\\
D5-6	&	9.0890	&	0.580	&	9.0918	&	0.452	\\
BP6-1	&	9.1666	&	1.291	&	8.4964	&	0.340	\\
BP6-2	&	9.1678	&	1.898	&	8.5008	&	0.056	\\
BP7-1	&	9.1749	&	1.123	&	9.2324	&	0.158	\\
BP7-2	&	9.1763	&	3.240	&	9.2325	&	0.092	\\
BP8-1	&	9.3283	&	0.880	&	9.4325	&	0.105	\\
BP8-2	&	9.3284	&	2.042	&	9.4330	&	0.020	\\
BP9-1	&	9.5379	&	4.064	&	9.5809	&	0.342	\\
BP9-2	&	9.5385	&	0.275	&	9.5818	&	0.015	\\
BP10-1  &                  &                &     9.6342	&	0.485  \\
BP10-2  &                  &                &     9.6346	&	0.848	\\
D6-1	&	9.6962	&	0.001	&	9.6896	&	0.013	\\
D6-2	&	9.7142	&	0.009	&	9.7103	&	0.025	\\
D6-3	&	9.7421	&	0.074	&	9.7415	&	0.341	\\
D6-4	&	9.7711	&	0.379	&	9.7776	&	1.133	\\
D6-5	&	9.7896	&	5.951	&	9.8134	&	0.345	\\
D6-6	&	9.8027	&	0.771	&	9.8440	&	0.008	\\
D6-7	&	9.8265	&	0.191	&	9.8648	&	0.019	\\
%D6-8	&	9.8532	&	0.014	&	
%D6-9	&	9.8732	&	0.004	&	
\bottomrule
\end{tabular}
\label{simu-table-dipole-2}
\end{table}
%\newpage

%% plot f vs. R/Q (dipole)
%\begin{figure}[h]
%\centering
%\subfigure[Dipole (electric boundary)]{
%\includegraphics[width=0.9\textwidth]{simu-dipole-ee}
%\label{simu-dipole-ee}
%}
%\subfigure[Dipole (magnetic boundary)]{
%\includegraphics[width=0.9\textwidth]{simu-dipole-mm}
%\label{simu-dipole-mm}
%}
%\subfigure[Dipole]{
%\includegraphics[width=0.9\textwidth]{simu-dipole}
%\label{simu-dipole}
%}
%\caption{The eigenmode frequency and $R/Q$ value of dipole modes with electric and magnetic boundaries.}
%\label{simu-dipole-ee-mm}
%\end{figure}

% mode list (Quadrupole)
%\subsection{\label{sec:mode-quad}Quadrupole Modes}
\begin{table}[h]
\setlength\tabcolsep{8pt}
\centering
\caption{Quadrupole modes with electric (EE) or magnetic (MM) boundaries.}
\medskip
\begin{tabular}{c|cc|cc}
\toprule
\multicolumn{1}{c}{} & \multicolumn{2}{|c|}{EE} & \multicolumn{2}{c}{MM} \\
\midrule
%\multirow{2}{*}{\textbf{Band}} & \multirow{2}{*}{\textbf{f}(GHz)} & \textbf{R/Q} & \multirow{2}{*}{\textbf{f}(GHz)} & \textbf{R/Q} \\ 
%& & ($\Omega$/cm$^2$) & & ($\Omega$/cm$^2$) \\
\textbf{Band} & \textbf{f}(GHz) & \textbf{R/Q}($\Omega$/cm$^2$)  & \textbf{f}(GHz) & \textbf{R/Q}($\Omega$/cm$^2$) \\
\midrule
BP1-1	&	6.2697	&	0.513	&	6.2697	&	0.513	\\
BP1-2	&	6.2698	&	0.742	&	6.2697	&	0.742	\\
Q1-1	&	6.5638	&	0.183	&	6.5638	&	0.183	\\
Q1-2	&	6.5843	&	3.734	&	6.5843	&	3.734	\\
Q1-3	&	6.6167	&	4.358	&	6.6167	&	4.359	\\
Q1-4	&	6.6583	&	0.183	&	6.6583	&	0.183	\\
Q1-5	&	6.7059	&	0.308	&	6.7059	&	0.307	\\
Q1-6	&	6.7546	&	0.002	&	6.7546	&	0.002	\\
Q1-7	&	6.7961	&	0.041	&	6.7961	&	0.041	\\
Q2-1	&	7.0005	&	0.135	&	7.0005	&	0.135	\\
Q2-2	&	7.0096	&	0.075	&	7.0096	&	0.075	\\
Q2-3	&	7.0456	&	0.152	&	7.0456	&	0.151	\\
Q2-4	&	7.0823	&	0.000	&	7.0823	&	0.000	\\
Q2-5	&	7.1158	&	0.221	&	7.1157	&	0.220	\\
Q2-6	&	7.1437	&	0.101	&	7.1436	&	0.103	\\
Q2-7	&	7.1653	&	0.579	&	7.1653	&	0.578	\\
Q2-8	&	7.1806	&	3.484	&	7.1806	&	3.483	\\
Q2-9	&	7.1897	&	2.125	&	7.1897	&	2.125	\\
BP2-1	&	7.4084	&	0.020	&	7.3248	&	0.008	\\
BP2-2	&	7.4085	&	0.020	&	7.3249	&	0.008	\\
BP3-1	&	7.7101	&	0.049	&	7.5397	&	0.030	\\
BP3-2	&	7.7102	&	0.033	&	7.5398	&	0.030	\\
BP4-1	&	8.1423	&	0.069	&	7.9134	&	0.052	\\
BP4-2	&	8.1431	&	0.069	&	7.9136	&	0.054	\\
BP5-1	&	8.6665	&	0.095	&	8.3961	&	0.086	\\
BP5-2	&	8.6665	&	0.108	&	8.3964	&	0.085	\\
Q3-1	&	9.1129	&	4.686	&	9.1147	&	3.176	\\
Q3-2	&	9.1133	&	0.053	&	9.1155	&	0.147	\\
Q3-3	&	9.1228	&	5.921	&	9.1232	&	7.555	\\
Q3-4	&	9.1340	&	11.102	&	9.1344	&	11.244	\\
Q3-5	&	9.1499	&	2.998	&	9.1501	&	2.608	\\
Q3-6	&	9.1692	&	0.003	&	9.1692	&	0.021	\\
Q3-7	&	9.1893	&	0.152	&	9.1890	&	0.152	\\
Q3-8	&	9.2053	&	0.000	&	9.2048	&	0.001	\\
BP6-1	&	9.2980	&	0.178	&	8.9709	&	0.167	\\
BP6-2	&	9.2980	&	0.179	&	8.9711	&	0.166	\\
\bottomrule
\end{tabular}
\label{simu-table-quad}
\end{table}
%\newpage

%% plot f vs. R/Q (Quadrupole)
%\begin{figure}[h]
%\centering
%\subfigure[Quadrupole (electric boundary)]{
%\includegraphics[width=0.9\textwidth]{simu-quad-ee}
%\label{simu-quad-ee}
%}
%\subfigure[Quadrupole (magnetic boundary)]{
%\includegraphics[width=0.9\textwidth]{simu-quad-mm}
%\label{simu-quad-mm}
%}
%\subfigure[Quadrupole]{
%\includegraphics[width=0.9\textwidth]{simu-quad}
%\label{simu-quad}
%}
%\caption{The eigenmode frequency and $R/Q$ value of quadrupole modes with electric and magnetic boundaries.}
%\label{simu-quad-ee-mm}
%\end{figure}

% mode list (Sextupole)
%\subsection{\label{sec:mode-quad}Sextupole Modes}
\begin{table}[h]
\setlength\tabcolsep{8pt}
\centering
\caption{Sextupole modes with electric (EE) or magnetic (MM) boundaries.}
\medskip
\begin{tabular}{c|cc|cc}
\toprule
\multicolumn{1}{c}{} & \multicolumn{2}{|c|}{EE} & \multicolumn{2}{c}{MM} \\
\midrule
%\multirow{2}{*}{\textbf{Band}} & \multirow{2}{*}{\textbf{f}(GHz)} & \textbf{R/Q} & \multirow{2}{*}{\textbf{f}(GHz)} & \textbf{R/Q} \\ 
%& & ($\Omega$/cm$^2$) & & ($\Omega$/cm$^2$) \\
\textbf{Band} & \textbf{f}(GHz) & \textbf{R/Q}($\Omega$/cm$^2$)  & \textbf{f}(GHz) & \textbf{R/Q}($\Omega$/cm$^2$) \\
\midrule
BP1-1	&	7.9214	&	0.069	&	7.9212	&	0.060	\\
BP1-2	&	7.9216	&	0.069	&	7.9215	&	0.068	\\
S1-1	&	8.1894	&	0.506	&	8.1894	&	0.506	\\
S1-2	&	8.1940	&	0.203	&	8.1940	&	0.203	\\
S1-3	&	8.2011	&	0.006	&	8.2011	&	0.006	\\
S1-4	&	8.2097	&	0.027	&	8.2097	&	0.027	\\
S1-5	&	8.2184	&	0.001	&	8.2184	&	0.001	\\
S1-6	&	8.2261	&	0.005	&	8.2261	&	0.005	\\
S1-7	&	8.2313	&	0.000	&	8.2313	&	0.000	\\
BP2-1	&	8.7611	&	0.015	&	8.7612	&	0.018	\\
BP2-2	&	8.7614	&	0.013	&	8.7615	&	0.017	\\
S2-1	&	8.8029	&	0.103	&	8.8029	&	0.000	\\
S2-2	&	8.8097	&	0.001	&	8.8097	&	0.001	\\
S2-3	&	8.8192	&	0.003	&	8.8191	&	0.002	\\
S2-4	&	8.8295	&	0.001	&	8.8294	&	0.412	\\
S2-5	&	8.8392	&	0.024	&	8.8391	&	0.028	\\
S2-6	&	8.8469	&	0.036	&	8.8468	&	0.035	\\
S2-7	&	8.8519	&	0.004	&	8.8519	&	0.004	\\
\bottomrule
\end{tabular}
\label{simu-table-sextu}
\end{table}

%% plot f vs. R/Q (Sextupole)
%\begin{figure}[h]
%\centering
%\subfigure[Sextupole (electric boundary)]{
%\includegraphics[width=0.9\textwidth]{simu-sextu-ee}
%\label{simu-sextu-ee}
%}
%\subfigure[Sextupole (magnetic boundary)]{
%\includegraphics[width=0.9\textwidth]{simu-sextu-mm}
%\label{simu-sextu-mm}
%}
%\subfigure[Sextupole]{
%\includegraphics[width=0.9\textwidth]{simu-sextu}
%\label{simu-sextu}
%}
%\caption{The eigenmode frequency and $R/Q$ value of sextupole modes with electric and magnetic boundaries.}
%\label{simu-sextu-ee-mm}
%\end{figure}
%
%% plot f vs. R/Q (HOM)
%\begin{figure}[h]
%\centering
%\subfigure[HOM (electric boundary)]{
%\includegraphics[width=0.9\textwidth]{simu-HOM-ee}
%\label{simu-HOM-ee}
%}
%\subfigure[HOM (magnetic boundary)]{
%\includegraphics[width=0.9\textwidth]{simu-HOM-mm}
%\label{simu-HOM-mm}
%}
%\caption{The eigenmode frequency and $R/Q$ value of HOMs with electric and magnetic boundaries.}
%\label{simu-HOM-ee-mm}
%\end{figure}

%%%%%%%%%%%%%%%%  Section 2  %%%%%%%%%%%%%%%%%%%%%%%%
%                                                                                                                                          %
%                                            Parameter Settings                                                                    %
%                                                                                                                                          %
%%%%%%%%%%%%%%%%%%%%%%%%%%%%%%%%%%%%%%%%%%%%%
\chapter{\label{sec:para-list}Parameter Settings used for Simulations}
The key parameters set in CST Microwave studio{\textregistered} for the eigenmode simulations shown in Appendix~\ref{sec:mode-list} are described in this section. In the``Mesh Type'' column, ``PBA'' denotes ``perfect boundary approximation'' while ``FPBA'' denotes ``fast PBA'' with ``Enhanced PBA accuracy''. The choice between ``PBA'' and ``FPBA'' is based on one principle in this study: ``FPBA'' was used only if ``PBA'' failed to generate a valid mesh. A comparison of performance between ``PBA'' and ``FPBA'' is discussed in \cite{pba-fpba}. In the ``Mesh cells (million)'' column, the number of mesh cells is for a quarter of the structure. The naming in the ``Band'' column follows the convention explained in Appendix~\ref{sec:mode-list}. 

%%%%% table simulation settings (mono, quad)(EE)  %%%%%%%%%
%\subsection{\label{sec:para-mono-ee}Parameter Settings for Monopole and Quadrupole Modes (EE)}
\begin{table}[h]
\setlength\tabcolsep{8pt}
\centering
\caption{Parameters setting for monopole and quadrupole modes with electric (EE) boundaries. }
\medskip
\begin{tabular}{m{2.3cm}m{2.2cm}m{1.2cm}m{2cm}m{2cm}m{2.7cm}}
%\begin{tabular}{ccccc}
\toprule
\textbf{Frequency range} (GHz) & \textbf{Lines per wavelength} & \textbf{Mesh type} & \textbf{Mesh cells} (million) & \textbf{Max mesh step} (mm)  & \textbf{Band}\\
\midrule
3.7-3.95  & 70  & PBA & 2.1  & 1.10  & M1 \\
\midrule
5.7-5.8  & 50  & PBA & 2.4  & 1.07  & MBP1 \\
\midrule
5.95-6.05  & 50  & PBA & 2.6  & 1.03  & MBP2 \\
\midrule
6.2-6.3  & 45  & PBA & 2.2  & 1.08  & QBP1 \\
\midrule
6.4-6.5  & 45  & PBA  & 2.4  & 1.04  & MBP3 \\
\midrule
6.5-6.8  & 40  & PBA  & 2.0  & 1.11  & Q1 \\
\midrule
6.85-7.6  & 25  & FPBA  & 1.4  & 0.91  & MBP4, M2(1-7), Q2, QBP2 \\
\midrule
7.5-8.25  & 20  & FPBA  & 1.1  & 1.06  & M2(8-9), M3, QBP3, QBP4 \\
\midrule
8.25-8.35  & 20  & FPBA  & 1.1  & 0.94  & MBP5 \\
\midrule
8.6-8.7  & 20  & PBA  & 1.2  & 0.94  & QBP5 \\
\midrule
8.9-9.35  & 24  & FPBA  & 2.3  & 0.82  & MBP6, Q3, QBP6 \\
\midrule
9.7-10.5  & 20  & FPBA  & 2.0  & 0.87  & M4 \\
\bottomrule
\end{tabular}
\label{simu-setting-mq-ee}
\end{table}
%\newpage

%%%%% table simulation settings (mono, quad)(MM) %%%%%%%%%%%
%\subsection{\label{sec:para-mono-ee}Parameter Settings for Monopole and Quadrupole Modes (MM)}
\begin{table}[h]
\setlength\tabcolsep{8pt}
\centering
\caption{Parameter setting for monopole and quadrupole modes with magnetic (MM) boundaries. }
\medskip
\begin{tabular}{m{2.3cm}m{2.2cm}m{1.2cm}m{2cm}m{2cm}m{2.7cm}}
%\begin{tabular}{ccccc}
\toprule
\textbf{Frequency range} (GHz) & \textbf{Lines per wavelength} & \textbf{Mesh type} & \textbf{Mesh cells} (million) & \textbf{Max mesh step} (mm)  & \textbf{Band}\\
\midrule
3.7-4.0  & 70  & PBA & 2.2  & 1.10  & M1 \\
\midrule
5.7-5.9  & 50  & PBA & 2.5  & 1.04  & MBP1 \\
\midrule
6.0-6.4  & 45  & PBA & 2.3  & 1.07  & MBP2, QBP1 \\
\midrule
6.4-6.8  & 40  & PBA  & 2.0  & 1.11  & MBP3, Q1 \\
\midrule
6.8-7.6  & 25  & FPBA  & 1.4  & 0.91  & M2, Q2, QBP2, QBP3 \\
\midrule
7.6-8.4  & 20  & FPBA  & 1.1  & 0.94  & M3, QBP4 \\
\midrule
8.2-8.4  & 20  & FPBA  & 1.1  & 0.94  & QBP5 \\
\midrule
8.4-8.85  & 20  & FPBA  & 1.3  & 0.94  & MBP5 \\
\midrule
8.85-9.3  & 20 & FPBA  & 1.4  & 0.91  & QBP6, Q3 \\
\midrule
9.3-9.6  & 20  & FPBA  & 1.5  & 0.87  & MBP6 \\
\midrule
9.7-10.0  & 20  & PBA  & 1.8  & 0.87 & M4(1-3) \\
\midrule
10.0-10.6  & 20  & PBA  & 2.0 & 0.87 & M4(4-9) \\
\bottomrule
\end{tabular}
\label{simu-setting-mq-mm}
\end{table}
%\newpage

%%%%%% table simulation settings (dipole, sextupole)(EE)  %%%%%%%
%\subsection{\label{sec:para-mono-ee}Parameter Settings for Dipole and Sextupole Modes (EE)}
\begin{table}[h]
\setlength\tabcolsep{8pt}
\centering
\caption{Parameters setting for dipole and sextupole modes with electric (EE) boundaries. }
\medskip
\begin{tabular}{m{2.3cm}m{2.2cm}m{1.2cm}m{2cm}m{2cm}m{2.7cm}}
%\begin{tabular}{ccccc}
\toprule
\textbf{Frequency range} (GHz) & \textbf{Lines per wavelength} & \textbf{Mesh type} & \textbf{Mesh cells} (million) & \textbf{Max mesh step} (mm)  & \textbf{Band}\\
\midrule
4.1-4.2  & 70  & PBA & 2.4  & 1.04  & DBP1 \\
\midrule
4.25-5.05  & 60  & PBA & 2.6  & 1.03  & D1(1-9) \\
\midrule
4.95-5.3  & 45  & PBA & 1.5  & 1.29  & D1(10), DBP2 \\
\midrule
5.3-5.6  & 50  & PBA & 2.2  & 1.10  & D2 \\
\midrule
5.8-5.9  & 45  & PBA & 2.0  & 1.18  & DBP3 \\
\midrule
6.5-6.6  & 40  & PBA & 1.9  & 1.18  & DBP4 \\
\midrule
6.75-7.7  & 40  & PBA & 3.0  & 0.99  & D3 \\
\midrule
7.7-8.1  & 20  & FPBA & 1.0  & 1.06  & SBP1 \\
\midrule
8.1-8.3  & 20  & FPBA & 1.1  & 0.94  & S1 \\
\midrule
8.2-8.4  & 20  & FPBA & 1.1  & 0.94  & DBP5 \\
\midrule
8.45-8.55  & 25  & FPBA & 2.0  & 0.87  & D4(1) \\
\midrule
8.4-9.05  & 20  & FPBA & 1.3  & 0.91  & D4(2-11), SBP2, S2 \\
\midrule
9.05-9.095  & 20  & PBA & 1.3  & 0.91  & D5 \\
\midrule
9.1-9.2  & 20  & FPBA & 1.3  & 0.91  & DBP6, DBP7 \\
\midrule
9.2-9.4  & 20  & FPBA & 1.4  & 0.91  & DBP8 \\
\midrule
9.4-9.6  & 20  & FPBA & 1.5  & 0.87  & DBP9 \\
\midrule
9.6-9.9  & 20  & FPBA & 1.7  & 0.87  & D6 \\
\bottomrule
\end{tabular}
\label{simu-setting-ds-ee}
\end{table}
%\newpage

%%%%%% table simulation settings (dipole, sextupole)(MM)  %%%%%%%%%
%\subsection{\label{sec:para-mono-ee}Parameter Settings for Dipole and Sextupole Modes (MM)}
\begin{table}[h]
\setlength\tabcolsep{8pt}
\centering
\caption{Parameters setting for dipole and sextupole modes with magnetic (MM) boundaries. }
\medskip
\begin{tabular}{m{2.3cm}m{2.2cm}m{1.2cm}m{2cm}m{2cm}m{2.7cm}}
%\begin{tabular}{ccccc}
\toprule
\textbf{Frequency range} (GHz) & \textbf{Lines per wavelength} & \textbf{Mesh type} & \textbf{Mesh cells} (million) & \textbf{Max mesh step} (mm)  & \textbf{Band}\\
\midrule
4.1-4.2  & 70  & PBA & 2.4  & 1.04  & DBP1 \\
\midrule
4.25-5.05  & 55  & PBA & 2.1  & 1.10  & D1, DBP2 \\
\midrule
5.3-5.7  & 50  & PBA & 2.3  & 1.07  & D2, DBP3 \\
\midrule
6.1-6.3  & 40  & PBA & 1.7  & 1.21  & DBP4 \\
\midrule
6.75-7.50  & 40  & PBA & 3.0  & 0.99  & D3 \\
\midrule
7.7-8.0  & 20  & FPBA & 1.0  & 1.06  & SBP1 \\
\midrule
7.85-7.95  & 22  & PBA & 1.2  & 0.94  & DBP5 \\
\midrule
8.0-8.3  & 20  & FPBA & 1.1  & 0.94  & S1 \\
\midrule
8.4-9.0  & 20  & FPBA & 1.3  & 0.91  & DBP6, D4(1-9), SBP2, S2 \\
\midrule
8.95-9.05  & 20  & FPBA & 1.3  & 0.91  & D4(10-11) \\
\midrule
9.05-9.1  & 16  & FPBA & 0.8  & 1.08  & D5 \\
\midrule
9.1-9.3  & 20  & FPBA & 1.4  & 0.91  & DBP7 \\
\midrule
9.3-9.5  & 20  & FPBA & 1.4  & 0.91  & DBP8 \\
\midrule
9.5-9.6  & 20  & FPBA & 1.5  & 0.87  & DBP9 \\
\midrule
9.6-9.9  & 20  & PBA & 1.7  & 0.87  & DBP10, D6 \\
\bottomrule
\end{tabular}
\label{simu-setting-ds-mm}
\end{table}
%\newpage

%%%%%%%%%%%%%%%%  Section 2  %%%%%%%%%%%%%%%%%%%%%%%%
%                                                                                                                                          %
%                                           Mode Distributions                                                                      %
%                                                                                                                                          %
%%%%%%%%%%%%%%%%%%%%%%%%%%%%%%%%%%%%%%%%%%%%%
\chapter{\label{sec:dist}Electric F{}ield Distributions of Modes}
Tables of parameter settings shown in this chapter are from Appendix~\ref{sec:para-list}.

%%%%%%%%%%%%%%%  Section 2  %%%%%%%%%%%%%%%%%
%                                                                                                                %
%                                           Monopole (EE)                                                %
%        																	%
%%%%%%%%%%%%%%%%%%%%%%%%%%%%%%%%%%%%%
\newpage
\section{\label{sec:mono-ee}Monopole (Electric Boundaries)}

% M1 (EE)
\subsection{M1 (EE)}
% parameter
\begin{table}[h]
\setlength\tabcolsep{8pt}
\centering
\caption{Parameters settings for M1 with electric (EE) boundaries. }
\medskip
\begin{tabular}{m{2.3cm}m{2.2cm}m{1.2cm}m{2.1cm}m{2cm}m{1.4cm}}
%\begin{tabular}{ccccc}
\toprule
\textbf{Frequency range} (GHz) & \textbf{Lines per wavelength} & \textbf{Mesh type} & \textbf{Number of Mesh cells} & \textbf{Max mesh step} (mm) & \textbf{Band} \\
\midrule
3.70-3.95  & 70  & PBA & 2,100,000  & 1.10 & M1\\
\bottomrule
\end{tabular}
\label{simu-setting-M1-EE}
\end{table}
% E filed
\begin{table}[h]
\setlength\tabcolsep{8pt}
\centering
\caption{Monopole modes in M1 with electric (EE) boundaries.}
\medskip
\begin{tabular}{cccc}
\toprule
 \textbf{E-f\mbox{}ield Amplitude} & \textbf{f}(GHz) & \textbf{R/Q}($\Omega$/cm$^2$)  & \textbf{Band}\\
\midrule
\includegraphics[width=0.45\textwidth]{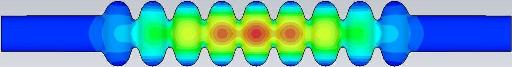}    & \raisebox{+1.8\height}{3.7466}    & \raisebox{+1.8\height}{0.008}  & \raisebox{+1.8\height}{M1-1}\\
\midrule
\includegraphics[width=0.45\textwidth]{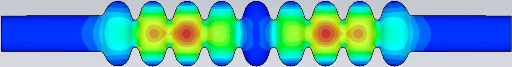}    & \raisebox{+1.8\height}{3.7601}    & \raisebox{+1.8\height}{0.061}  & \raisebox{+1.8\height}{M1-2}\\
\midrule
\includegraphics[width=0.45\textwidth]{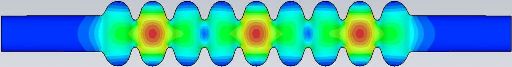}    & \raisebox{+1.8\height}{3.7808}    & \raisebox{+1.8\height}{0.090}  & \raisebox{+1.8\height}{M1-3}\\
\midrule
\includegraphics[width=0.45\textwidth]{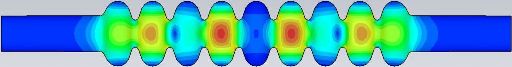}    & \raisebox{+1.8\height}{3.8065}    & \raisebox{+1.8\height}{0.170}  & \raisebox{+1.8\height}{M1-4}\\
\midrule
\includegraphics[width=0.45\textwidth]{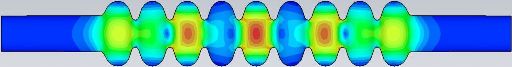}    & \raisebox{+1.8\height}{3.8340}    & \raisebox{+1.8\height}{0.307}  & \raisebox{+1.8\height}{M1-5}\\
\midrule
\includegraphics[width=0.45\textwidth]{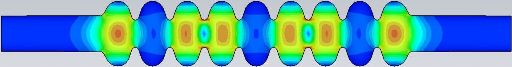}    & \raisebox{+1.8\height}{3.8602}    & \raisebox{+1.8\height}{0.203}  & \raisebox{+1.8\height}{M1-6}\\
\midrule
\includegraphics[width=0.45\textwidth]{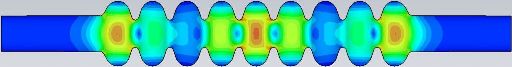}    & \raisebox{+1.8\height}{3.8817}    & \raisebox{+1.8\height}{0.468}  & \raisebox{+1.8\height}{M1-7}\\
\midrule
\includegraphics[width=0.45\textwidth]{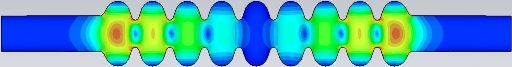}    & \raisebox{+1.8\height}{3.8958}    & \raisebox{+1.8\height}{0.195}  & \raisebox{+1.8\height}{M1-8}\\
\midrule
\includegraphics[width=0.45\textwidth]{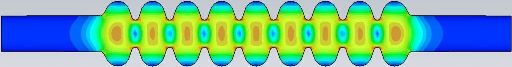}    & \raisebox{+1.8\height}{3.9008}    & \raisebox{+1.8\height}{373.113}  & \raisebox{+1.8\height}{M1-9}\\
\bottomrule
\end{tabular}
\label{simu-table-M1-EE}
\end{table}
\newpage

% MBP1,2,3,4 (EE)
\subsection{MBP1, MBP2, MBP3 and MBP4 (EE)}
% parameter
\begin{table}[h]
\setlength\tabcolsep{8pt}
\centering
\caption{Parameters settings for MBP1, MBP2, MBP3 and MBP4 with electric (EE) boundaries. }
\medskip
\begin{tabular}{m{2.3cm}m{2.2cm}m{1.2cm}m{2.1cm}m{2cm}m{1.4cm}}
%\begin{tabular}{ccccc}
\toprule
\textbf{Frequency range} (GHz) & \textbf{Lines per wavelength} & \textbf{Mesh type} & \textbf{Number of Mesh cells} & \textbf{Max mesh step} (mm) & \textbf{Band} \\
\midrule
5.7-5.8  & 50  & PBA & 2,400,000  & 1.07  & MBP1 \\
\midrule
5.95-6.05  & 50  & PBA & 2,600,000  & 1.03  & MBP2 \\
\midrule
6.4-6.5  & 45  & PBA  & 2,400,000  & 1.04  & MBP3 \\
\midrule
6.85-7.60  & 25  & FPBA  & 1,400,000  & 0.91  & MBP4 \\
\bottomrule
\end{tabular}
\label{simu-setting-MBP1-4-EE}
\end{table}
% E field
\begin{table}[h]
\setlength\tabcolsep{8pt}
\centering
\caption{Monopole modes in MBP1, MBP2, MBP3 and MBP4 with electric (EE) boundaries.}
\medskip
\begin{tabular}{cccc}
\toprule
 \textbf{E-f\mbox{}ield Amplitude} & \textbf{f}(GHz) & \textbf{R/Q}($\Omega$/cm$^2$)  & \textbf{Band}\\
\midrule
% MBP1
\includegraphics[width=0.45\textwidth]{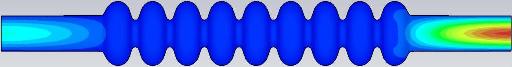}    & \raisebox{+1.8\height}{5.7685}    & \raisebox{+1.8\height}{6.967}  & \raisebox{+1.8\height}{MBP1-1}\\
\midrule
\includegraphics[width=0.45\textwidth]{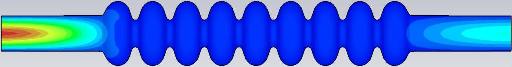}    & \raisebox{+1.8\height}{5.7685}    & \raisebox{+1.8\height}{2.443}  & \raisebox{+1.8\height}{MBP1-2}\\
\midrule
% MBP2
\includegraphics[width=0.45\textwidth]{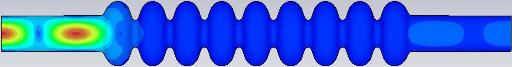}    & \raisebox{+1.8\height}{6.0123}    & \raisebox{+1.8\height}{2.867}  & \raisebox{+1.8\height}{MBP2-1}\\
\midrule
\includegraphics[width=0.45\textwidth]{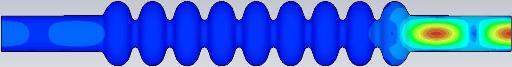}    & \raisebox{+1.8\height}{6.0123}    & \raisebox{+1.8\height}{3.335}  & \raisebox{+1.8\height}{MBP2-2}\\
\midrule
% MBP3
\includegraphics[width=0.45\textwidth]{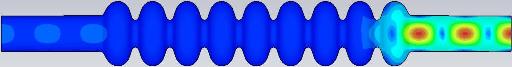}    & \raisebox{+1.8\height}{6.4403}    & \raisebox{+1.8\height}{5.212}  & \raisebox{+1.8\height}{MBP3-1}\\
\midrule
\includegraphics[width=0.45\textwidth]{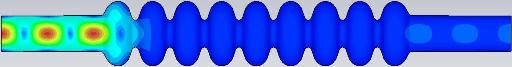}    & \raisebox{+1.8\height}{6.4403}    & \raisebox{+1.8\height}{6.116}  & \raisebox{+1.8\height}{MBP3-2}\\
\midrule
% MBP4
\includegraphics[width=0.45\textwidth]{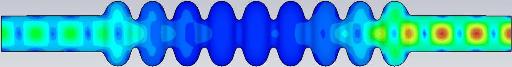}    & \raisebox{+1.8\height}{6.9393}    & \raisebox{+1.8\height}{8.454}  & \raisebox{+1.8\height}{MBP4-1}\\
\midrule
\includegraphics[width=0.45\textwidth]{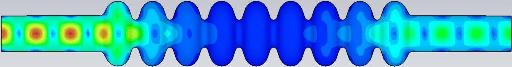}    & \raisebox{+1.8\height}{6.9394}    & \raisebox{+1.8\height}{3.971}  & \raisebox{+1.8\height}{MBP4-2}\\
\bottomrule
\end{tabular}
\label{simu-table-MBP1-4-EE}
\end{table}
\newpage

% M2
\subsection{M2 (EE)}
% parameter
\begin{table}[h]
\setlength\tabcolsep{8pt}
\centering
\caption{Parameters settings for M2 with electric (EE) boundaries. }
\medskip
\begin{tabular}{m{2.3cm}m{2.2cm}m{1.2cm}m{2.1cm}m{2cm}m{1.4cm}}
%\begin{tabular}{ccccc}
\toprule
\textbf{Frequency range} (GHz) & \textbf{Lines per wavelength} & \textbf{Mesh type} & \textbf{Number of Mesh cells} & \textbf{Max mesh step} (mm) & \textbf{Band} \\
\midrule
6.85-7.6  & 25  & FPBA  & 1,400,000  & 0.91  & M2(1-7) \\
\midrule
7.5-8.25  & 20  & FPBA  & 1,100,000  & 1.06  & M2(8-9) \\
\bottomrule
\end{tabular}
\label{simu-setting-M2-EE}
\end{table}
% E filed
\begin{table}[h]
\setlength\tabcolsep{8pt}
\centering
\caption{Monopole modes in M2 with electric (EE) boundaries.}
\medskip
\begin{tabular}{cccc}
\toprule
 \textbf{E-f\mbox{}ield Amplitude} & \textbf{f}(GHz) & \textbf{R/Q}($\Omega$/cm$^2$)  & \textbf{Band}\\
\midrule
\includegraphics[width=0.45\textwidth]{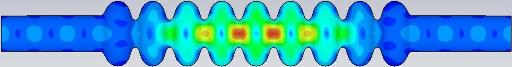}    & \raisebox{+1.8\height}{7.0483}    & \raisebox{+1.8\height}{0.127}  & \raisebox{+1.8\height}{M2-1}\\
\midrule
\includegraphics[width=0.45\textwidth]{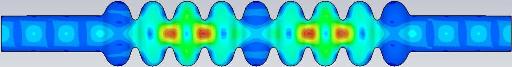}    & \raisebox{+1.8\height}{7.0863}    & \raisebox{+1.8\height}{1.012}  & \raisebox{+1.8\height}{M2-2}\\
\midrule
\includegraphics[width=0.45\textwidth]{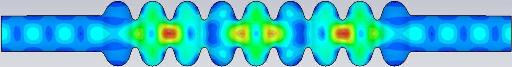}    & \raisebox{+1.8\height}{7.1424}    & \raisebox{+1.8\height}{0.088}  & \raisebox{+1.8\height}{M2-3}\\
\midrule
\includegraphics[width=0.45\textwidth]{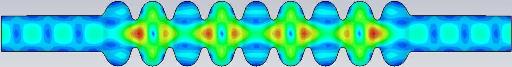}    & \raisebox{+1.8\height}{7.2113}    & \raisebox{+1.8\height}{2.914}  & \raisebox{+1.8\height}{M2-4}\\
\midrule
\includegraphics[width=0.45\textwidth]{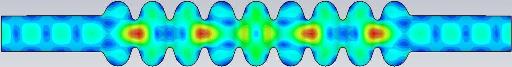}    & \raisebox{+1.8\height}{7.2877}    & \raisebox{+1.8\height}{0.677}  & \raisebox{+1.8\height}{M2-5}\\
\midrule
\includegraphics[width=0.45\textwidth]{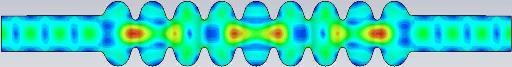}    & \raisebox{+1.8\height}{7.3662}    & \raisebox{+1.8\height}{2.727}  & \raisebox{+1.8\height}{M2-6}\\
\midrule
\includegraphics[width=0.45\textwidth]{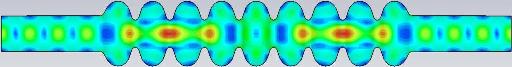}    & \raisebox{+1.8\height}{7.4418}    & \raisebox{+1.8\height}{10.686}  & \raisebox{+1.8\height}{M2-7}\\
\midrule
\includegraphics[width=0.45\textwidth]{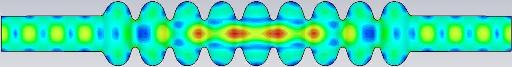}    & \raisebox{+1.8\height}{7.5118}    & \raisebox{+1.8\height}{18.963}  & \raisebox{+1.8\height}{M2-8}\\
\midrule
\includegraphics[width=0.45\textwidth]{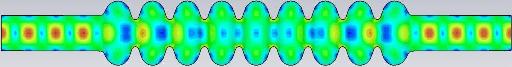}    & \raisebox{+1.8\height}{7.5843}    & \raisebox{+1.8\height}{47.909}  & \raisebox{+1.8\height}{M2-9}\\
\bottomrule
\end{tabular}
\label{simu-table-M2-EE}
\end{table}
\newpage

% M3
\subsection{M3 (EE)}
% parameter
\begin{table}[h]
\setlength\tabcolsep{8pt}
\centering
\caption{Parameters settings for M3 with electric (EE) boundaries. }
\medskip
\begin{tabular}{m{2.3cm}m{2.2cm}m{1.2cm}m{2.1cm}m{2cm}m{1.4cm}}
%\begin{tabular}{ccccc}
\toprule
\textbf{Frequency range} (GHz) & \textbf{Lines per wavelength} & \textbf{Mesh type} & \textbf{Number of Mesh cells} & \textbf{Max mesh step} (mm) & \textbf{Band} \\
\midrule
7.50-8.25  & 20  & FPBA  & 1,100,000  & 1.06 & M3 \\
\bottomrule
\end{tabular}
\label{simu-setting-M3-EE}
\end{table}
% E filed
\begin{table}[h]
\setlength\tabcolsep{8pt}
\centering
\caption{Monopole modes in M3 with electric (EE) boundaries.}
\medskip
\begin{tabular}{cccc}
\toprule
 \textbf{E-f\mbox{}ield Amplitude} & \textbf{f}(GHz) & \textbf{R/Q}($\Omega$/cm$^2$)  & \textbf{Band}\\
\midrule
\includegraphics[width=0.45\textwidth]{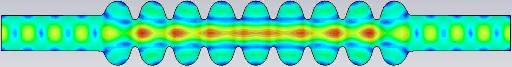}    & \raisebox{+1.8\height}{7.6443}    & \raisebox{+1.8\height}{0.593}  & \raisebox{+1.8\height}{M3-1}\\
\midrule
\includegraphics[width=0.45\textwidth]{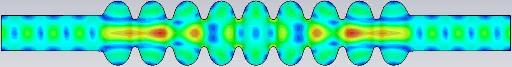}    & \raisebox{+1.8\height}{7.7248}    & \raisebox{+1.8\height}{21.212}  & \raisebox{+1.8\height}{M3-2}\\
\midrule
\includegraphics[width=0.45\textwidth]{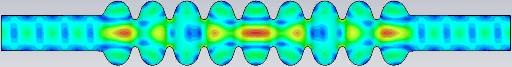}    & \raisebox{+1.8\height}{7.8036}    & \raisebox{+1.8\height}{0.245}  & \raisebox{+1.8\height}{M3-3}\\
\midrule
\includegraphics[width=0.45\textwidth]{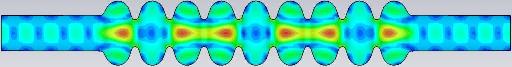}    & \raisebox{+1.8\height}{7.8809}    & \raisebox{+1.8\height}{1.707}  & \raisebox{+1.8\height}{M3-4}\\
\midrule
\includegraphics[width=0.45\textwidth]{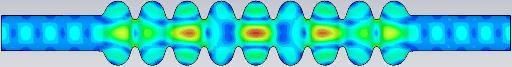}    & \raisebox{+1.8\height}{7.9547}    & \raisebox{+1.8\height}{0.640}  & \raisebox{+1.8\height}{M3-5}\\
\midrule
\includegraphics[width=0.45\textwidth]{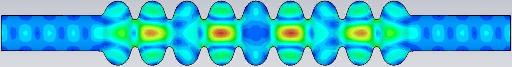}    & \raisebox{+1.8\height}{8.0229}    & \raisebox{+1.8\height}{0.015}  & \raisebox{+1.8\height}{M3-6}\\
\midrule
\includegraphics[width=0.45\textwidth]{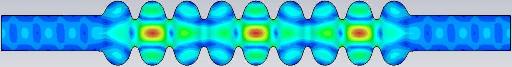}    & \raisebox{+1.8\height}{8.0829}    & \raisebox{+1.8\height}{0.113}  & \raisebox{+1.8\height}{M3-7}\\
\midrule
\includegraphics[width=0.45\textwidth]{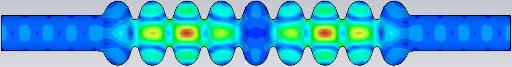}    & \raisebox{+1.8\height}{8.1311}    & \raisebox{+1.8\height}{0.014}  & \raisebox{+1.8\height}{M3-8}\\
\midrule
\includegraphics[width=0.45\textwidth]{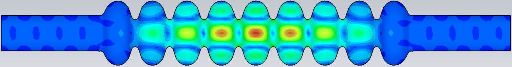}    & \raisebox{+1.8\height}{8.1631}    & \raisebox{+1.8\height}{0.001}  & \raisebox{+1.8\height}{M3-9}\\
\bottomrule
\end{tabular}
\label{simu-table-M3-EE}
\end{table}
\newpage

% MBP5,6 (EE)
\subsection{MBP5 and MBP6 (EE)}
% parameter
\begin{table}[h]
\setlength\tabcolsep{8pt}
\centering
\caption{Parameters settings for MBP5 and MBP6 with electric (EE) boundaries. }
\medskip
\begin{tabular}{m{2.3cm}m{2.2cm}m{1.2cm}m{2.1cm}m{2cm}m{1.4cm}}
%\begin{tabular}{ccccc}
\toprule
\textbf{Frequency range} (GHz) & \textbf{Lines per wavelength} & \textbf{Mesh type} & \textbf{Number of Mesh cells} & \textbf{Max mesh step} (mm) & \textbf{Band} \\
\midrule
8.25-8.35  & 20  & FPBA  & 1,100,000  & 0.94  & MBP5 \\
\midrule
8.9-9.35  & 24  & FPBA  & 2,300,000  & 0.82  & MBP6 \\
\bottomrule
\end{tabular}
\label{simu-setting-MBP5-6-EE}
\end{table}
% E filed
\begin{table}[h]
\setlength\tabcolsep{8pt}
\centering
\caption{Monopole modes in MBP5 and MBP6 with electric (EE) boundaries.}
\medskip
\begin{tabular}{cccc}
\toprule
\textbf{E-f\mbox{}ield Amplitude} & \textbf{f}(GHz) & \textbf{R/Q}($\Omega$/cm$^2$)  & \textbf{Band}\\
\midrule
% MBP5
\includegraphics[width=0.45\textwidth]{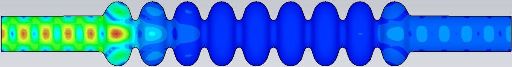}    & \raisebox{+1.8\height}{8.3376}    & \raisebox{+1.8\height}{0.765}  & \raisebox{+1.8\height}{MBP5-1}\\
\midrule
\includegraphics[width=0.45\textwidth]{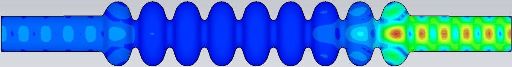}    & \raisebox{+1.8\height}{8.3376}    & \raisebox{+1.8\height}{1.045}  & \raisebox{+1.8\height}{MBP5-2}\\
\midrule
% MBP6
\includegraphics[width=0.45\textwidth]{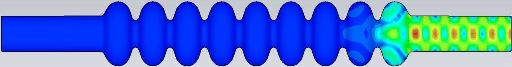}    & \raisebox{+1.8\height}{9.1202}    & \raisebox{+1.8\height}{1.670}  & \raisebox{+1.8\height}{MBP6-1}\\
\midrule
\includegraphics[width=0.45\textwidth]{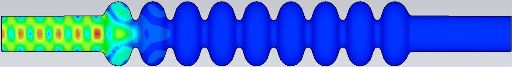}    & \raisebox{+1.8\height}{9.1202}    & \raisebox{+1.8\height}{1.755}  & \raisebox{+1.8\height}{MBP6-2}\\
\bottomrule
\end{tabular}
\label{simu-table-MBP5-6-EE}
\end{table}
\newpage

% M4
\subsection{M4 (EE)}
% parameter
\begin{table}[h]
\setlength\tabcolsep{8pt}
\centering
\caption{Parameters settings for M4 with electric (EE) boundaries. }
\medskip
\begin{tabular}{m{2.3cm}m{2.2cm}m{1.2cm}m{2.1cm}m{2cm}m{1.4cm}}
%\begin{tabular}{ccccc}
\toprule
\textbf{Frequency range} (GHz) & \textbf{Lines per wavelength} & \textbf{Mesh type} & \textbf{Number of Mesh cells} & \textbf{Max mesh step} (mm) & \textbf{Band} \\
\midrule
9.7-10.5  & 20  & FPBA & 2,000,000  & 0.87 & M4 \\
\bottomrule
\end{tabular}
\label{simu-setting-M4-EE}
\end{table}
% E filed
\begin{table}[h]
\setlength\tabcolsep{8pt}
\centering
\caption{Monopole modes in M4 with electric (EE) boundaries.}
\medskip
\begin{tabular}{cccc}
\toprule
\textbf{E-f\mbox{}ield Amplitude} & \textbf{f}(GHz) & \textbf{R/Q}($\Omega$/cm$^2$)  & \textbf{Band}\\
\midrule
\includegraphics[width=0.45\textwidth]{E-abs-Xaxis-M4-1-mode18-EE}    & \raisebox{+1.8\height}{9.7966}    & \raisebox{+1.8\height}{0.000}  & \raisebox{+1.8\height}{M4-1}\\
\midrule
\includegraphics[width=0.45\textwidth]{E-abs-Xaxis-M4-2-mode15-EE}    & \raisebox{+1.8\height}{9.8340}    & \raisebox{+1.8\height}{1.511}  & \raisebox{+1.8\height}{M4-2}\\
\midrule
\includegraphics[width=0.45\textwidth]{E-abs-Xaxis-M4-3-mode12-EE}    & \raisebox{+1.8\height}{9.8868}    & \raisebox{+1.8\height}{0.395}  & \raisebox{+1.8\height}{M4-3}\\
\midrule
\includegraphics[width=0.45\textwidth]{E-abs-Xaxis-M4-4-mode8-EE}    & \raisebox{+1.8\height}{9.9384}    & \raisebox{+1.8\height}{7.388}  & \raisebox{+1.8\height}{M4-4}\\
\midrule
\includegraphics[width=0.45\textwidth]{E-abs-Xaxis-M4-5-mode6-EE}    & \raisebox{+1.8\height}{9.9886}    & \raisebox{+1.8\height}{0.015}  & \raisebox{+1.8\height}{M4-5}\\
\midrule
\includegraphics[width=0.45\textwidth]{E-abs-Xaxis-M4-6-mode2-EE}    & \raisebox{+1.8\height}{10.0619}    & \raisebox{+1.8\height}{0.921}  & \raisebox{+1.8\height}{M4-6}\\
\midrule
\includegraphics[width=0.45\textwidth]{E-abs-Xaxis-M4-7-mode3-EE}    & \raisebox{+1.8\height}{10.1692}    & \raisebox{+1.8\height}{1.125}  & \raisebox{+1.8\height}{M4-7}\\
\midrule
\includegraphics[width=0.45\textwidth]{E-abs-Xaxis-M4-8-mode10-EE}    & \raisebox{+1.8\height}{10.3015}    & \raisebox{+1.8\height}{0.778}  & \raisebox{+1.8\height}{M4-8}\\
\midrule
\includegraphics[width=0.45\textwidth]{E-abs-Xaxis-M4-9-mode20-EE}    & \raisebox{+1.8\height}{10.4485}    & \raisebox{+1.8\height}{0.277}  & \raisebox{+1.8\height}{M4-9}\\
\bottomrule
\end{tabular}
\label{simu-table-M4-EE}
\end{table}
\newpage

%%%%%%%%%%%%%%%  Section 3  %%%%%%%%%%%%%%%%%
%                                                                                                                %
%                                           Monopole (MM)                                                %
%        																	%
%%%%%%%%%%%%%%%%%%%%%%%%%%%%%%%%%%%%%
\section{\label{sec:mono-mm}Monopole (Magnetic Boundaries)}

% M1
\subsection{M1 (MM)}
% parameter
\begin{table}[h]
\setlength\tabcolsep{8pt}
\centering
\caption{Parameters settings for M1 with magnetic (MM) boundaries. }
\medskip
\begin{tabular}{m{2.3cm}m{2.2cm}m{1.2cm}m{2.1cm}m{2cm}m{1.4cm}}
%\begin{tabular}{ccccc}
\toprule
\textbf{Frequency range} (GHz) & \textbf{Lines per wavelength} & \textbf{Mesh type} & \textbf{Number of Mesh cells} & \textbf{Max mesh step} (mm) & \textbf{Band} \\
\midrule
3.7-4.0  & 70  & PBA & 2,200,000  & 1.10  & M1 \\
\bottomrule
\end{tabular}
\label{simu-setting-M1-MM}
\end{table}
% E filed
\begin{table}[h]
\setlength\tabcolsep{8pt}
\centering
\caption{Monopole modes in M1 with magnetic (MM) boundaries.}
\medskip
\begin{tabular}{cccc}
\toprule
\textbf{E-f\mbox{}ield Amplitude} & \textbf{f}(GHz) & \textbf{R/Q}($\Omega$/cm$^2$)  & \textbf{Band}\\
\midrule
\includegraphics[width=0.45\textwidth]{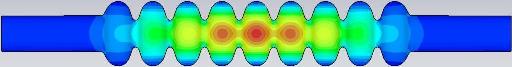}    & \raisebox{+1.8\height}{3.7466}    & \raisebox{+1.8\height}{0.008}  & \raisebox{+1.8\height}{M1-1}\\
\midrule
\includegraphics[width=0.45\textwidth]{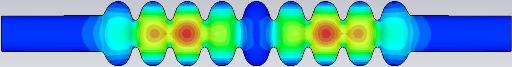}    & \raisebox{+1.8\height}{3.7601}    & \raisebox{+1.8\height}{0.061}  & \raisebox{+1.8\height}{M1-2}\\
\midrule
\includegraphics[width=0.45\textwidth]{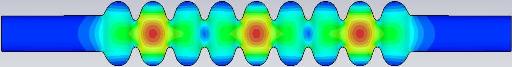}    & \raisebox{+1.8\height}{3.7808}    & \raisebox{+1.8\height}{0.090}  & \raisebox{+1.8\height}{M1-3}\\
\midrule
\includegraphics[width=0.45\textwidth]{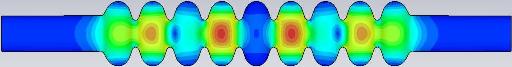}    & \raisebox{+1.8\height}{3.8065}    & \raisebox{+1.8\height}{0.170}  & \raisebox{+1.8\height}{M1-4}\\
\midrule
\includegraphics[width=0.45\textwidth]{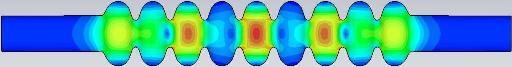}    & \raisebox{+1.8\height}{3.8341}    & \raisebox{+1.8\height}{0.309}  & \raisebox{+1.8\height}{M1-5}\\
\midrule
\includegraphics[width=0.45\textwidth]{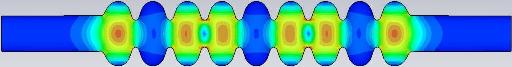}    & \raisebox{+1.8\height}{3.8602}    & \raisebox{+1.8\height}{0.204}  & \raisebox{+1.8\height}{M1-6}\\
\midrule
\includegraphics[width=0.45\textwidth]{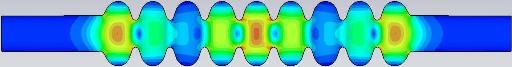}    & \raisebox{+1.8\height}{3.8817}    & \raisebox{+1.8\height}{0.481}  & \raisebox{+1.8\height}{M1-7}\\
\midrule
\includegraphics[width=0.45\textwidth]{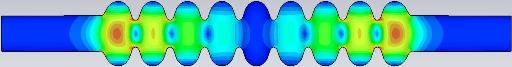}    & \raisebox{+1.8\height}{3.8958}    & \raisebox{+1.8\height}{0.197}  & \raisebox{+1.8\height}{M1-8}\\
\midrule
\includegraphics[width=0.45\textwidth]{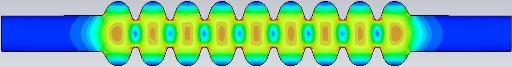}    & \raisebox{+1.8\height}{3.9008}    & \raisebox{+1.8\height}{373.097}  & \raisebox{+1.8\height}{M1-9}\\
\bottomrule
\end{tabular}
\label{simu-table-M1-MM}
\end{table}
\newpage

% MBP1, MBP2 and MBP3 (MM)
\subsection{MBP1, MBP2 and MBP3 (MM)}
% parameter
\begin{table}[h]
\setlength\tabcolsep{8pt}
\centering
\caption{Parameters settings for MBP1, MBP2 and MBP3 with magnetic (MM) boundaries. }
\medskip
\begin{tabular}{m{2.3cm}m{2.2cm}m{1.2cm}m{2.1cm}m{2cm}m{1.4cm}}
%\begin{tabular}{ccccc}
\toprule
\textbf{Frequency range} (GHz) & \textbf{Lines per wavelength} & \textbf{Mesh type} & \textbf{Number of Mesh cells} & \textbf{Max mesh step} (mm) & \textbf{Band} \\
\midrule
5.7-5.9  & 50  & PBA & 2,500,000  & 1.04  & MBP1 \\
\midrule
6.0-6.4  & 45  & PBA & 2,300,000  & 1.07  & MBP2 \\
\midrule
6.4-6.8  & 40  & PBA  & 2,000,000  & 1.11  & MBP3 \\
\bottomrule
\end{tabular}
\label{simu-setting-MBP1-3-MM}
\end{table}
% E filed
\begin{table}[h]
\setlength\tabcolsep{8pt}
\centering
\caption{Monopole modes in MBP1, MBP2 and MBP3 with magnetic (MM) boundaries.}
\medskip
\begin{tabular}{cccc}
\toprule
\textbf{E-f\mbox{}ield Amplitude} & \textbf{f}(GHz) & \textbf{R/Q}($\Omega$/cm$^2$)  & \textbf{Band}\\
\midrule
% MBP1
\includegraphics[width=0.45\textwidth]{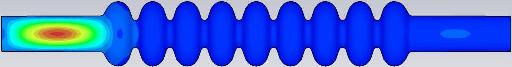}    & \raisebox{+1.8\height}{5.8624}    & \raisebox{+1.8\height}{0.051}  & \raisebox{+1.8\height}{MBP1-1}\\
\midrule
\includegraphics[width=0.45\textwidth]{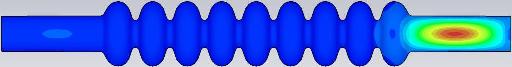}    & \raisebox{+1.8\height}{5.8624}    & \raisebox{+1.8\height}{0.051}  & \raisebox{+1.8\height}{MBP1-2}\\
\midrule
% MBP2
\includegraphics[width=0.45\textwidth]{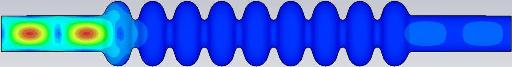}    & \raisebox{+1.8\height}{6.2095}    & \raisebox{+1.8\height}{0.879}  & \raisebox{+1.8\height}{MBP2-1}\\
\midrule
\includegraphics[width=0.45\textwidth]{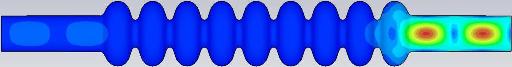}    & \raisebox{+1.8\height}{6.2095}    & \raisebox{+1.8\height}{0.665}  & \raisebox{+1.8\height}{MBP2-2}\\
\midrule
% MBP3
\includegraphics[width=0.45\textwidth]{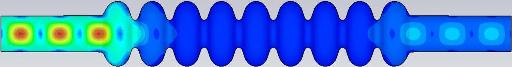}    & \raisebox{+1.8\height}{6.6886}    & \raisebox{+1.8\height}{4.119}  & \raisebox{+1.8\height}{MBP3-1}\\
\midrule
\includegraphics[width=0.45\textwidth]{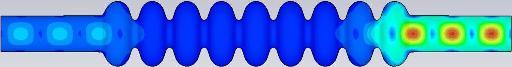}    & \raisebox{+1.8\height}{6.6886}    & \raisebox{+1.8\height}{2.302}  & \raisebox{+1.8\height}{MBP3-2}\\
\bottomrule
\end{tabular}
\label{simu-table-MBP1-3-MM}
\end{table}
\newpage

% M2
\subsection{M2 (MM)}
% parameter
\begin{table}[h]
\setlength\tabcolsep{8pt}
\centering
\caption{Parameters settings for M2 with magnetic (MM) boundaries. }
\medskip
\begin{tabular}{m{2.3cm}m{2.2cm}m{1.2cm}m{2.1cm}m{2cm}m{1.4cm}}
%\begin{tabular}{ccccc}
\toprule
\textbf{Frequency range} (GHz) & \textbf{Lines per wavelength} & \textbf{Mesh type} & \textbf{Number of Mesh cells} & \textbf{Max mesh step} (mm) & \textbf{Band} \\
\midrule
6.8-7.6  & 25  & FPBA  & 1,400,000  & 0.91  & M2 \\
\bottomrule
\end{tabular}
\label{simu-setting-M2-MM}
\end{table}
% E filed
\begin{table}[h]
\setlength\tabcolsep{8pt}
\centering
\caption{Monopole modes in M2 with magnetic (MM) boundaries.}
\medskip
\begin{tabular}{cccc}
\toprule
\textbf{E-f\mbox{}ield Amplitude} & \textbf{f}(GHz) & \textbf{R/Q}($\Omega$/cm$^2$)  & \textbf{Band}\\
\midrule
\includegraphics[width=0.45\textwidth]{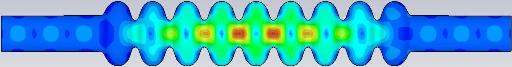}    & \raisebox{+1.8\height}{7.0449}    & \raisebox{+1.8\height}{0.007}  & \raisebox{+1.8\height}{M2-1}\\
\midrule
\includegraphics[width=0.45\textwidth]{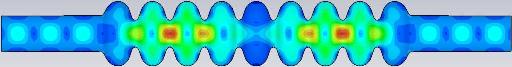}    & \raisebox{+1.8\height}{7.0738}    & \raisebox{+1.8\height}{0.034}  & \raisebox{+1.8\height}{M2-2}\\
\midrule
\includegraphics[width=0.45\textwidth]{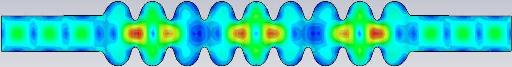}    & \raisebox{+1.8\height}{7.1145}    & \raisebox{+1.8\height}{0.127}  & \raisebox{+1.8\height}{M2-3}\\
\midrule
\includegraphics[width=0.45\textwidth]{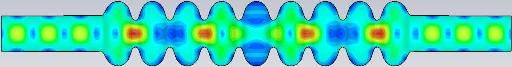}    & \raisebox{+1.8\height}{7.1589}    & \raisebox{+1.8\height}{1.012}  & \raisebox{+1.8\height}{M2-4}\\
\midrule
\includegraphics[width=0.45\textwidth]{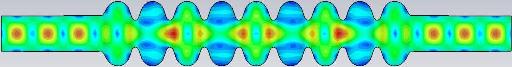}    & \raisebox{+1.8\height}{7.2058}    & \raisebox{+1.8\height}{0.088}  & \raisebox{+1.8\height}{M2-5}\\
\midrule
\includegraphics[width=0.45\textwidth]{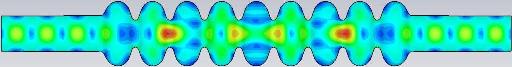}    & \raisebox{+1.8\height}{7.2625}    & \raisebox{+1.8\height}{2.914}  & \raisebox{+1.8\height}{M2-6}\\
\midrule
\includegraphics[width=0.45\textwidth]{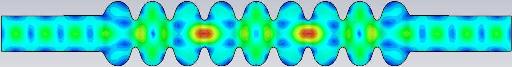}    & \raisebox{+1.8\height}{7.3331}    & \raisebox{+1.8\height}{0.677}  & \raisebox{+1.8\height}{M2-7}\\
\midrule
\includegraphics[width=0.45\textwidth]{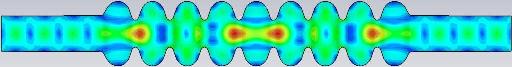}    & \raisebox{+1.8\height}{7.4140}    & \raisebox{+1.8\height}{2.727}  & \raisebox{+1.8\height}{M2-8}\\
\midrule
\includegraphics[width=0.45\textwidth]{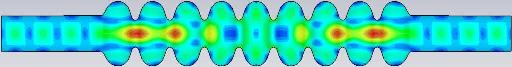}    & \raisebox{+1.8\height}{7.4995}    & \raisebox{+1.8\height}{10.686}  & \raisebox{+1.8\height}{M2-9}\\
\midrule
\includegraphics[width=0.45\textwidth]{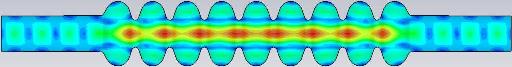}    & \raisebox{+1.8\height}{7.5810}    & \raisebox{+1.8\height}{18.963}  & \raisebox{+1.8\height}{M2-10}\\
\bottomrule
\end{tabular}
\label{simu-table-M2-MM}
\end{table}
\newpage

% M3
\subsection{M3 (MM)}
% parameter
\begin{table}[h]
\setlength\tabcolsep{8pt}
\centering
\caption{Parameters settings for M3 with magnetic (MM) boundaries. }
\medskip
\begin{tabular}{m{2.3cm}m{2.2cm}m{1.2cm}m{2.1cm}m{2cm}m{1.4cm}}
%\begin{tabular}{ccccc}
\toprule
\textbf{Frequency range} (GHz) & \textbf{Lines per wavelength} & \textbf{Mesh type} & \textbf{Number of Mesh cells} & \textbf{Max mesh step} (mm) & \textbf{Band} \\
\midrule
7.6-8.4  & 20  & FPBA  & 1,100,000  & 0.94 & M3 \\
\bottomrule
\end{tabular}
\label{simu-setting-M3-MM}
\end{table}
% E filed
\begin{table}[h]
\setlength\tabcolsep{8pt}
\centering
\caption{Monopole modes in M3 with magnetic (MM) boundaries.}
\medskip
\begin{tabular}{cccc}
\toprule
\textbf{E-f\mbox{}ield Amplitude} & \textbf{f}(GHz) & \textbf{R/Q}($\Omega$/cm$^2$)  & \textbf{Band}\\
\midrule
\includegraphics[width=0.45\textwidth]{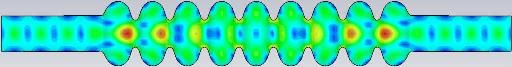}    & \raisebox{+1.8\height}{7.6780}    & \raisebox{+1.8\height}{47.909}  & \raisebox{+1.8\height}{M3-1}\\
\midrule
\includegraphics[width=0.45\textwidth]{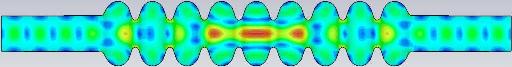}    & \raisebox{+1.8\height}{7.7577}    & \raisebox{+1.8\height}{0.593}  & \raisebox{+1.8\height}{M3-2}\\
\midrule
\includegraphics[width=0.45\textwidth]{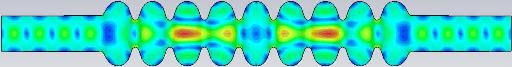}    & \raisebox{+1.8\height}{7.8333}    & \raisebox{+1.8\height}{21.212}  & \raisebox{+1.8\height}{M3-3}\\
\midrule
\includegraphics[width=0.45\textwidth]{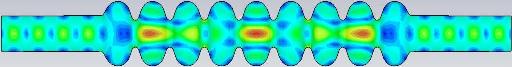}    & \raisebox{+1.8\height}{7.9021}    & \raisebox{+1.8\height}{0.245}  & \raisebox{+1.8\height}{M3-4}\\
\midrule
\includegraphics[width=0.45\textwidth]{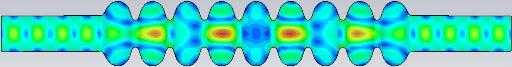}    & \raisebox{+1.8\height}{7.9625}    & \raisebox{+1.8\height}{1.707}  & \raisebox{+1.8\height}{M3-5}\\
\midrule
\includegraphics[width=0.45\textwidth]{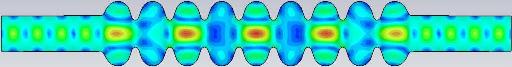}    & \raisebox{+1.8\height}{8.0148}    & \raisebox{+1.8\height}{0.640}  & \raisebox{+1.8\height}{M3-6}\\
\midrule
\includegraphics[width=0.45\textwidth]{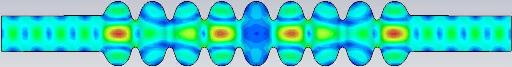}    & \raisebox{+1.8\height}{8.0615}    & \raisebox{+1.8\height}{0.015}  & \raisebox{+1.8\height}{M3-7}\\
\midrule
\includegraphics[width=0.45\textwidth]{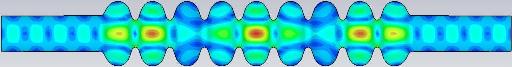}    & \raisebox{+1.8\height}{8.1044}    & \raisebox{+1.8\height}{0.113}  & \raisebox{+1.8\height}{M3-8}\\
\midrule
\includegraphics[width=0.45\textwidth]{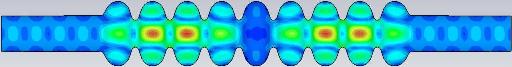}    & \raisebox{+1.8\height}{8.1408}    & \raisebox{+1.8\height}{0.014}  & \raisebox{+1.8\height}{M3-9}\\
\midrule
\includegraphics[width=0.45\textwidth]{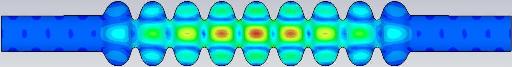}    & \raisebox{+1.8\height}{8.1656}    & \raisebox{+1.8\height}{0.001}  & \raisebox{+1.8\height}{M3-10}\\
\bottomrule
\end{tabular}
\label{simu-table-M3-MM}
\end{table}
\newpage

% MBP5 and MBP6 (MM)
\subsection{MBP5 and MBP6 (MM)}
% parameter
\begin{table}[h]
\setlength\tabcolsep{8pt}
\centering
\caption{Parameters settings for MBP5 and MBP6 with magnetic (MM) boundaries. }
\medskip
\begin{tabular}{m{2.3cm}m{2.2cm}m{1.2cm}m{2.1cm}m{2cm}m{1.4cm}}
%\begin{tabular}{ccccc}
\toprule
\textbf{Frequency range} (GHz) & \textbf{Lines per wavelength} & \textbf{Mesh type} & \textbf{Number of Mesh cells} & \textbf{Max mesh step} (mm) & \textbf{Band} \\
\midrule
8.40-8.85  & 20  & FPBA  & 1,300,000  & 0.94  & MBP5 \\
\midrule
9.3-9.6  & 20  & FPBA  & 1,500,000  & 0.87  & MBP6 \\
\bottomrule
\end{tabular}
\label{simu-setting-MBP5-6-MM}
\end{table}
% E filed
\begin{table}[h]
\setlength\tabcolsep{8pt}
\centering
\caption{Monopole modes in MBP5 and MBP6 with magnetic (MM) boundaries.}
\medskip
\begin{tabular}{cccc}
\toprule
\textbf{E-f\mbox{}ield Amplitude} & \textbf{f}(GHz) & \textbf{R/Q}($\Omega$/cm$^2$)  & \textbf{Band}\\
\midrule
% MBP5
\includegraphics[width=0.45\textwidth]{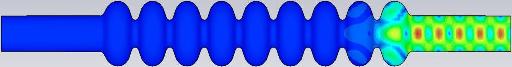}    & \raisebox{+1.8\height}{8.7111}    & \raisebox{+1.8\height}{0.521}  & \raisebox{+1.8\height}{MBP5-1}\\
\midrule
\includegraphics[width=0.45\textwidth]{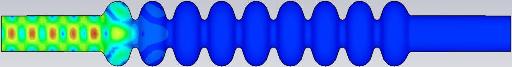}    & \raisebox{+1.8\height}{8.7113}    & \raisebox{+1.8\height}{0.525}  & \raisebox{+1.8\height}{MBP5-2}\\
\midrule
% MBP6
\includegraphics[width=0.45\textwidth]{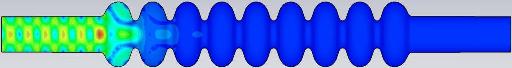}    & \raisebox{+1.8\height}{9.5377}    & \raisebox{+1.8\height}{1.912}  & \raisebox{+1.8\height}{MBP6-1}\\
\midrule
\includegraphics[width=0.45\textwidth]{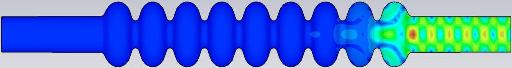}    & \raisebox{+1.8\height}{9.5378}    & \raisebox{+1.8\height}{1.945}  & \raisebox{+1.8\height}{MBP6-2}\\
\bottomrule
\end{tabular}
\label{simu-table-MBP5-6-MM}
\end{table}
\newpage

% M4
\subsection{M4 (MM)}
% parameter
\begin{table}[h]
\setlength\tabcolsep{8pt}
\centering
\caption{Parameters settings for M4 with magnetic (MM) boundaries. }
\medskip
\begin{tabular}{m{2.3cm}m{2.2cm}m{1.2cm}m{2.1cm}m{2cm}m{1.4cm}}
%\begin{tabular}{ccccc}
\toprule
\textbf{Frequency range} (GHz) & \textbf{Lines per wavelength} & \textbf{Mesh type} & \textbf{Number of Mesh cells} & \textbf{Max mesh step} (mm) & \textbf{Band} \\
\midrule
9.7-10.0  & 20  & PBA  & 1,800,000  & 0.87 & M4(1-3) \\
\midrule
10.0-10.6  & 20  & PBA  & 2,000,000  & 0.87 & M4(4-9) \\
\bottomrule
\end{tabular}
\label{simu-setting-M4-MM}
\end{table}
% E filed
\begin{table}[h]
\setlength\tabcolsep{8pt}
\centering
\caption{Monopole modes in M4 with magnetic (MM) boundaries.}
\medskip
\begin{tabular}{cccc}
\toprule
\textbf{E-f\mbox{}ield Amplitude} & \textbf{f}(GHz) & \textbf{R/Q}($\Omega$/cm$^2$)  & \textbf{Band}\\
\midrule
\includegraphics[width=0.45\textwidth]{E-abs-Xaxis-M4-1-mode4-MM}    & \raisebox{+1.8\height}{9.7907}    & \raisebox{+1.8\height}{0.000}  & \raisebox{+1.8\height}{M4-1}\\
\midrule
\includegraphics[width=0.45\textwidth]{E-abs-Xaxis-M4-2-mode2-MM}    & \raisebox{+1.8\height}{9.8379}    & \raisebox{+1.8\height}{0.072}  & \raisebox{+1.8\height}{M4-2}\\
\midrule
\includegraphics[width=0.45\textwidth]{E-abs-Xaxis-M4-3-mode5-MM}    & \raisebox{+1.8\height}{9.9124}    & \raisebox{+1.8\height}{0.360}  & \raisebox{+1.8\height}{M4-3}\\
\midrule
\includegraphics[width=0.45\textwidth]{E-abs-Xaxis-M4-4-mode23-MM}    & \raisebox{+1.8\height}{10.0099}    & \raisebox{+1.8\height}{5.067}  & \raisebox{+1.8\height}{M4-4}\\
\midrule
\includegraphics[width=0.45\textwidth]{E-abs-Xaxis-M4-5-mode18-MM}    & \raisebox{+1.8\height}{10.1270}    & \raisebox{+1.8\height}{0.264}  & \raisebox{+1.8\height}{M4-5}\\
\midrule
\includegraphics[width=0.45\textwidth]{E-abs-Xaxis-M4-6-mode3-MM}    & \raisebox{+1.8\height}{10.2547}    & \raisebox{+1.8\height}{0.723}  & \raisebox{+1.8\height}{M4-6}\\
\midrule
\includegraphics[width=0.45\textwidth]{E-abs-Xaxis-M4-7-mode4-MM}    & \raisebox{+1.8\height}{10.3692}    & \raisebox{+1.8\height}{2.342}  & \raisebox{+1.8\height}{M4-7}\\
\midrule
\includegraphics[width=0.45\textwidth]{E-abs-Xaxis-M4-8-mode8-MM}    & \raisebox{+1.8\height}{10.4406}    & \raisebox{+1.8\height}{0.312}  & \raisebox{+1.8\height}{M4-8}\\
\midrule
\includegraphics[width=0.45\textwidth]{E-abs-Xaxis-M4-9-mode20-MM}    & \raisebox{+1.8\height}{10.5080}    & \raisebox{+1.8\height}{0.719}  & \raisebox{+1.8\height}{M4-9}\\
\bottomrule
\end{tabular}
\label{simu-table-M4-MM}
\end{table}
\newpage

%%%%%%%%%%%%%%%  Section 4  %%%%%%%%%%%%%%%%%
%                                                                                                                %
%                                             Dipole (EE)                                                   %
%        																	%
%%%%%%%%%%%%%%%%%%%%%%%%%%%%%%%%%%%%%
\section{\label{sec:dipole-ee}Dipole (Electric Boundaries)}

% DBP1
\subsection{DBP1 (EE)}
% parameter
\begin{table}[h]
\setlength\tabcolsep{8pt}
\centering
\caption{Parameters settings for DBP1 with electric (EE) boundaries. }
\medskip
\begin{tabular}{m{2.3cm}m{2.2cm}m{1.2cm}m{2.1cm}m{2cm}m{1.4cm}}
\toprule
\textbf{Frequency range} (GHz) & \textbf{Lines per wavelength} & \textbf{Mesh type} & \textbf{Number of Mesh cells} & \textbf{Max mesh step} (mm) & \textbf{Band} \\
\midrule
4.1-4.2  & 70  & PBA & 2,400,000  & 1.04  & DBP1 \\
\bottomrule
\end{tabular}
\label{simu-setting-DBP1-EE}
\end{table}
% E filed
\begin{table}[h]
\setlength\tabcolsep{8pt}
\centering
\caption{Dipole modes in DBP1 with electric (EE) boundaries.}
\medskip
\begin{tabular}{cccc}
\toprule
\textbf{E-f\mbox{}ield Amplitude} & \textbf{f}(GHz) & \textbf{R/Q}($\Omega$/cm$^2$)  & \textbf{Band}\\
\midrule
\includegraphics[width=0.45\textwidth]{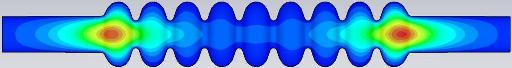}    & \raisebox{+1.8\height}{4.1489}    & \raisebox{+1.8\height}{0.234}  & \raisebox{+1.8\height}{DBP1-1}\\
\midrule
\includegraphics[width=0.45\textwidth]{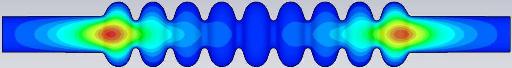}    & \raisebox{+1.8\height}{4.1491}    & \raisebox{+1.8\height}{1.318}  & \raisebox{+1.8\height}{DBP1-2}\\
\bottomrule
\end{tabular}
\label{simu-table-DBP1-EE}
\end{table}
\newpage

% D1
\subsection{D1 (EE)}
% parameter
\begin{table}[h]
\setlength\tabcolsep{8pt}
\centering
\caption{Parameters settings for D1 with electric (EE) boundaries. }
\medskip
\begin{tabular}{m{2.3cm}m{2.2cm}m{1.2cm}m{2.1cm}m{2cm}m{1.4cm}}
\toprule
\textbf{Frequency range} (GHz) & \textbf{Lines per wavelength} & \textbf{Mesh type} & \textbf{Number of Mesh cells} & \textbf{Max mesh step} (mm) & \textbf{Band} \\
\midrule
4.25-5.05  & 60  & PBA & 2,600,000  & 1.03  & D1(1-9) \\
\midrule
4.95-5.30  & 45  & PBA & 1,500,000  & 1.29  & D1(10) \\
\bottomrule
\end{tabular}
\label{simu-setting-D1-EE}
\end{table}
% E filed
\begin{table}[h]
\setlength\tabcolsep{8pt}
\centering
\caption{Dipole modes in D1 with electric (EE) boundaries.}
\medskip
\begin{tabular}{cccc}
\toprule
\textbf{E-f\mbox{}ield Amplitude} & \textbf{f}(GHz) & \textbf{R/Q}($\Omega$/cm$^2$)  & \textbf{Band}\\
\midrule
\includegraphics[width=0.44\textwidth]{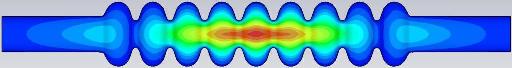}    & \raisebox{+1.8\height}{4.2982}    & \raisebox{+1.8\height}{0.001}  & \raisebox{+1.8\height}{D1-1}\\
\midrule
\includegraphics[width=0.44\textwidth]{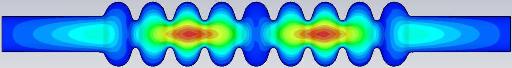}    & \raisebox{+1.8\height}{4.3607}    & \raisebox{+1.8\height}{0.292}  & \raisebox{+1.8\height}{D1-2}\\
\midrule
\includegraphics[width=0.44\textwidth]{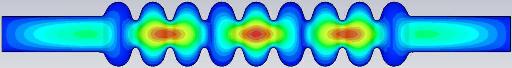}    & \raisebox{+1.8\height}{4.4485}    & \raisebox{+1.8\height}{0.002}  & \raisebox{+1.8\height}{D1-3}\\
\midrule
\includegraphics[width=0.44\textwidth]{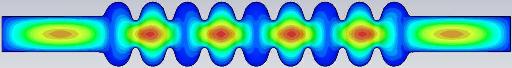}    & \raisebox{+1.8\height}{4.5410}    & \raisebox{+1.8\height}{1.076}  & \raisebox{+1.8\height}{D1-4}\\
\midrule
\includegraphics[width=0.44\textwidth]{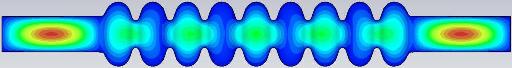}    & \raisebox{+1.8\height}{4.5989}    & \raisebox{+1.8\height}{0.784}  & \raisebox{+1.8\height}{D1-5}\\
\midrule
\includegraphics[width=0.44\textwidth]{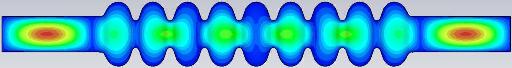}    & \raisebox{+1.8\height}{4.6415}    & \raisebox{+1.8\height}{0.165}  & \raisebox{+1.8\height}{D1-6}\\
\midrule
\includegraphics[width=0.44\textwidth]{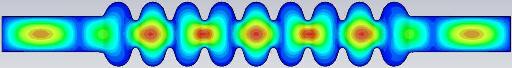}    & \raisebox{+1.8\height}{4.7245}    & \raisebox{+1.8\height}{10.572}  & \raisebox{+1.8\height}{D1-7}\\
\midrule
\includegraphics[width=0.44\textwidth]{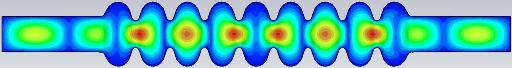}    & \raisebox{+1.8\height}{4.8327}    & \raisebox{+1.8\height}{50.307}  & \raisebox{+1.8\height}{D1-8}\\
\midrule
\includegraphics[width=0.44\textwidth]{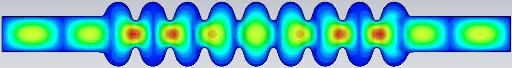}    & \raisebox{+1.8\height}{4.9270}    & \raisebox{+1.8\height}{30.174}  & \raisebox{+1.8\height}{D1-9}\\
\midrule
\includegraphics[width=0.44\textwidth]{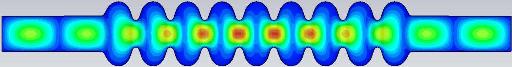}    & \raisebox{+1.8\height}{4.9899}    & \raisebox{+1.8\height}{0.000}  & \raisebox{+1.8\height}{D1-10}\\
\bottomrule
\end{tabular}
\label{simu-table-D1-EE}
\end{table}
\newpage

% DBP2 and D2 (EE)
\subsection{DBP2 and D2 (EE)}
% parameter
\begin{table}[h]
\setlength\tabcolsep{8pt}
\centering
\caption{Parameters settings for DBP2 and D2 with electric (EE) boundaries. }
\medskip
\begin{tabular}{m{2.3cm}m{2.2cm}m{1.2cm}m{2.1cm}m{2cm}m{1.4cm}}
\toprule
\textbf{Frequency range} (GHz) & \textbf{Lines per wavelength} & \textbf{Mesh type} & \textbf{Number of Mesh cells} & \textbf{Max mesh step} (mm) & \textbf{Band} \\
\midrule
4.95-5.30  & 45  & PBA & 1,500,000  & 1.29  & DBP2 \\
\midrule
5.3-5.6  & 50  & PBA & 2,200,000  & 1.10  & D2 \\
\bottomrule
\end{tabular}
\label{simu-setting-DBP2-D2-EE}
\end{table}
% E filed
\begin{table}[h]
\setlength\tabcolsep{8pt}
\centering
\caption{Dipole modes in DBP2 and D2 with electric (EE) boundaries.}
\medskip
\begin{tabular}{cccc}
\toprule
\textbf{E-f\mbox{}ield Amplitude} & \textbf{f}(GHz) & \textbf{R/Q}($\Omega$/cm$^2$)  & \textbf{Band}\\
\midrule
% DBP2
\includegraphics[width=0.44\textwidth]{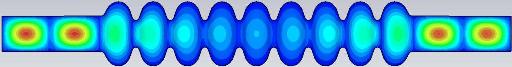}    & \raisebox{+1.8\height}{5.2014}    & \raisebox{+1.8\height}{0.300}  & \raisebox{+1.8\height}{DBP2-1}\\
\midrule
\includegraphics[width=0.44\textwidth]{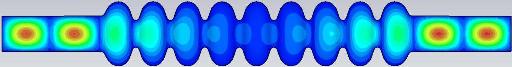}    & \raisebox{+1.8\height}{5.2040}    & \raisebox{+1.8\height}{2.036}  & \raisebox{+1.8\height}{DBP2-2}\\
% D2
\includegraphics[width=0.44\textwidth]{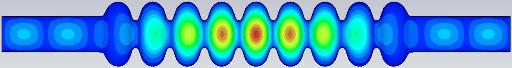}    & \raisebox{+1.8\height}{5.3581}    & \raisebox{+1.8\height}{0.041}  & \raisebox{+1.8\height}{D2-1}\\
\midrule
\includegraphics[width=0.44\textwidth]{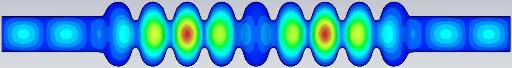}    & \raisebox{+1.8\height}{5.4050}    & \raisebox{+1.8\height}{5.057}  & \raisebox{+1.8\height}{D2-2}\\
\midrule
\includegraphics[width=0.44\textwidth]{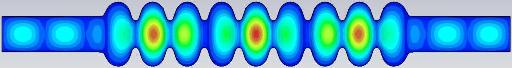}    & \raisebox{+1.8\height}{5.4427}    & \raisebox{+1.8\height}{20.877}  & \raisebox{+1.8\height}{D2-3}\\
\midrule
\includegraphics[width=0.44\textwidth]{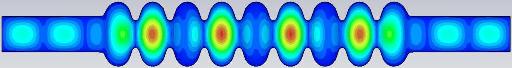}    & \raisebox{+1.8\height}{5.4678}    & \raisebox{+1.8\height}{15.776}  & \raisebox{+1.8\height}{D2-4}\\
\midrule
\includegraphics[width=0.44\textwidth]{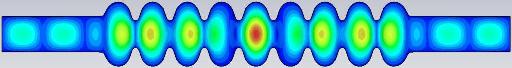}    & \raisebox{+1.8\height}{5.4829}    & \raisebox{+1.8\height}{0.895}  & \raisebox{+1.8\height}{D2-5}\\
\midrule
\includegraphics[width=0.44\textwidth]{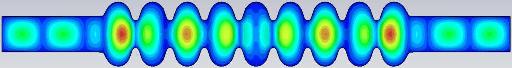}    & \raisebox{+1.8\height}{5.4911}    & \raisebox{+1.8\height}{1.261}  & \raisebox{+1.8\height}{D2-6}\\
\midrule
\includegraphics[width=0.44\textwidth]{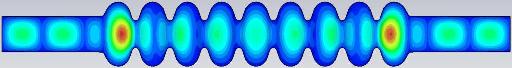}    & \raisebox{+1.8\height}{5.4950}    & \raisebox{+1.8\height}{0.307}  & \raisebox{+1.8\height}{D2-7}\\
\midrule
\includegraphics[width=0.44\textwidth]{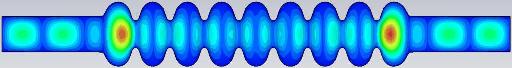}    & \raisebox{+1.8\height}{5.4958}    & \raisebox{+1.8\height}{0.549}  & \raisebox{+1.8\height}{D2-8}\\
\bottomrule
\end{tabular}
\label{simu-table-DBP2-D2-EE}
\end{table}
\newpage

% DBP3 and DBP4 (EE)
\subsection{DBP3 and DBP4 (EE)}
% parameter
\begin{table}[h]
\setlength\tabcolsep{8pt}
\centering
\caption{Parameters settings for DBP3 and DBP4 with electric (EE) boundaries. }
\medskip
\begin{tabular}{m{2.3cm}m{2.2cm}m{1.2cm}m{2.1cm}m{2cm}m{1.4cm}}
\toprule
\textbf{Frequency range} (GHz) & \textbf{Lines per wavelength} & \textbf{Mesh type} & \textbf{Number of Mesh cells} & \textbf{Max mesh step} (mm) & \textbf{Band} \\
\midrule
5.8-5.9  & 45  & PBA & 2,000,000  & 1.18  & DBP3 \\
\midrule
6.5-6.6  & 40  & PBA & 1,900,000  & 1.18  & DBP4 \\
\bottomrule
\end{tabular}
\label{simu-setting-DBP3-4-EE}
\end{table}
% E filed
\begin{table}[h]
\setlength\tabcolsep{8pt}
\centering
\caption{Dipole modes in DBP3 and DBP4 with electric (EE) boundaries.}
\medskip
\begin{tabular}{cccc}
\toprule
\textbf{E-f\mbox{}ield Amplitude} & \textbf{f}(GHz) & \textbf{R/Q}($\Omega$/cm$^2$)  & \textbf{Band}\\
\midrule
% DBP3
\includegraphics[width=0.45\textwidth]{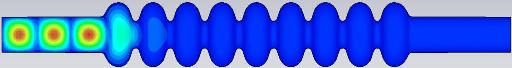}    & \raisebox{+1.8\height}{5.8644}    & \raisebox{+1.8\height}{1.028}  & \raisebox{+1.8\height}{DBP3-1}\\
\midrule
\includegraphics[width=0.45\textwidth]{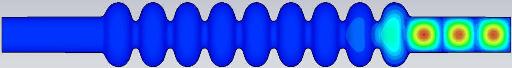}    & \raisebox{+1.8\height}{5.8644}    & \raisebox{+1.8\height}{1.026}  & \raisebox{+1.8\height}{DBP3-2}\\
\midrule
% DBP4
\includegraphics[width=0.45\textwidth]{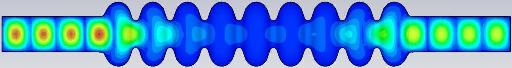}    & \raisebox{+1.8\height}{6.5593}    & \raisebox{+1.8\height}{0.344}  & \raisebox{+1.8\height}{DBP4-1}\\
\midrule
\includegraphics[width=0.45\textwidth]{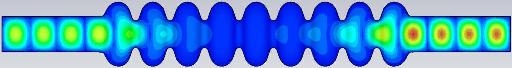}    & \raisebox{+1.8\height}{6.5594}    & \raisebox{+1.8\height}{0.397}  & \raisebox{+1.8\height}{DBP4-2}\\
\bottomrule
\end{tabular}
\label{simu-table-DBP3-4-EE}
\end{table}
\newpage

% D3
\subsection{D3 (EE)}
% parameter
\begin{table}[h]
\setlength\tabcolsep{8pt}
\centering
\caption{Parameters settings for D3 with electric (EE) boundaries. }
\medskip
\begin{tabular}{m{2.3cm}m{2.2cm}m{1.2cm}m{2.1cm}m{2cm}m{1.4cm}}
\toprule
\textbf{Frequency range} (GHz) & \textbf{Lines per wavelength} & \textbf{Mesh type} & \textbf{Number of Mesh cells} & \textbf{Max mesh step} (mm) & \textbf{Band} \\
\midrule
6.75-7.7  & 40  & PBA & 3,000,000  & 0.99  & D3 \\
\bottomrule
\end{tabular}
\label{simu-setting-D3-EE}
\end{table}
% E filed
\begin{table}[h]
\setlength\tabcolsep{8pt}
\centering
\caption{Dipole modes in D3 with electric (EE) boundaries.}
\medskip
\begin{tabular}{cccc}
\toprule
\textbf{E-f\mbox{}ield Amplitude} & \textbf{f}(GHz) & \textbf{R/Q}($\Omega$/cm$^2$)  & \textbf{Band}\\
\midrule
\includegraphics[width=0.44\textwidth]{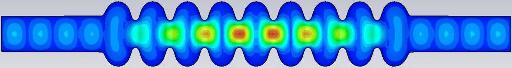}    & \raisebox{+1.8\height}{6.8238}    & \raisebox{+1.8\height}{0.011}  & \raisebox{+1.8\height}{D3-1}\\
\midrule
\includegraphics[width=0.44\textwidth]{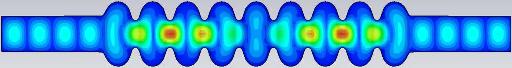}    & \raisebox{+1.8\height}{6.9003}    & \raisebox{+1.8\height}{0.035}  & \raisebox{+1.8\height}{D3-2}\\
\midrule
\includegraphics[width=0.44\textwidth]{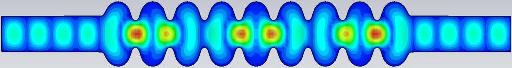}    & \raisebox{+1.8\height}{7.0027}    & \raisebox{+1.8\height}{0.058}  & \raisebox{+1.8\height}{D3-3}\\
\midrule
\includegraphics[width=0.44\textwidth]{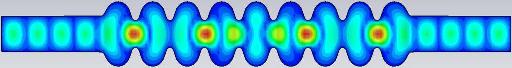}    & \raisebox{+1.8\height}{7.1225}    & \raisebox{+1.8\height}{0.189}  & \raisebox{+1.8\height}{D3-4}\\
\midrule
\includegraphics[width=0.44\textwidth]{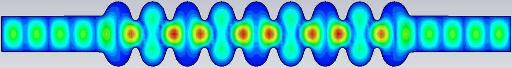}    & \raisebox{+1.8\height}{7.2541}    & \raisebox{+1.8\height}{0.549}  & \raisebox{+1.8\height}{D3-5}\\
\midrule
\includegraphics[width=0.44\textwidth]{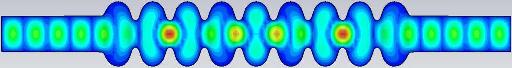}    & \raisebox{+1.8\height}{7.3833}    & \raisebox{+1.8\height}{0.014}  & \raisebox{+1.8\height}{D3-6}\\
\midrule
\includegraphics[width=0.44\textwidth]{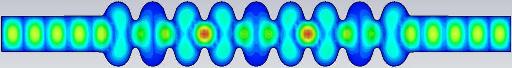}    & \raisebox{+1.8\height}{7.4889}    & \raisebox{+1.8\height}{0.455}  & \raisebox{+1.8\height}{D3-7}\\
\midrule
\includegraphics[width=0.44\textwidth]{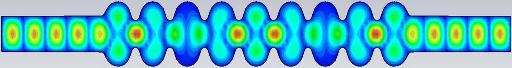}    & \raisebox{+1.8\height}{7.5621}    & \raisebox{+1.8\height}{0.269}  & \raisebox{+1.8\height}{D3-8}\\
\midrule
\includegraphics[width=0.44\textwidth]{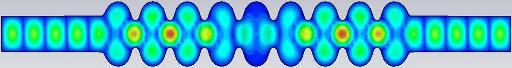}    & \raisebox{+1.8\height}{7.6196}    & \raisebox{+1.8\height}{1.354}  & \raisebox{+1.8\height}{D3-9}\\
\midrule
\includegraphics[width=0.44\textwidth]{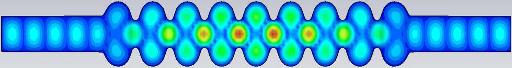}    & \raisebox{+1.8\height}{7.6680}    & \raisebox{+1.8\height}{28.926}  & \raisebox{+1.8\height}{D3-10}\\
\bottomrule
\end{tabular}
\label{simu-table-D3-EE}
\end{table}
\newpage

% DBP5 (EE)
\subsection{DBP5 (EE)}
% parameter
\begin{table}[h]
\setlength\tabcolsep{8pt}
\centering
\caption{Parameters settings for DBP5 with electric (EE) boundaries. }
\medskip
\begin{tabular}{m{2.3cm}m{2.2cm}m{1.2cm}m{2.1cm}m{2cm}m{1.4cm}}
\toprule
\textbf{Frequency range} (GHz) & \textbf{Lines per wavelength} & \textbf{Mesh type} & \textbf{Number of Mesh cells} & \textbf{Max mesh step} (mm) & \textbf{Band} \\
\midrule
8.2-8.4  & 20  & FPBA & 1,100,000  & 0.94  & DBP5 \\
\bottomrule
\end{tabular}
\label{simu-setting-DBP5-EE}
\end{table}
% E filed
\begin{table}[h]
\setlength\tabcolsep{8pt}
\centering
\caption{Dipole modes in DBP5 with electric (EE) boundaries.}
\medskip
\begin{tabular}{cccc}
\toprule
\textbf{E-f\mbox{}ield Amplitude} & \textbf{f}(GHz) & \textbf{R/Q}($\Omega$/cm$^2$)  & \textbf{Band}\\
\midrule
\includegraphics[width=0.45\textwidth]{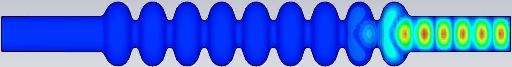}    & \raisebox{+1.8\height}{8.3033}    & \raisebox{+1.8\height}{1.543}  & \raisebox{+1.8\height}{DBP5-1}\\
\midrule
\includegraphics[width=0.45\textwidth]{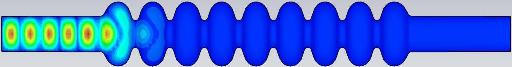}    & \raisebox{+1.8\height}{8.3039}    & \raisebox{+1.8\height}{1.537}  & \raisebox{+1.8\height}{DBP5-2}\\
\bottomrule
\end{tabular}
\label{simu-table-DBP5-EE}
\end{table}
\newpage

% D4 (EE)
\subsection{D4 (EE)}
% parameter
\begin{table}[h]
\setlength\tabcolsep{8pt}
\centering
\caption{Parameters settings for D4 with electric (EE) boundaries. }
\medskip
\begin{tabular}{m{2.3cm}m{2.2cm}m{1.2cm}m{2.1cm}m{2cm}m{1.5cm}}
\toprule
\textbf{Frequency range} (GHz) & \textbf{Lines per wavelength} & \textbf{Mesh type} & \textbf{Number of Mesh cells} & \textbf{Max mesh step} (mm) & \textbf{Band} \\
\midrule
8.45-8.55  & 25  & FPBA & 2,000,000  & 0.87  & D4(1) \\
\midrule
8.40-9.05  & 20  & FPBA & 1,300,000  & 0.91  & D4(2-11) \\
\bottomrule
\end{tabular}
\label{simu-setting-D4-EE}
\end{table}
% E filed
\begin{table}[h]
\setlength\tabcolsep{8pt}
\centering
\caption{Dipole modes in D4 with electric (EE) boundaries.}
\medskip
\begin{tabular}{cccc}
\toprule
\textbf{E-f\mbox{}ield Amplitude} & \textbf{f}(GHz) & \textbf{R/Q}($\Omega$/cm$^2$)  & \textbf{Band}\\
\midrule
\includegraphics[width=0.38\textwidth]{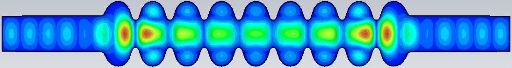}    & \raisebox{+1.5\height}{8.5002}    & \raisebox{+1.5\height}{0.130}  & \raisebox{+1.5\height}{D4-1}\\
\midrule
\includegraphics[width=0.38\textwidth]{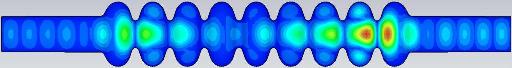}    & \raisebox{+1.5\height}{8.5042}    & \raisebox{+1.5\height}{0.096}  & \raisebox{+1.5\height}{D4-2}\\
\midrule
\includegraphics[width=0.38\textwidth]{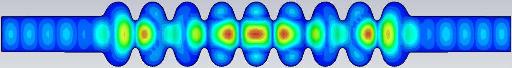}    & \raisebox{+1.5\height}{8.5322}    & \raisebox{+1.5\height}{0.152}  & \raisebox{+1.5\height}{D4-3}\\
\midrule
\includegraphics[width=0.38\textwidth]{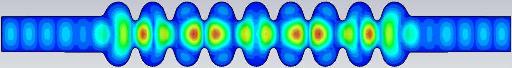}    & \raisebox{+1.5\height}{8.5763}    & \raisebox{+1.5\height}{0.115}  & \raisebox{+1.5\height}{D4-4}\\
\midrule
\includegraphics[width=0.38\textwidth]{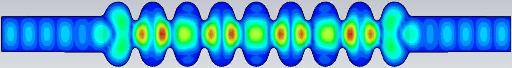}    & \raisebox{+1.5\height}{8.6397}    & \raisebox{+1.5\height}{0.415}  & \raisebox{+1.5\height}{D4-5}\\
\midrule
\includegraphics[width=0.38\textwidth]{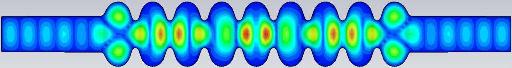}    & \raisebox{+1.5\height}{8.7205}    & \raisebox{+1.5\height}{1.038}  & \raisebox{+1.5\height}{D4-6}\\
\midrule
\includegraphics[width=0.38\textwidth]{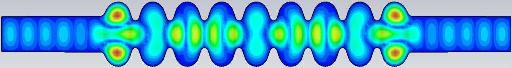}    & \raisebox{+1.5\height}{8.8033}    & \raisebox{+1.5\height}{10.205}  & \raisebox{+1.5\height}{D4-7}\\
\bottomrule
\end{tabular}
\label{simu-table-D4-EE}
\end{table}
\newpage

% D5 (EE)
\subsection{D5 (EE)}
% parameter
\begin{table}[h]
\setlength\tabcolsep{8pt}
\centering
\caption{Parameters settings for D5 with electric (EE) boundaries. }
\medskip
\begin{tabular}{m{2.3cm}m{2.2cm}m{1.2cm}m{2.1cm}m{2cm}m{1.4cm}}
\toprule
\textbf{Frequency range} (GHz) & \textbf{Lines per wavelength} & \textbf{Mesh type} & \textbf{Number of Mesh cells} & \textbf{Max mesh step} (mm) & \textbf{Band} \\
\midrule
9.05-9.095  & 20  & PBA & 1,300,000  & 0.91  & D5 \\
\bottomrule
\end{tabular}
\label{simu-setting-D5-EE}
\end{table}
% E filed
\begin{table}[h]
\setlength\tabcolsep{8pt}
\centering
\caption{Dipole modes in D5 with electric (EE) boundaries.}
\medskip
\begin{tabular}{cccc}
\toprule
\textbf{E-f\mbox{}ield Amplitude} & \textbf{f}(GHz) & \textbf{R/Q}($\Omega$/cm$^2$)  & \textbf{Band}\\
\midrule
\includegraphics[width=0.45\textwidth]{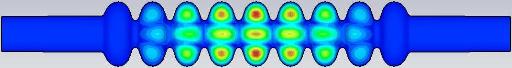}    & \raisebox{+1.8\height}{9.0523}    & \raisebox{+1.8\height}{0.002}  & \raisebox{+1.8\height}{D5-1}\\
\midrule
\includegraphics[width=0.45\textwidth]{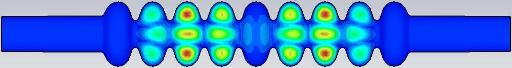}    & \raisebox{+1.8\height}{9.0530}    & \raisebox{+1.8\height}{0.053}  & \raisebox{+1.8\height}{D5-2}\\
\midrule
\includegraphics[width=0.45\textwidth]{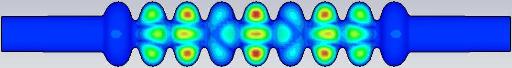}    & \raisebox{+1.8\height}{9.0546}    & \raisebox{+1.8\height}{0.058}  & \raisebox{+1.8\height}{D5-3}\\
\midrule
\includegraphics[width=0.45\textwidth]{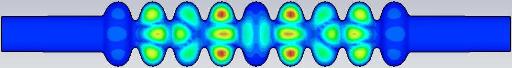}    & \raisebox{+1.8\height}{9.0581}    & \raisebox{+1.8\height}{2.171}  & \raisebox{+1.8\height}{D5-4}\\
\midrule
\includegraphics[width=0.45\textwidth]{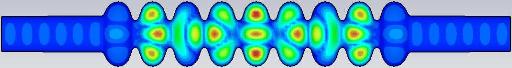}    & \raisebox{+1.8\height}{9.0664}    & \raisebox{+1.8\height}{4.116}  & \raisebox{+1.8\height}{D5-5}\\
\midrule
\includegraphics[width=0.45\textwidth]{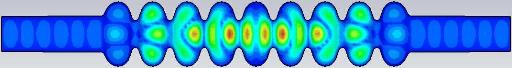}    & \raisebox{+1.8\height}{9.0890}    & \raisebox{+1.8\height}{0.580}  & \raisebox{+1.8\height}{D5-6}\\
\bottomrule
\end{tabular}
\label{simu-table-D5-EE}
\end{table}
\newpage

% DBP6, DBP7, DBP8 and DBP9 (EE)
\subsection{DBP6, DBP7, DBP8 and DBP9 (EE)}
% parameter
\begin{table}[h]
\setlength\tabcolsep{8pt}
\centering
\caption{Parameters settings for DBP6, DBP7, DBP8 and DBP9 with electric (EE) boundaries. }
\medskip
\begin{tabular}{m{2.3cm}m{2.2cm}m{1.2cm}m{2.1cm}m{2cm}m{2.3cm}}
\toprule
\textbf{Frequency range} (GHz) & \textbf{Lines per wavelength} & \textbf{Mesh type} & \textbf{Number of Mesh cells} & \textbf{Max mesh step} (mm) & \textbf{Band} \\
\midrule
9.1-9.2  & 20  & FPBA & 1,300,000  & 0.91  & DBP6, DBP7 \\
\midrule
9.2-9.4  & 20  & FPBA & 1,400,000  & 0.91  & DBP8 \\
\midrule
9.4-9.6  & 20  & FPBA & 1,500,000  & 0.87  & DBP9 \\
\bottomrule
\end{tabular}
\label{simu-setting-DBP6-9-EE}
\end{table}
% E filed
\begin{table}[h]
\setlength\tabcolsep{8pt}
\centering
\caption{Dipole modes in DBP6, DBP7, DBP8 and DBP9 with electric (EE) boundaries.}
\medskip
\begin{tabular}{cccc}
\toprule
\textbf{E-f\mbox{}ield Amplitude} & \textbf{f}(GHz) & \textbf{R/Q}($\Omega$/cm$^2$)  & \textbf{Band}\\
\midrule
% DBP6
\includegraphics[width=0.45\textwidth]{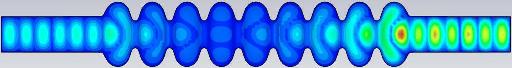}    & \raisebox{+1.8\height}{9.1666}    & \raisebox{+1.8\height}{1.291}  & \raisebox{+1.8\height}{DBP6-1}\\
\midrule
\includegraphics[width=0.45\textwidth]{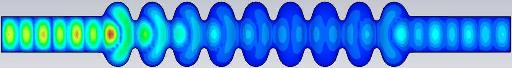}    & \raisebox{+1.8\height}{9.1678}    & \raisebox{+1.8\height}{1.898}  & \raisebox{+1.8\height}{DBP6-2}\\
\midrule
% DBP7
\includegraphics[width=0.45\textwidth]{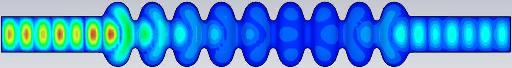}    & \raisebox{+1.8\height}{9.1749}    & \raisebox{+1.8\height}{1.123}  & \raisebox{+1.8\height}{DBP7-1}\\
\midrule
\includegraphics[width=0.45\textwidth]{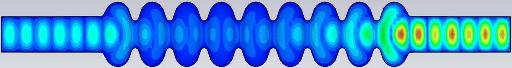}    & \raisebox{+1.8\height}{9.1763}    & \raisebox{+1.8\height}{3.240}  & \raisebox{+1.8\height}{DBP7-2}\\
\midrule
% DBP8
\includegraphics[width=0.45\textwidth]{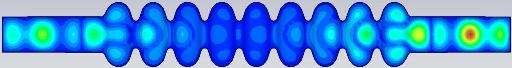}    & \raisebox{+1.8\height}{9.3283}    & \raisebox{+1.8\height}{0.880}  & \raisebox{+1.8\height}{DBP8-1}\\
\midrule
\includegraphics[width=0.45\textwidth]{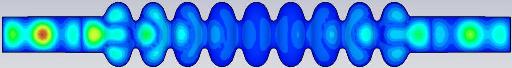}    & \raisebox{+1.8\height}{9.3284}    & \raisebox{+1.8\height}{2.042}  & \raisebox{+1.8\height}{DBP8-2}\\
\midrule
% DBP9
\includegraphics[width=0.45\textwidth]{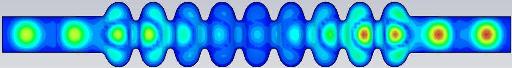}    & \raisebox{+1.8\height}{9.5379}    & \raisebox{+1.8\height}{4.064}  & \raisebox{+1.8\height}{DBP9-1}\\
\midrule
\includegraphics[width=0.45\textwidth]{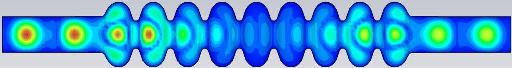}    & \raisebox{+1.8\height}{9.5385}    & \raisebox{+1.8\height}{0.275}  & \raisebox{+1.8\height}{DBP9-2}\\
\bottomrule
\end{tabular}
\label{simu-table-DBP6-9-EE}
\end{table}
\newpage

% D6 (EE)
\subsection{D6 (EE)}
% parameter
\begin{table}[h]
\setlength\tabcolsep{8pt}
\centering
\caption{Parameters settings for D6 with electric (EE) boundaries. }
\medskip
\begin{tabular}{m{2.3cm}m{2.2cm}m{1.2cm}m{2.1cm}m{2cm}m{1.4cm}}
\toprule
\textbf{Frequency range} (GHz) & \textbf{Lines per wavelength} & \textbf{Mesh type} & \textbf{Number of Mesh cells} & \textbf{Max mesh step} (mm) & \textbf{Band} \\
\midrule
9.6-9.9  & 20  & FPBA & 1,700,000  & 0.87  & D6 \\
\bottomrule
\end{tabular}
\label{simu-setting-D6-EE}
\end{table}
% E filed
\begin{table}[h]
\setlength\tabcolsep{8pt}
\centering
\caption{Dipole modes in D6 with electric (EE) boundaries.}
\medskip
\begin{tabular}{cccc}
\toprule
\textbf{E-f\mbox{}ield Amplitude} & \textbf{f}(GHz) & \textbf{R/Q}($\Omega$/cm$^2$)  & \textbf{Band}\\
\midrule
\includegraphics[width=0.45\textwidth]{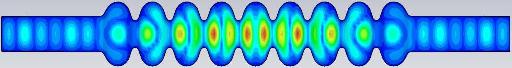}    & \raisebox{+1.8\height}{9.6962}    & \raisebox{+1.8\height}{0.001}  & \raisebox{+1.8\height}{D6-1}\\
\midrule
\includegraphics[width=0.45\textwidth]{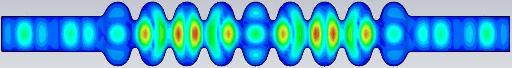}    & \raisebox{+1.8\height}{9.7142}    & \raisebox{+1.8\height}{0.009}  & \raisebox{+1.8\height}{D6-2}\\
\midrule
\includegraphics[width=0.45\textwidth]{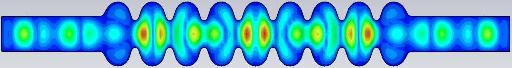}    & \raisebox{+1.8\height}{9.7421}    & \raisebox{+1.8\height}{0.074}  & \raisebox{+1.8\height}{D6-3}\\
\midrule
\includegraphics[width=0.45\textwidth]{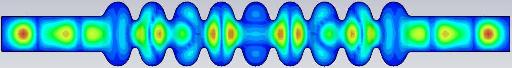}    & \raisebox{+1.8\height}{9.7711}    & \raisebox{+1.8\height}{0.379}  & \raisebox{+1.8\height}{D6-4}\\
\midrule
\includegraphics[width=0.45\textwidth]{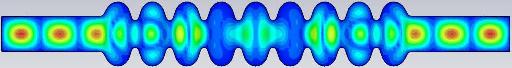}    & \raisebox{+1.8\height}{9.7896}    & \raisebox{+1.8\height}{5.951}  & \raisebox{+1.8\height}{D6-5}\\
\midrule
\includegraphics[width=0.45\textwidth]{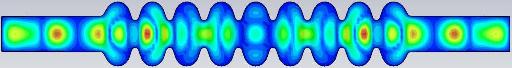}    & \raisebox{+1.8\height}{9.8027}    & \raisebox{+1.8\height}{0.771}  & \raisebox{+1.8\height}{D6-6}\\
\midrule
\includegraphics[width=0.45\textwidth]{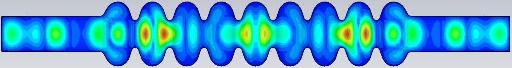}    & \raisebox{+1.8\height}{9.8265}    & \raisebox{+1.8\height}{0.191}  & \raisebox{+1.8\height}{D6-7}\\
%\midrule
%\includegraphics[width=0.45\textwidth]{E-abs-Xaxis-D6-8-mode8-EE}    & \raisebox{+1.8\height}{9.8532}    & \raisebox{+1.8\height}{0.014}  & \raisebox{+1.8\height}{D6-8}\\
%\midrule
%\includegraphics[width=0.45\textwidth]{E-abs-Xaxis-D6-9-mode9-EE}    & \raisebox{+1.8\height}{9.8732}    & \raisebox{+1.8\height}{0.004}  & \raisebox{+1.8\height}{D6-9}\\
\bottomrule
\end{tabular}
\label{simu-table-D6-EE}
\end{table}
\newpage

%%%%%%%%%%%%%%%  Section 5  %%%%%%%%%%%%%%%%%
%                                                                                                                %
%                                             Dipole (MM)                                                   %
%        																	%
%%%%%%%%%%%%%%%%%%%%%%%%%%%%%%%%%%%%%
\section{\label{sec:dipole-mm}Dipole (Magnetic Boundaries)}

% DBP1
\subsection{DBP1 (MM)}
% parameter
\begin{table}[h]
\setlength\tabcolsep{8pt}
\centering
\caption{Parameters settings for DBP1 with magnetic (MM) boundaries. }
\medskip
\begin{tabular}{m{2.3cm}m{2.2cm}m{1.2cm}m{2.1cm}m{2cm}m{1.4cm}}
\toprule
\textbf{Frequency range} (GHz) & \textbf{Lines per wavelength} & \textbf{Mesh type} & \textbf{Number of Mesh cells} & \textbf{Max mesh step} (mm) & \textbf{Band} \\
\midrule
4.1-4.2  & 70  & PBA & 2,400,000  & 1.04  & DBP1 \\
\bottomrule
\end{tabular}
\label{simu-setting-DBP1-MM}
\end{table}
% E filed
\begin{table}[h]
\setlength\tabcolsep{8pt}
\centering
\caption{Dipole modes in DBP1 with magnetic (MM) boundaries.}
\medskip
\begin{tabular}{cccc}
\toprule
\textbf{E-f\mbox{}ield Amplitude} & \textbf{f}(GHz) & \textbf{R/Q}($\Omega$/cm$^2$)  & \textbf{Band}\\
\midrule
\includegraphics[width=0.45\textwidth]{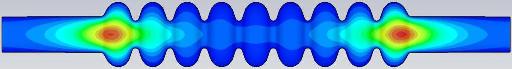}    & \raisebox{+1.8\height}{4.1474}    & \raisebox{+1.8\height}{0.241}  & \raisebox{+1.8\height}{DBP1-1}\\
\midrule
\includegraphics[width=0.45\textwidth]{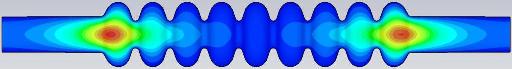}    & \raisebox{+1.8\height}{4.1475}    & \raisebox{+1.8\height}{1.299}  & \raisebox{+1.8\height}{DBP1-2}\\
\bottomrule
\end{tabular}
\label{simu-table-DBP1-MM}
\end{table}
\newpage

% D1 (MM)
\subsection{D1 (MM)}
% parameter
\begin{table}[h]
\setlength\tabcolsep{8pt}
\centering
\caption{Parameters settings for D1 with magnetic (MM) boundaries. }
\medskip
\begin{tabular}{m{2.3cm}m{2.2cm}m{1.2cm}m{2.1cm}m{2cm}m{1.4cm}}
\toprule
\textbf{Frequency range} (GHz) & \textbf{Lines per wavelength} & \textbf{Mesh type} & \textbf{Number of Mesh cells} & \textbf{Max mesh step} (mm) & \textbf{Band} \\
\midrule
4.25-5.05  & 55  & PBA & 2,100,000  & 1.10  & D1 \\
\bottomrule
\end{tabular}
\label{simu-setting-D1-MM}
\end{table}
% E filed
\begin{table}[h]
\setlength\tabcolsep{8pt}
\centering
\caption{Dipole modes in D1 with magnetic (MM) boundaries.}
\medskip
\begin{tabular}{cccc}
\toprule
\textbf{E-f\mbox{}ield Amplitude} & \textbf{f}(GHz) & \textbf{R/Q}($\Omega$/cm$^2$)  & \textbf{Band}\\
\midrule
\includegraphics[width=0.45\textwidth]{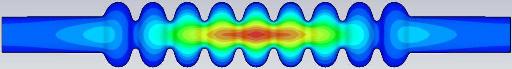}    & \raisebox{+1.8\height}{4.2979}    & \raisebox{+1.8\height}{0.001}  & \raisebox{+1.8\height}{D1-1}\\
\midrule
\includegraphics[width=0.45\textwidth]{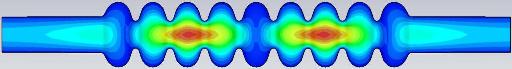}    & \raisebox{+1.8\height}{4.3592}    & \raisebox{+1.8\height}{0.263}  & \raisebox{+1.8\height}{D1-2}\\
\midrule
\includegraphics[width=0.45\textwidth]{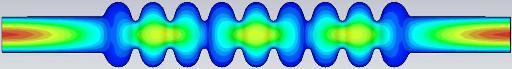}    & \raisebox{+1.8\height}{4.4306}    & \raisebox{+1.8\height}{0.072}  & \raisebox{+1.8\height}{D1-3}\\
\midrule
\includegraphics[width=0.45\textwidth]{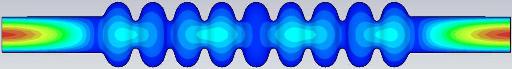}    & \raisebox{+1.8\height}{4.4516}    & \raisebox{+1.8\height}{0.000}  & \raisebox{+1.8\height}{D1-4}\\
\midrule
\includegraphics[width=0.45\textwidth]{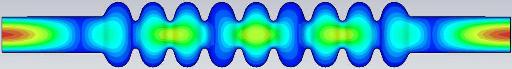}    & \raisebox{+1.8\height}{4.4770}    & \raisebox{+1.8\height}{0.327}  & \raisebox{+1.8\height}{D1-5}\\
\midrule
\includegraphics[width=0.45\textwidth]{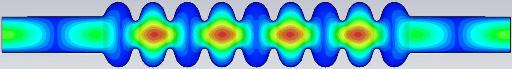}    & \raisebox{+1.8\height}{4.5703}    & \raisebox{+1.8\height}{1.213}  & \raisebox{+1.8\height}{D1-6}\\
\midrule
\includegraphics[width=0.45\textwidth]{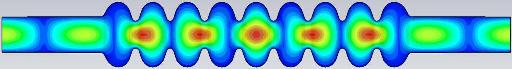}    & \raisebox{+1.8\height}{4.6804}    & \raisebox{+1.8\height}{1.586}  & \raisebox{+1.8\height}{D1-7}\\
\midrule
\includegraphics[width=0.45\textwidth]{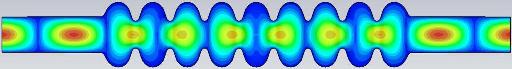}    & \raisebox{+1.8\height}{4.7749}    & \raisebox{+1.8\height}{27.165}  & \raisebox{+1.8\height}{D1-8}\\
\midrule
\includegraphics[width=0.45\textwidth]{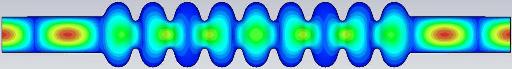}    & \raisebox{+1.8\height}{4.8455}    & \raisebox{+1.8\height}{32.124}  & \raisebox{+1.8\height}{D1-9}\\
\midrule
\includegraphics[width=0.45\textwidth]{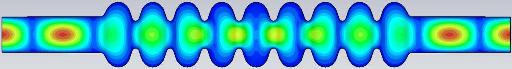}    & \raisebox{+1.8\height}{4.9162}    & \raisebox{+1.8\height}{21.833}  & \raisebox{+1.8\height}{D1-10}\\
\bottomrule
\end{tabular}
\label{simu-table-D1-MM}
\end{table}
\newpage

% DBP2 and D2 (MM)
\subsection{DBP2 and D2 (MM)}
% parameter
\begin{table}[h]
\setlength\tabcolsep{8pt}
\centering
\caption{Parameters settings for DBP2 and D2 with magnetic (MM) boundaries. }
\medskip
\begin{tabular}{m{2.3cm}m{2.2cm}m{1.2cm}m{2.1cm}m{2cm}m{1.4cm}}
\toprule
\textbf{Frequency range} (GHz) & \textbf{Lines per wavelength} & \textbf{Mesh type} & \textbf{Number of Mesh cells} & \textbf{Max mesh step} (mm) & \textbf{Band} \\
\midrule
4.25-5.05  & 55  & PBA & 2,100,000  & 1.10  & DBP2 \\
\midrule
5.3-5.7  & 50  & PBA & 2,300,000  & 1.07  & D2 \\
\bottomrule
\end{tabular}
\label{simu-setting-DBP2-D2-MM}
\end{table}
% E filed
\begin{table}[h]
\setlength\tabcolsep{8pt}
\centering
\caption{Dipole modes in DBP2 and D2 with magnetic (MM) boundaries.}
\medskip
\begin{tabular}{cccc}
\toprule
\textbf{E-f\mbox{}ield Amplitude} & \textbf{f}(GHz) & \textbf{R/Q}($\Omega$/cm$^2$)  & \textbf{Band}\\
\midrule
% DBP2
\includegraphics[width=0.44\textwidth]{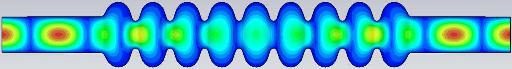}    & \raisebox{+1.8\height}{4.9945}    & \raisebox{+1.8\height}{7.376}  & \raisebox{+1.8\height}{DBP2-1}\\
\midrule
\includegraphics[width=0.44\textwidth]{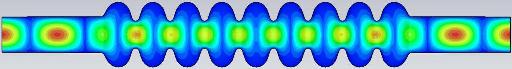}    & \raisebox{+1.8\height}{5.0233}    & \raisebox{+1.8\height}{3.844}  & \raisebox{+1.8\height}{DBP2-2}\\
\midrule
% D2
\includegraphics[width=0.44\textwidth]{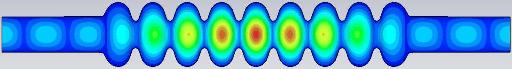}    & \raisebox{+1.8\height}{5.3518}    & \raisebox{+1.8\height}{0.055}  & \raisebox{+1.8\height}{D2-1}\\
\midrule
\includegraphics[width=0.44\textwidth]{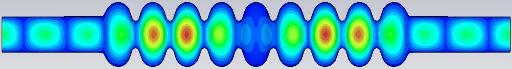}    & \raisebox{+1.8\height}{5.3923}    & \raisebox{+1.8\height}{2.114}  & \raisebox{+1.8\height}{D2-2}\\
\midrule
\includegraphics[width=0.44\textwidth]{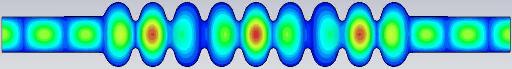}    & \raisebox{+1.8\height}{5.4272}    & \raisebox{+1.8\height}{10.770}  & \raisebox{+1.8\height}{D2-3}\\
\midrule
\includegraphics[width=0.44\textwidth]{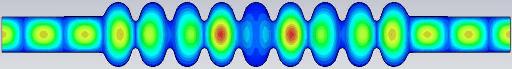}    & \raisebox{+1.8\height}{5.4528}    & \raisebox{+1.8\height}{17.024}  & \raisebox{+1.8\height}{D2-4}\\
\midrule
\includegraphics[width=0.44\textwidth]{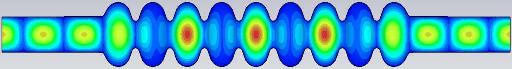}    & \raisebox{+1.8\height}{5.4711}    & \raisebox{+1.8\height}{9.368}  & \raisebox{+1.8\height}{D2-5}\\
\midrule
\includegraphics[width=0.44\textwidth]{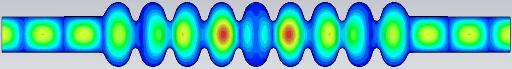}    & \raisebox{+1.8\height}{5.4834}    & \raisebox{+1.8\height}{0.409}  & \raisebox{+1.8\height}{D2-6}\\
\midrule
\includegraphics[width=0.44\textwidth]{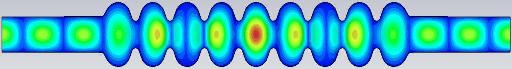}    & \raisebox{+1.8\height}{5.4908}    & \raisebox{+1.8\height}{0.343}  & \raisebox{+1.8\height}{D2-7}\\
\midrule
\includegraphics[width=0.44\textwidth]{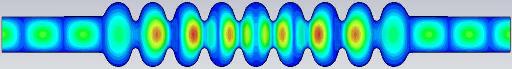}    & \raisebox{+1.8\height}{5.4944}    & \raisebox{+1.8\height}{0.033}  & \raisebox{+1.8\height}{D2-8}\\
\bottomrule
\end{tabular}
\label{simu-table-DBP2-D2-MM}
\end{table}
\newpage

% DBP3 and DBP4 (MM)
\subsection{DBP3 and DBP4 (MM)}
% parameter
\begin{table}[h]
\setlength\tabcolsep{8pt}
\centering
\caption{Parameters settings for DBP3 and DBP4 with magnetic (MM) boundaries. }
\medskip
\begin{tabular}{m{2.3cm}m{2.2cm}m{1.2cm}m{2.1cm}m{2cm}m{1.4cm}}
\toprule
\textbf{Frequency range} (GHz) & \textbf{Lines per wavelength} & \textbf{Mesh type} & \textbf{Number of Mesh cells} & \textbf{Max mesh step} (mm) & \textbf{Band} \\
\midrule
5.3-5.7  & 50  & PBA & 2,300,000  & 1.07  & DBP3 \\
\midrule
6.1-6.3  & 40  & PBA & 1,700,000  & 1.21  & DBP4 \\
\bottomrule
\end{tabular}
\label{simu-setting-DBP3-4-MM}
\end{table}
% E filed
\begin{table}[h]
\setlength\tabcolsep{8pt}
\centering
\caption{Dipole modes in DBP3 and DBP4 with magnetic (MM) boundaries.}
\medskip
\begin{tabular}{cccc}
\toprule
\textbf{E-f\mbox{}ield Amplitude} & \textbf{f}(GHz) & \textbf{R/Q}($\Omega$/cm$^2$)  & \textbf{Band}\\
\midrule
% DBP3
\includegraphics[width=0.45\textwidth]{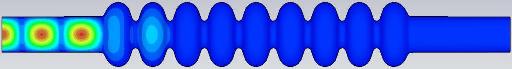}    & \raisebox{+1.8\height}{5.5532}    & \raisebox{+1.8\height}{2.994}  & \raisebox{+1.8\height}{DBP3-1}\\
\hline \\
\includegraphics[width=0.45\textwidth]{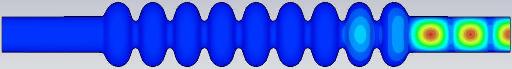}    & \raisebox{+1.8\height}{5.5532}    & \raisebox{+1.8\height}{2.995}  & \raisebox{+1.8\height}{DBP3-2}\\
\midrule
% DBP4
\includegraphics[width=0.45\textwidth]{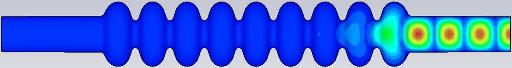}    & \raisebox{+1.8\height}{6.2123}    & \raisebox{+1.8\height}{0.595}  & \raisebox{+1.8\height}{DBP4-1}\\
\midrule
\includegraphics[width=0.45\textwidth]{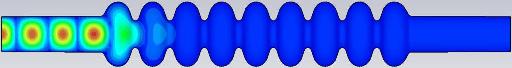}    & \raisebox{+1.8\height}{6.2144}    & \raisebox{+1.8\height}{0.636}  & \raisebox{+1.8\height}{DBP4-2}\\
\bottomrule
\end{tabular}
\label{simu-table-DBP3-4-MM}
\end{table}
\newpage

% D3 (MM)
\subsection{D3 (MM)}
% parameter
\begin{table}[h]
\setlength\tabcolsep{8pt}
\centering
\caption{Parameters settings for D3 with magnetic (MM) boundaries. }
\medskip
\begin{tabular}{m{2.3cm}m{2.2cm}m{1.2cm}m{2.1cm}m{2cm}m{1.4cm}}
\toprule
\textbf{Frequency range} (GHz) & \textbf{Lines per wavelength} & \textbf{Mesh type} & \textbf{Number of Mesh cells} & \textbf{Max mesh step} (mm) & \textbf{Band} \\
\midrule
6.75-7.50  & 40  & PBA & 3,000,000  & 0.99  & D3 \\
\bottomrule
\end{tabular}
\label{simu-setting-D3-MM}
\end{table}
% E filed
\begin{table}[h]
\setlength\tabcolsep{8pt}
\centering
\caption{Dipole modes in D3 with magnetic (MM) boundaries.}
\medskip
%
\label{simu-table-D3-MM}
\end{table}
\newpage

% DBP5 and DBP6 (MM)
\subsection{DBP5 and DBP6 (MM)}
% parameter
\begin{table}[h]
\setlength\tabcolsep{8pt}
\centering
\caption{Parameters settings for DBP5 and DBP6 with magnetic (MM) boundaries. }
\medskip
\begin{tabular}{m{2.3cm}m{2.2cm}m{1.2cm}m{2.1cm}m{2cm}m{1.4cm}}
\toprule
\textbf{Frequency range} (GHz) & \textbf{Lines per wavelength} & \textbf{Mesh type} & \textbf{Number of Mesh cells} & \textbf{Max mesh step} (mm) & \textbf{Band} \\
\midrule
7.85-7.95  & 22  & PBA & 1,200,000  & 0.94  & DBP5 \\
\midrule
8.4-9.0  & 20  & FPBA & 1,300,000  & 0.91  & DBP6\\
\bottomrule
\end{tabular}
\label{simu-setting-DBP5-6-MM}
\end{table}
% E filed
\begin{table}[h]
\setlength\tabcolsep{8pt}
\centering
\caption{Dipole modes in DBP5 and DBP6 with magnetic (MM) boundaries.}
\medskip
%
\label{simu-table-DBP5-6-MM}
\end{table}
\newpage

% D4 (MM)
\subsection{D4 (MM)}
% parameter
\begin{table}[h]
\setlength\tabcolsep{8pt}
\centering
\caption{Parameters settings for D4 with magnetic (MM) boundaries. }
\medskip
%
\label{simu-setting-D4-MM}
\end{table}
% E filed
\begin{table}[h]
\setlength\tabcolsep{8pt}
\centering
\caption{Dipole modes in D4 with magnetic (MM) boundaries.}
\medskip
%
\label{simu-table-D4-MM}
\end{table}
\newpage

% D5 (MM)
\subsection{D5 (MM)}
% parameter
\begin{table}[h]
\setlength\tabcolsep{8pt}
\centering
\caption{Parameters settings for D5 with magnetic (MM) boundaries. }
\medskip
%
\label{simu-table-D5-MM}
\end{table}
\newpage

% DBP7, DBP8 and DBP9 (MM)
\subsection{DBP7, DBP8 and DBP9 (MM)}
% parameter
\begin{table}[h]
\setlength\tabcolsep{8pt}
\centering
\caption{Parameters settings for DBP7, DBP8 and DBP9 with magnetic (MM) boundaries. }
\medskip
%
\label{simu-setting-DBP7-9-MM}
\end{table}
% E filed
\begin{table}[h]
\setlength\tabcolsep{8pt}
\centering
\caption{Dipole modes in DBP7, DBP8 and DBP9 with magnetic (MM) boundaries.}
\medskip
%
\label{simu-table-DBP7-9-MM}
\end{table}
\newpage

% DBP10 and D6 (MM)
\subsection{DBP10 and D6 (MM)}
% parameter
\begin{table}[h]
\setlength\tabcolsep{8pt}
\centering
\caption{Parameters settings for DBP10 and D6 with magnetic (MM) boundaries. }
\medskip
%
\label{simu-setting-D6-MM}
\end{table}
% E filed
\begin{table}[h]
\setlength\tabcolsep{8pt}
\centering
\caption{Dipole modes in DBP10 and D6 with magnetic (MM) boundaries.}
\medskip
%
\label{simu-table-DBP10-D6-MM}
\end{table}
\newpage

%%%%%%%%%%%%%%%  Section 6  %%%%%%%%%%%%%%%%%
%                                                                                                                %
%                                        Quadrupole (EE)                                                 %
%        																	%
%%%%%%%%%%%%%%%%%%%%%%%%%%%%%%%%%%%%%
\section{\label{sec:quad-ee}Quadrupole (Electric Boundaries)}

% QBP1 and Q1 (EE)
\subsection{QBP1 and Q1 (EE)}
% parameter
\begin{table}[h]
\setlength\tabcolsep{8pt}
\centering
\caption{Parameters settings for QBP1 and Q1 with electric (EE) boundaries. }
\medskip
%
\label{simu-setting-QBP1-Q1-EE}
\end{table}
% E filed
\begin{table}[h]
\setlength\tabcolsep{8pt}
\centering
\caption{Quadrupole modes in QBP1 and Q1 with electric (EE) boundaries.}
\medskip
%
\label{simu-table-QBP1-Q1-EE}
\end{table}
\newpage

% Q2 (EE)
\subsection{Q2 (EE)}
% parameter
\begin{table}[h]
\setlength\tabcolsep{8pt}
\centering
\caption{Parameters settings for Q2 with electric (EE) boundaries. }
\medskip
%
\label{simu-setting-Q2-EE}
\end{table}
% E filed
\begin{table}[h]
\setlength\tabcolsep{8pt}
\centering
\caption{Quadrupole modes in Q2 with electric (EE) boundaries.}
\medskip
%
\label{simu-table-Q2-EE}
\end{table}
\newpage

% QBP2, QBP3, QBP4 and QBP5 (EE)
\subsection{QBP2, QBP3, QBP4 and QBP5 (EE)}
% parameter
\begin{table}[h]
\setlength\tabcolsep{8pt}
\centering
\caption{Parameters settings for QBP2, QBP3, QBP4 and QBP5 with electric (EE) boundaries. }
\medskip
%
\label{simu-setting-QBP2-5-EE}
\end{table}
% E filed
\begin{table}[h]
\setlength\tabcolsep{8pt}
\centering
\caption{Quadrupole modes in QBP2, QBP3, QBP4 and QBP5 with electric (EE) boundaries.}
\medskip
%
\label{simu-table-QBP2-5-EE}
\end{table}
\newpage

% Q3 and QBP6 (EE)
\subsection{Q3 and QBP6 (EE)}
% parameter
\begin{table}[h]
\setlength\tabcolsep{8pt}
\centering
\caption{Parameters settings for Q3 and QBP6 with electric (EE) boundaries. }
\medskip
%
\label{simu-setting-Q3-QBP6-EE}
\end{table}
% E filed
\begin{table}[h]
\setlength\tabcolsep{8pt}
\centering
\caption{Quadrupole modes in Q3 and QBP6 with electric (EE) boundaries.}
\medskip
%
\label{simu-table-Q3-QBP6-EE}
\end{table}
\newpage

%%%%%%%%%%%%%%%  Section 7  %%%%%%%%%%%%%%%%%
%                                                                                                                %
%                                        Quadrupole (MM)                                                 %
%        																	%
%%%%%%%%%%%%%%%%%%%%%%%%%%%%%%%%%%%%%
\section{\label{sec:quad-mm}Quadrupole (Magnetic Boundaries)}

% QBP1 and Q1 (MM)
\subsection{QBP1 and Q1 (MM)}
% parameter
\begin{table}[h]
\setlength\tabcolsep{8pt}
\centering
\caption{Parameters settings for QBP1 and Q1 with magnetic (MM) boundaries. }
\medskip
%
\label{simu-setting-QBP1-Q1-MM}
\end{table}
% E filed
\begin{table}[h]
\setlength\tabcolsep{8pt}
\centering
\caption{Quadrupole modes in QBP1 and Q1 with magnetic (MM) boundaries.}
\medskip
%
\label{simu-table-QBP1-Q1-MM}
\end{table}
\newpage

% Q2 (MM)
\subsection{Q2 (MM)}
% parameter
\begin{table}[h]
\setlength\tabcolsep{8pt}
\centering
\caption{Parameters settings for Q2 with magnetic (MM) boundaries. }
\medskip
%
\label{simu-setting-Q2-MM}
\end{table}
% E filed
\begin{table}[h]
\setlength\tabcolsep{8pt}
\centering
\caption{Quadrupole modes in Q2 with magnetic (MM) boundaries.}
\medskip
%
\label{simu-table-Q2-MM}
\end{table}
\newpage

% QBP2, QBP3, QBP4 and QBP5 (MM)
\subsection{QBP2, QBP3, QBP4, QBP5 and QBP6 (MM)}
% parameter
\begin{table}[h]
\setlength\tabcolsep{8pt}
\centering
\caption{Parameters settings for QBP2, QBP3, QBP4, QBP5 and QBP6 with magnetic (MM) boundaries. }
\medskip
%
\label{simu-setting-QBP2-6-MM}
\end{table}
% E filed
\begin{table}[h]
\setlength\tabcolsep{8pt}
\centering
\caption{Quadrupole modes in QBP2, QBP3, QBP4, QBP5 and QBP6 with magnetic (MM) boundaries.}
\medskip
%
\label{simu-table-QBP2-6-MM}
\end{table}
\newpage

% Q3 (MM)
\subsection{Q3 (MM)}
% parameter
\begin{table}[h]
\setlength\tabcolsep{8pt}
\centering
\caption{Parameters settings for Q3 with magnetic (MM) boundaries. }
\medskip
%
\label{simu-setting-Q3-MM}
\end{table}
% E filed
\begin{table}[h]
\setlength\tabcolsep{8pt}
\centering
\caption{Quadrupole modes in Q3 with magnetic (MM) boundaries.}
\medskip
%
\label{simu-table-Q3-MM}
\end{table}
\newpage

%%%%%%%%%%%%%%%  Section 8  %%%%%%%%%%%%%%%%%
%                                                                                                                %
%                                        Sextupole (EE)                                                 %
%        																	%
%%%%%%%%%%%%%%%%%%%%%%%%%%%%%%%%%%%%%
\section{\label{sec:sextu-ee}Sextupole (Electric Boundaries)}

% SBP1 and S1 (EE)
\subsection{SBP1 and S1 (EE)}
% parameter
\begin{table}[h]
\setlength\tabcolsep{8pt}
\centering
\caption{Parameters settings for SBP1 and S1 with electric (EE) boundaries.}
\medskip
%
\label{simu-setting-SBP1-S1-EE}
\end{table}
% E filed
\begin{table}[h]
\setlength\tabcolsep{8pt}
\centering
\caption{Quadrupole modes in SBP1 and S1 with electric (EE) boundaries.}
\medskip
%
\label{simu-table-SBP1-S1-EE}
\end{table}
\newpage

% SBP2 and S2 (EE)
\subsection{SBP2 and S2 (EE)}
% parameter
\begin{table}[h]
\setlength\tabcolsep{8pt}
\centering
\caption{Parameters settings for SBP2 and S2 with electric (EE) boundaries.}
\medskip
%
\label{simu-setting-SBP2-S2-EE}
\end{table}
% E filed
\begin{table}[h]
\setlength\tabcolsep{8pt}
\centering
\caption{Quadrupole modes in SBP2 and S2 with electric (EE) boundaries.}
\medskip
%
\label{simu-table-SBP2-S2-EE}
\end{table}
\newpage

%%%%%%%%%%%%%%%  Section 9  %%%%%%%%%%%%%%%%%
%                                                                                                                %
%                                        Sextupole (MM)                                                 %
%        																	%
%%%%%%%%%%%%%%%%%%%%%%%%%%%%%%%%%%%%%
\section{\label{sec:sextu-mm}Sextupole (Magnetic Boundaries)}

% SBP1 and S1 (MM)
\subsection{SBP1 and S1 (MM)}
% parameter
\begin{table}[h]
\setlength\tabcolsep{8pt}
\centering
\caption{Parameters settings for SBP1 and S1 with magnetic (MM) boundaries.}
\medskip
%
\label{simu-setting-SBP1-S1-MM}
\end{table}
% E filed
\begin{table}[h]
\setlength\tabcolsep{8pt}
\centering
\caption{Quadrupole modes in SBP1 and S1 with magnetic (MM) boundaries.}
\medskip
%
\label{simu-table-SBP1-S1-MM}
\end{table}
\newpage

% SBP2 and S2 (MM)
\subsection{SBP2 and S2 (MM)}
% parameter
\begin{table}[h]
\setlength\tabcolsep{8pt}
\centering
\caption{Parameters settings for SBP2 and S2 with magnetic (MM) boundaries.}
\medskip
%
\label{simu-setting-SBP2-S2-MM}
\end{table}
% E filed
\begin{table}[h]
\setlength\tabcolsep{8pt}
\centering
\caption{Quadrupole modes in SBP2 and S2 with magnetic (MM) boundaries.}
\medskip
%
\label{simu-table-SBP2-S2-MM}
\end{table}